\documentclass[journal,article,submit,pdftex,moreauthors]{Definitions/mdpi} 
\usepackage{physics}
\usepackage[normalem]{ulem}
\firstpage{1} 
\pubvolume{1}
\issuenum{1}
\articlenumber{0}
\pubyear{2025}
\copyrightyear{2025}
\datereceived{ } 
\daterevised{ } 
\dateaccepted{ } 
\datepublished{ } 
\hreflink{https://doi.org/} 

\Title{Unified Description of Pseudoscalar Meson Structure from Light to Heavy Quarks}

\TitleCitation{Unified Description of Pseudoscalar Meson Structure from Light to Heavy Quarks}

\usepackage{ulem}

\Author{B. Almeida-Zamora$^{1,2}$\orcidA{}, L. Albino$^{1,2,3}$\orcidB{}, A. Bashir$^{3,4}$\orcidC{}, J.J. Cobos-Martínez$^{1}$\orcidD{}, J. Segovia$^{2*}$\orcidE{}}

\AuthorNames{B. Almeida-Zamora, L. Albino, A. Bashir, J.J. Cobos-Martínez and J. Segovia}

\AuthorCitation{B. Almeida-Zamora, L. Albino, A. Bashir, J.J. Cobos-Martínez and J. Segovia}

\address{%
$^{1}$ \quad Departamento de Física, Universidad de Sonora,
Boulevard Luis Encinas J. y Rosales, 83000, Hermosillo, Sonora, Mexico.\\
$^{2}$ \quad Departamento de Sistemas Físicos, Químicos y Naturales,
Universidad Pablo de Olavide, E-41013 Sevilla, Spain.\\
$^{3}$ \quad Departmento de Ciencias Integradas, Universidad de Huelva, E-21071 Huelva, Spain.\\
$^{4}$ \quad Instituto de Física y Matemáticas, Universidad Michoacana de San Nicolás de Hidalgo, Morelia, Michoacán 58040, México.
}

\corres{Correspondence: jsegovia@upo.es}

\abstract{We present a comprehensive review of the structure of pseudoscalar mesons within an algebraic model formulated in the light-front framework. The approach provides a unified description of leading-twist parton distribution amplitudes, light-front wave functions, generalized parton distributions, parton distribution functions, elastic electromagnetic form factors, charge radii, and impact-parameter GPDs, all derived consistently from the same underlying Bethe–Salpeter amplitudes. Results are discussed for light ($\pi$, $K$), heavy–light ($D$, $D_s$, $B$, $B_s$, $B_c$), and heavy–heavy ($\eta_c$, $\eta_b$) pseudoscalar mesons, allowing for a systematic analysis of the role played by quark-mass asymmetry and heavy-quark dynamics. Wherever possible, comparisons with experiments, lattice QCD, Dyson–Schwinger equation studies, contact-interaction models, effective quark models and holographic QCD are presented. Overall, the algebraic model offers a transparent and symmetry-preserving framework to explore the three-dimensional momentum and spatial structure of pseudoscalar mesons across all quark-mass regimes.}


\begin{document}

\section{Introduction}
The quest to understand the structure of hadrons—the strongly interacting bound states of quarks and gluons—remains one of the deepest challenges in modern physics. Quantum Chromodynamics (QCD), the fundamental theory of strong interactions, has been firmly established since the mid-1970s as part of the Standard Model \cite{Yang:1954ek,Fritzsch:1973pi}. As a non-Abelian gauge theory with local $SU(3)_c$ symmetry, QCD displays two hallmark phenomena: asymptotic freedom, whereby the interaction strength decreases at short distances, and confinement, the empirical fact that no color non-singlet combinations of quarks and gluons are ever observed as free particles \cite{Wilson:1974sk,Greensite:2011zz}. These features, combined with the emergence of dynamical chiral symmetry breaking (DCSB), make QCD a remarkably rich and complex theory, governing the bulk of visible matter in the universe 
\cite{Wilczek:2000it,Brodsky:2024zev}.

The discovery of asymptotic freedom in 1973 laid the foundation for a perturbative treatment of QCD at high energies \cite{Gross:1973id}. Yet, at low energies—the regime that controls the mass spectrum, form factors, and distribution functions of hadrons—perturbative techniques fail \cite{Politzer:1973fx}. The strong coupling grows large, the expansion in Feynman diagrams becomes meaningless, and purely non-perturbative phenomena of confinement and DCSB dominate the emergent infrared dynamics. This transition from the elegance of QCD at the perturbative level to the difficulty of extracting predictions in the non-perturbative domain has motivated the development of a wide range of theoretical and computational strategies \cite{Weinberg:1996kr}.

Lattice QCD, formulated in the late 1970s, provides a discretized path-integral approach to QCD that is systematically improvable and non-perturbative by construction \cite{Wilson:1983xri}. It has yielded impressive results for hadron spectroscopy, static properties, and certain partonic observables \cite{Gattringer:2010zz,FlavourLatticeAveragingGroup:2019iem}. However, the challenges persist in its direct access to real-time dynamics, light-front quantities, and the full kinematical range of parton distributions. The continuum approaches based upon fundamental field theoretic equations offer alternative approach which span all regions of phase space including the ones that remain inaccessible to lattice calculations \cite{Cloet:2013jya}.

These functional methods which are based on Dyson–Schwinger equations (DSEs) and Bethe–Salpeter equations (BSEs), have become a cornerstone of such continuum approaches \cite{Alkofer:2000wg,Maris:2003vk,Chang:2009zb}. The DSEs are an infinite tower of coupled integral equations for the Green’s functions of QCD, while the BSEs describe the relativistic bound states of quarks and gluons. When truncated in symmetry-preserving ways, these equations provide a powerful framework that connects the QCD Lagrangian to hadronic observables \cite{Maris:1999nt}. Importantly, they naturally incorporate phenomena such as DCSB and allow direct studies of quark and gluon dressing, propagators, and amplitudes \cite{Bender:1996bb,Bhagwat:2004hn}.

Yet, the complexity of solving the DSE–BSE system in full generality remains formidable. This is where effective models inspired by QCD become essential. Among them, the Nambu–Jona-Lasinio (NJL) model has long served as a paradigm for studying chiral symmetry breaking and meson phenomenology \cite{Nambu:1961tp}. However, its lack of confinement and reliance on a hard cutoff limit its applicability \cite{Hatsuda:1994pi}. Similarly, contact-interaction (CI) model provide a useful first approximation within the DSE–BSE framework, yielding tractable equations and symmetry-preserving solutions \cite{Gutierrez-Guerrero:2010waf}. Nevertheless, their momentum-independent interaction oversimplifies the dynamics, suppressing important features of hadron structure such as the correct asymptotic scaling of form factors \cite{Cloet:2008re}.

The algebraic model emerges as an intermediate step between these approaches and full QCD calculations. By incorporating analytic parametrizations of dressed quark propagators and Bethe–Salpeter amplitudes, the algebraic model captures essential aspects of confinement and DCSB while retaining analytic tractability. Its modern implementation is inspired by the Nakanishi representation of the Bethe–Salpeter amplitude~\cite{Nakanishi:1963zz,Nakanishi:1969ph}, which in turn enables a closed-form projection of the Bethe-Salpeter wave function onto the light-front. Unlike NJL or CI models, the algebraic model provides momentum-dependent structures and light-front wave functions (LFWFs) that have been employed in the consistent calculation of parton distribution amplitudes (PDAs)~\cite{Chang:2013pq,Chang:2013epa,Gao:2014bca}, parton distribution functions (PDFs)~\cite{Chang:2014lva,Chang:2014gga}, electromagnetic form factors (EFFs)~\cite{Higuera-Angulo:2024oui}, and generalized parton distributions (GPDs)~\cite{Mezrag:2014jka,Mezrag:2016hnp} within a single unified framework, with further developments and applications reported in Refs.~\cite{Shi:2015esa,Raya:2015gva,Chen:2016sno,Bedolla:2016yxq,Li:2016mah,Chouika:2017rzs,Zhang:2021mtn,Raya:2021zrz,Raya:2022eqa}. Compared with lattice QCD, the algebraic model cannot claim first-principles derivations; however, its analytic accessibility makes it an invaluable complement, especially for exploring correlations among different hadronic observables and for providing intuitive insight in regimes beyond the reach of current lattice simulations.

In this review, we present a comprehensive account of the algebraic model for meson structure, based on a series of recent works in which this framework has been applied to pseudoscalar mesons~\cite{Albino:2022gzs,Almeida-Zamora:2023bqb,Almeida-Zamora:2024mdc,Albino:2026hhx}, and subsequently extended to light vector mesons~\cite{Almeida-Zamora:2023rwg}. We begin by outlining its theoretical foundations and construction, emphasizing its relationship to the DSE–BSE formalism. We then survey its main applications to meson observables, including PDFs, EFFs, and GPDs. Finally, we discuss the strengths and limitations of the approach, and its potential role in the broader landscape of non-perturbative QCD and hadron phenomenology.

\section{The Algebraic Model: Formalism}

Within phenomenological descriptions of strongly interacting bound states in QCD, the algebraic model (AM) provides a framework in which meson observables can be computed analytically while preserving essential features of QCD dynamics. Its construction is motivated by solutions of Dyson--Schwinger and Bethe--Salpeter equations, which reveal the role of dynamical chiral symmetry breaking and confinement in shaping hadron properties \cite{Roberts:1994dr,Maris:2003vk,Eichmann:2009qa}. 

A central ingredient in any QCD-inspired framework is the quark propagator, as it encodes nonperturbative phenomena sucha as DCSB and exhibits features of confinement through its analytical structure. In the DSE formalism, the fully dressed quark propagator in Euclidean space is typically expressed as
\begin{equation}
S_{q}(p)^{-1} = i \gamma \cdot p A(p^2) + B(p^2) ,
\end{equation}
where $A(p^2)$ is the wave-function renormalization and $B(p^2)$ is the scalar self-energy related to the mass function $M(p^2) = B(p^2)/A(p^2)$. In realistic numerical DSE studies, $A(p^2)$ and $B(p^2)$ are obtained by solving the gap equation with model-dependent gluon interaction kernels. This procedure, while systematic, is computationally intensive and less suitable for closed-form analyses of hadronic observables.

In the AM, the analytic complexity of the quark propagator is effectively encoded in a parametrization that preserves the essential features of the propagator. A widely used form is
\begin{equation}
S_{q}(p) = \frac{-i  \gamma \cdot p + M_{q}}{p^2 + M_{q}^2} ,
\end{equation}
where $M_{q}$ is interpreted as a constituent-like mass scale dynamically generated by DCSB. Even though $M_{q}$ is treated as a parameter, it effectively encodes the dynamical mass that quarks acquire in the infrared due to strong interactions. Notably, it is not a current quark mass but rather a manifestation of DCSB. 

With this parametrization of the quark propagator in hand, one can proceed to the description of strongly interacting bound states within the algebraic model. In this context, mesons occupy a privileged position in hadron physics, since they are the simplest realization of color-singlet bound states in Quantum Chromodynamics. Their dual nature---as quark--antiquark composites and as emergent collective excitations of the QCD vacuum---makes them essential for understanding the interplay between confinement and DCSB.  

In the DSE--BSE formalism, mesons appear as poles in the quark--antiquark scattering matrix. The existence of these bound-state poles with factorizable residues leads to the homogeneous BSE, an eigenvalue equation whose solutions determine the bound-state masses. The corresponding eigenvectors define the Bethe-Salpeter amplitudes (BSAs), which encode the momentum and Dirac structure of the quark–antiquark correlation and thereby provide information about the internal structure of the meson. In terms of the quark–antiquark relative momentum $k$ and total bound-state momentum $P$, the homogeneous Bethe–Salpeter equation for a meson ${\mathrm{M}}$ reads~\cite{Salpeter:1951sz}
\begin{equation}
\Gamma_{\mathrm{M}}(k;P) = \int \frac{d^4q}{(2\pi)^4} \, K(k,q;P) \, \big[ S_{q}(q_+) \, \Gamma_{\mathrm{M}}(q;P) \, S_{\bar{h}}(q_-) \big] \,, \label{BSE - General definition} 
\end{equation}
where $\Gamma_{\mathrm{M}}$ is the meson's BSA, with Dirac indices ommited for simplicity, $S_{q(\bar{h})}$ are the fully dressed quark (antiquark) propagators carrying momenta $q_{\pm} = q \pm \eta_{\pm} P$, with $\eta_{+}-\eta_{-} = 1$, and $K$ is the quark--antiquark scattering kernel. In principle, solving this equation with kernels derived from QCD would yield the complete meson spectrum and structure. However, the nonperturbative complexity of QCD renders exact solutions impractical.  

To retain predictive power while preserving symmetries, one adopts controlled truncations of the DSE--BSE system. The most widely used is the rainbow--ladder truncation, which ensures that the axial-vector Ward--Takahashi identity is satisfied. As a consequence, pseudoscalar mesons, such as the pion and kaon, emerge as the Goldstone bosons associated with DCSB, being massless in the chiral limit while remaining as tightly bound quark--antiquark states \cite{Munczek:1994zz,Bender:1996bb}. 

A key technical advantage of the algebraic model is its reliance on the Nakanishi Integral Representation (NIR) \cite{Nakanishi:1963zz} to express the BSAs. Originally introduced in the context of perturbation theory for relativistic bound states, the NIR provides a way to cast the BSA in terms of an integral over a weight function, thereby simplifying the analytic continuation to light-front variables \cite{Nakanishi:1963zz}. This innovation allows the BSA to be written as
\begin{equation}
\Gamma_{\mathrm{M}}(k;P) = \int_{-1}^1 dz \int_0^\infty d\gamma \; \frac{\rho_\mathrm{M}(\gamma,z)}{\left[ k^2 + z \, k \cdot P + \tfrac{1}{4} P^2 + \gamma + M_q^2 \right]^\mathrm{n} } \, \mathcal{O}_\mathrm{M} \,, \label{BSA - NIR representation}
\end{equation}
where $\rho_\mathrm{M}(\gamma,z)$ is the Nakanishi weight function, $\mathcal{O}_\mathrm{M}$ the Dirac operator appropriate to the channel (\textit{e.g.} $\gamma_5$ for pseudoscalars, $\gamma_\mu$ for vectors), and $\mathrm{n}$ a positive integer specifying the power of the denominator in the representation. By adopting simple but physically motivated forms for $\rho_\mathrm{M}$, the BSE reduces to an algebraic problem, thereby enabling closed-form analytic solutions \cite{Chang:2013pq,Mezrag:2014jka}.  

In traditional approaches, distinct techniques are often required to extract PDAs, PDFs and GPDs, each involving nontrivial projections of the BSA onto the light-front. In contrast, the NIR allows these projections to be carried out in a unified and analytic fashion, since the entire LFWF can be reconstructed from the same weight function that defines the covariant BSA \cite{Carbonell:2010zw,Mezrag:2016hnp}, as we shall discuss in Sec. \ref{Subsection - Applications to Hadron Observables}.

At the very heart of the connection between the BSA and the LFWF lies the Bethe--Salpeter wave function (BSWF), which combines the dressed quark propagators with the BSA and provides the natural object for light-front projections. For a meson $\mathrm{M}$, the corresponding BSWF is defined as
\begin{equation}
\chi_\mathrm{M}(k;P) = S_q(k_+) \, \Gamma_{\mathrm{M}}(k;P) \, S_{\bar{h}}(k_-) \,, \label{BSWF - General definition}
\end{equation} 
where the momenta $k_{\pm}$ carried by the quark and antiquark propagators are defined consistently with Eq.~\ref{BSE - General definition}, namely $k_{\pm} = k \pm \eta_{\pm} P$, with $\eta_{+} + \eta_{-} = 1$.

In particular, the BSWF provides the natural starting point for light-front projections of meson partonic distributions. For instance, the leading-twist PDA $\varphi_\mathrm{M}(x)$ of a meson $\mathrm{M}$ is obtained through the light-front projection of the BSWF, i.e.
\begin{equation}
f_\mathrm{M} \, \varphi_\mathrm{M}(x) = \int \frac{d^4k}{(2\pi)^4} \, \delta\!\left( n \cdot k - x \, n \cdot P \right) \, 
\text{Tr} \left[ \gamma_5 \gamma \cdot n \, \chi_\mathrm{M}(k;P) \right] \,, \label{PDA - light-front projection of BSWF}
\end{equation}
where $n$ is a lightlike vector ($n^2=0$). In the algebraic model, the NIR yields a representation of the PDA projection in terms of integrals over the Nakanishi weight function, \emph{cf.} Eqs. (\ref{BSA - NIR representation}-\ref{PDA - light-front projection of BSWF}). With further physically motivated simplifications of $\rho_\mathrm{M}(\gamma,z)$, this representation enables the derivation of closed analytic forms for $\varphi_\mathrm{M}(x)$ that are consistent with both symmetry constraints and lattice QCD benchmarks \cite{Chang:2013pq,Mezrag:2016hnp}. Therefore, the algebraic model, supported by the NIR, serves as a bridge between the fully covariant Bethe--Salpeter formalism and the light-front description of mesons, providing a unified, symmetry-preserving, and analytically tractable framework that connects confinement, DCSB, and partonic structure across different kinematic regimes. The explicit realization of these simplifications will be discussed in the next subsection.

\subsection{Algebraic Ansatz for the Pion Bethe–Salpeter Amplitude}

Within the algebraic model, the BSA of the pion is represented through an analytic ansatz inspired by the NIR~\cite{Nakanishi:1963zz,Nakanishi:1969ph}. This parametrization preserves the essential features of chiral symmetry and dynamical mass generation, while providing analytic tractability in both Euclidean and light-front formulations. In its original implementation~\cite{Chang:2013pq,Chang:2013epa}, the algebraic ansatz employed for the quark (antiquark) propagator and pion BSA reads
\begin{eqnarray}
S_q(k) &=& \;(- i \gamma \cdot k + M_q) \Delta(k^2,M_q^2), \label{eq:propagator_AM} \\
n_{\mathrm{M}} \Gamma_\mathrm{M}(k,P) &=&\; i \gamma_5  \int_{-1}^1 \mathrm{d}z \rho_\mathrm{M}^\nu(z) \left[ \hat{\Delta}(k^2_{+z},M_q^2)\right]^\nu, \label{AM-BSA for pseudoscalar mesons}
\end{eqnarray}
where the mass scale $M_q$ is interpreted as a constituent quark mass, $n_\mathrm{M}$ is a normalization constant, and the denominators are expressed as  $\Delta(s,t) = \frac{1}{s + t}$ and $\hat{\Delta}(s,t) =t\Delta(s,t)$. Moreover, $k_{\pm z} =k - \left( \frac{1\mp z}{2} - \eta \right) P$, with $P^2=-m_\mathrm{M}^2$ and $\eta\in\left[0,1 \right]$. On the other hand, the function $\rho_\mathrm{M}^\nu(z)$ can be regarded as a spectral density, whose analytical structure has a crucial impact on the BSA and meson observables, and $\nu > -1$ is a model-parameter which controls the asymptotic behavior of the BSA.

As an illustrative example, the particular choice $2 \rho_\mathrm{M}^\nu(z) = \delta(1+z) + \delta(1-z)$ assigns equal probability to configurations in which either the valence quark carries all the meson’s momentum and the antiquark none, or vice versa, thereby corresponding to a point-like meson~\cite{Chang:2013pq}. Remarkably, when combined with $\eta = 0$, this spectral density yields $\Gamma_{\mathrm{M}}(k,P) \sim 1/k^2$. In contrast, the asymptotic meson's profile $\Gamma_{\mathrm{M}}(k,P) \sim 1/(k^2)^{\nu}$ is obtained if
\begin{align}
\rho_\mathrm{M}^\nu(z) =&\; \frac{1}{\sqrt{\pi}} \frac{\Gamma(\nu + \frac{3}{2})}{\Gamma(\nu + 1)} (1-z^2)^\nu \,.
\end{align}

The algebraic model has been used in modern Dyson-Schwinger and Bethe-Salpeter studies of mesons, see \textit{e.g.} Refs. \cite{Gao:2014bca,Chang:2014lva,Chang:2014gga, Mezrag:2014jka,Shi:2015esa,Raya:2015gva,Chen:2016sno,Bedolla:2016yxq,Li:2016mah,Chouika:2017rzs,Zhang:2021mtn,Raya:2021zrz,Raya:2022eqa}. Notably, in some of these works, \textit{e.g.} \cite{Zhang:2021mtn,Raya:2021zrz,Raya:2022eqa}, a mass-scale parameter $\Lambda$ is introduced in the BSA, replacing $M_q \rightarrow \Lambda$ in the denominator of Eq.~\eqref{AM-BSA for pseudoscalar mesons}. As a further step, and following Ref. \cite{Albino:2022gzs}, we have promoted $\Lambda \rightarrow \Lambda(w)$, with $w=z-1$, in order to incorporate and arbitrary yet convenient $z$-dependence. This allows us to rewrite the BSA, Eq.~\eqref{AM-BSA for pseudoscalar mesons}, as
\begin{equation}\label{eq:BSA_AM}
    n_\mathrm{M} \Gamma_\mathrm{M} (k,P) = i\gamma_5 \int_{-1}^{1} \mathrm{d}w \; \rho_\mathrm{M}^\nu(w) \left[ \hat{\Delta}(k^2_{w}, \Lambda_w^2)\right]^{\nu},
\end{equation}
where $k_w = k + \frac{w}{2} P$ and $\Lambda_w^2 \equiv \Lambda^2(w)$. Remarkably, for any admissible choice of the spectral density $\rho_\mathrm{M}^\nu(z)$, Poincaré covariance ensures that physical observables do not depend on the momentum partitioning parameter $\eta$. Accordingly, and for convenience, we have set $\eta=0$ in the above Eq.~\eqref{eq:BSA_AM}, which in turn allows us to cast the BSWF as
\begin{equation}\label{bswf1}
    \chi_\mathrm{M} (k_{-}, P) = S_q(k) \Gamma_\mathrm{M} (k_{-}, P) S_{\bar{h}}(p),
\end{equation}
with $p=k-P$ and $k_{-}=k - P/2$, where we have correspondingly set $\eta_{-}=\eta=0$, cf. Eq.~(\ref{BSWF - General definition}). Hence, combining Eqs. \eqref{eq:propagator_AM}, \eqref{eq:BSA_AM}, and \eqref{bswf1}, the BSWF becomes
\begin{eqnarray}
n_\mathrm{M} \chi(k_{-},P) = \; \mathcal{M}_{q,\bar{h}}(k,P)\int_{-1}^{1} \mathrm{d}w \; \mathcal{D}_{q,\bar{h}}^\nu(k,P) \Lambda_w^{2\nu} \rho_\mathrm{M}^\nu(w)
\,, 
\label{eq:AM_BSWF}
\end{eqnarray}
where $\mathcal{D}_{q,\bar{h}}^\nu(k,P)$ is a product of quadratic denominators and $\mathcal{M}_{q,\bar{h}}(k,P)$ contains a tensor structure that characterizes the Bethe-Salpeter wave function, i.e.
\begin{eqnarray}
\mathcal{D}_{q,\bar{h}}^\nu(k,P) &=& \Delta(k^2, M_q^2) \left[ \Delta(k^2_{w-1}, \Lambda_w^2)\right]^{\nu}\Delta(p^2, M_{\bar{h}}^2) \,, \label{eq:AM_denominator} \\
\mathcal{M}_{q,\bar{h}}(k,P) &=& -\gamma_5 \left[M_q\gamma\cdot P + \gamma\cdot k (M_{\bar{h}}^2 - M_q^2) \right.\nonumber \\
&& \left.+ \sigma_{\mu\nu}k_\mu P_\nu+ i ( k \cdot p -M_q M_{\bar{h}}) \right] \,. 
\end{eqnarray}
On the other hand, Feynman parametrization enables the denominators in Eq. \eqref{eq:AM_denominator} to be combined into a single one, yielding
\begin{align}
    \mathcal{D}_{q,\bar{h}}^\nu &= \frac{\Gamma(\nu + 2)}{\Gamma(\nu)} \int_0^1\mathrm{d}a\int_0^1\mathrm{d}b\int_0^1\mathrm{d}c \frac{a^{\nu-1} \delta(1-a-b-c)}{D^{\nu+2}},
\end{align}
where $a$, $b$ and $c$ are the corresponding Feynman parameters, and
\begin{equation}
    D = a(k_{w-1}^2 + \Lambda_w^2) + b (p^2 + M_h^2) + c(k^2+M_q^2).
\end{equation}
After performing algebraic manipulations, together with a suitable change of variables and rearrangements, we obtain
\begin{align}
    D=& (k - \alpha P)^2 + \Omega_\mathrm{M}^2(\alpha,\beta,w),
\end{align}
where we have defined the $\Omega_\mathrm{M}^2(\alpha,\beta,w)$ function as
\begin{align}\label{eq:AM_omega}
\Omega_\mathrm{M}^2(\alpha,\beta,w) =& \beta M_q^2 + (1-\beta)\Lambda_w^2 + \left[ \alpha - \frac{1}{2}(1-\beta)(1-w) \right] (M_h^2 - M_q^2) \nonumber \\ 
\;& - \left[ \alpha(\alpha-1) + \frac{1}{4} (1-\beta)(1-w^2)\right] P^2,
\end{align}
with $\beta = 1- a$ and $\alpha = b -  \frac{1}{2}a(w-1)$. Therefore, the product of quadratic denominators becomes
\begin{align}
    \mathcal{D}_{q,\bar{h}}^\nu &= \nu(\nu+1) \int_0^1\mathrm{d}\beta\int_{\frac{1}{2}(\beta-1)(w-1)}^{\frac{1}{2}\left[(w+1)\beta-(w-1)\right]} \mathrm{d}\alpha\;  \frac{(1-\beta)^{\nu-1}}{\left[(k-\alpha P)^2 + \Omega_\mathrm{M}^2(\alpha,\beta,w)\right]^{\nu+2} 
    } \,. \label{product of quadratic denominators}
\end{align}
Substituting the above Eq.~\eqref{product of quadratic denominators} into Eq.~\eqref{eq:AM_BSWF}, and performing a subsequent rearrangement in the order of integration between $\alpha$ and $w$, the BSWF can be written as
\begin{align}
n_\mathrm{M} \chi(k_{-},P)=& \;  \mathcal{M}_{q,\bar{h}}(k,P)\nu(\nu+1)\int_{0}^{1} \mathrm{d}\alpha \; \left[ \int_{-1}^{1-2\alpha}\mathrm{d}w \int_{\frac{2\alpha}{w-1}+1}^{1} \mathrm{d}\beta \right. \nonumber \\
&\left. + \int_{1-2\alpha}^1\mathrm{d}w \int_{\frac{2\alpha + w-1}{w+1}}^{1} \mathrm{d}\beta \right] \;  \frac{(1-\beta)^{\nu-1}\tilde{\rho}_\mathrm{M}^\nu (w)}{\left[(k-\alpha P)^2 + \Omega_\mathrm{M}^2(\alpha,\beta,w)\right]^{\nu+2}} \,, \label{BSWF - explicit beta dependence}
\end{align}
where $\tilde{\rho}_\mathrm{M}^\nu (w) = \Lambda_w^{2\nu}\rho_\mathrm{M}^\nu(w)$. From this point, the analytical calculation cannot be carried out further, since the integration over $\beta$ cannot be performed due to the simultaneous presence of two $\beta$-dependent terms, namely $(1-\beta)^\nu$ and $\Omega_\mathrm{M}^2(\alpha,\beta,w)$, Eq.~\eqref{eq:AM_omega}. In order to retain an analytically tractable formulation, it is necessary to eliminate the explicit $\beta$-dependence from the denominator of the above Eq.~\eqref{BSWF - explicit beta dependence}. To this end, we specify the a priori arbitrary functional form of $\Lambda_w^2$ as
\begin{equation}\label{eq:lambdaw}
    \Lambda_w^2=  M_q^2 - \frac{1}{4} (1 - w^2) m_{\mathrm{M}}^2 + \frac{1}{2} (1-w) (M_{\bar{h}}^2 - M_q^2),
\end{equation}
such that the function $\Omega_\mathrm{M}^2$, Eq.~\eqref{eq:AM_omega}, is now simplified to
\begin{align}\label{eq:AM_omega2}
\Omega_\mathrm{M}^2(\alpha) =& M_q^2 +\alpha(\alpha-1) m_{\mathrm{M}}^2 + \alpha (M_{\bar{h}}^2 - M_q^2) = \Lambda^2_{1-2\alpha} \,,
\end{align}
and the BSWF is thus expressed as
\begin{align}\label{eq:BSWF-3.119}
n_\mathrm{M} \chi(k_{-},P)=& \;  \mathcal{M}_{q,\bar{h}}(k,P)\nu(\nu+1)\int_{0}^{1} \mathrm{d}\alpha \; \left[ \int_{-1}^{1-2\alpha}\mathrm{d}w \int_{\frac{2\alpha}{w-1}+1}^{1} \mathrm{d}\beta \right. \nonumber \\
&\left. + \int_{1-2\alpha}^1\mathrm{d}w \int_{\frac{2\alpha + w-1}{w+1}}^{1} \mathrm{d}\beta \right] \;  \frac{(1-\beta)^{\nu-1}\tilde{\rho}_\mathrm{M}^\nu (w)}{\left[(k-\alpha P)^2 + \Lambda^2_{1-2\alpha}\right]^{\nu+2}} \,,
\end{align}
where the only $\beta$ dependence appears in the numerator through the factor $(1-\beta)^\nu$, allowing the $\beta$-integral to be performed analytically. This yields the following expression for the BSWF,
\begin{align}\label{BSWF}
 n_\mathrm{M} \chi(k_{-},P)=&\; 2^\nu(\nu+1)\mathcal{M}_{q,\bar{h}}(k,P) \int_{0}^{1} \mathrm{d}\alpha\left[ \int_{-1}^{1-2\alpha}\mathrm{d}w \left( \frac{\alpha}{1- w}\right)^\nu \right. \nonumber \\
 &\left. + \int_{1-2\alpha}^1\mathrm{d}w \left(\frac{1- \alpha}{1-w}\right)^\nu \right] \frac{\tilde{\rho}_\mathrm{M}^\nu (w)}{\left[(k-\alpha P)^2  + \Lambda^2_{1-2\alpha}\right]^{\nu+2}}.
\end{align}

It is important to highlight that the parameter $\nu > -1$ controls the ultraviolet behavior of the Bethe--Salpeter amplitude, ensuring that it remains finite at large momenta, consistent with its interpretation as a bound-state wave function \cite{Maris:1999nt}. Notably, $\nu$ does not serve as a regulator of divergences; instead, it governs the asymptotic falloff of the meson’s Bethe--Salpeter wave function. In this study, we adopt $\nu = 1$, which leads to the expected power-law behavior for mesons \cite{Roberts:1994dr}, and reproduces the results found in Refs. \cite{Chouika:2017rzs, Mezrag:2016hnp, Mezrag:2014jka}.

On the other hand, the functional dependence of $\Lambda$ on $\omega$ leads to a simplification of the integrals, making it possible to derive closed-form algebraic expressions that connect different structural distributions. Eq. \eqref{eq:lambdaw} also displays several key features. First, it includes a constant term $M_q^2$, carried over from earlier formulations \cite{Raya:2022eqa, Raya:2021zrz, Zhang:2021mtn, Chouika:2017dhe, Chouika:2017rzs, Mezrag:2016hnp, Mezrag:2014jka}, which have successfully described a broad set of observables related to GPDs. Second, a linear term in $\omega$ is incorporated to account for mesons formed by quark–antiquark pairs of different flavors. Lastly, the coefficients in Eq. \eqref{eq:lambdaw} are chosen to guarantee that $\Lambda_\omega^2$ remains strictly positive. This requirement imposes the condition $|M_{\bar{h}} - M_q| < m_{\mathrm{M}} < M_{\bar{h}} + M_q$ on the constituent quark masses.


\section{Applications to Meson Observables}

\subsection{Light-Front Wave Function}
\label{Subsection - Applications to Hadron Observables}
Following the formalism developed in Refs.~\cite{Albino:2022gzs,Almeida-Zamora:2023bqb}, the algebraic model has demonstrated remarkable versatility in the computation of light-front wave functions (LFWFs) for pseudoscalar mesons spanning from the light to the heavy sector. In what follows, we summarize the general procedure for obtaining the pseudoscalar meson's LFWF and highlight its connection with PDA.

In the light-front formalism, hadronic bound states are described by wave functions that depend explicitly on the longitudinal momentum fraction carried by each constituent and their transverse momenta. These LFWFs encode the full dynamical information of the system and serve as a direct bridge between the Bethe–Salpeter amplitude and experimentally accessible observables such as PDAs, PDFs, and GPDs \cite{Brodsky:1997de,Diehl:2003ny}. Within the algebraic model, the NIR provides a natural framework to project the covariant BSA onto the light-front, ensuring that the resulting LFWFs retain the analytic structure and symmetry properties of the underlying Dyson–Schwinger and Bethe–Salpeter framework \cite{Carbonell:2010tz,Frederico:2011ws}. By expressing the BSA as an integral over the Nakanishi weight function, one obtains closed-form expressions for the LFWFs that are consistent with confinement and chiral symmetry breaking, while remaining computationally tractable \cite{dePaula:2016oct,Mezrag:2016hnp}. This formalism thus enables a unified, model-independent approach to describe hadron structure both in momentum space and on the light-front.

The LFWFs encapsulates the internal structure of mesons in terms of quark and antiquark momentum fractions and transverse momenta. It can be obtained from the covariant Bethe–Salpeter amplitude through its projection onto the light-front hypersurface, which effectively isolates the valence component of the bound state. For a pseudscalar meson $\mathrm{M}$, the corresponsing LFWF is obtained through the projection
\begin{equation}\label{eq:ap_lfwf_pseudo}
\psi^q_\mathrm{M} (x,k_\perp^2) = \mathrm{Tr} \int \frac{\mathrm{d}^2 k_\parallel}{\pi} \,
\delta (n\cdot k - x\, n\cdot P)\, \gamma_5\, \gamma \cdot n\, \chi_\mathrm{M} (k_{-}, P),
\end{equation}
where the four-momentum integral has been decomposed as
\begin{equation}\label{ap.momentumseparation}
\int \frac{\mathrm{d}^4 k }{(2\pi)^4} = \frac{1}{16\pi^3} \int \mathrm{d}^2 k_\perp \cdot \frac{1}{\pi} \int \mathrm{d}^2 k_\parallel \,,
\end{equation}
and only the longitudinal momentum $k_\parallel$ is to be integrated over in the LFWF projection, Eq.~\eqref{eq:ap_lfwf_pseudo}. Here, the light-like vector $n^\mu$ satisfies $n^2 = 0$ and $n \cdot P = -m_\mathrm{M}$. Notably, the LFWF can also be characterized in terms of its Mellin moments, defined as
\begin{equation}
    \langle x^m \rangle_{\psi^q_\mathrm{M}} = \int_0^1 \mathrm{d}x \; x^m \, \psi^q_\mathrm{M}(x,k_\perp^2) \,. \label{Mellin moments}
\end{equation}
Hence, substituting Eq.~\eqref{eq:ap_lfwf_pseudo} in the above Eq.~\eqref{Mellin moments}, and integrating over $x$, one obtains
\begin{align}
\langle x^m \rangle_{\psi^q_\mathrm{M}} &=
\mathrm{Tr} \int_0^1 \mathrm{d}x \int \frac{\mathrm{d}^2 k_\parallel}{\pi}
x^m \, \delta(n\cdot k - x n\cdot P)\, \gamma_5\, \gamma \cdot n\, \chi_\mathrm{M}(k_{-}, P), \nonumber \\
&= \mathrm{Tr} \int \frac{\mathrm{d}^2 k_\parallel}{\pi} \frac{1}{n \cdot P}
\left( \frac{n \cdot k}{n \cdot P}\right)^m
\gamma_5\, \gamma \cdot n\, \chi_\mathrm{M} (k_{-}, P) \,,
\end{align}
Using the Bethe--Salpeter wave function definition, Eq.~\eqref{BSWF}, this expression becomes
\begin{align}
\langle x^m \rangle_{\psi^q_\mathrm{M}}
=&\; \frac{\nu(\nu+1)}{n\cdot P} \int \frac{\mathrm{d}^2 k_\parallel}{\pi}
\left( \frac{n \cdot k}{n \cdot P}\right)^m
\mathrm{Tr} \left\{ \gamma_5 \gamma \cdot n\, \mathcal{M}_{q,h} (k,P)
\int_0^1 \mathrm{d}\alpha
\left[ \int_{-1}^{1-2\alpha} \mathrm{d}w
\int_{\frac{2 \alpha}{w-1}+1}^1 \mathrm{d}\beta \right. \right. \nonumber \\
&\left. \left. + \int_{1-2\alpha}^1 \mathrm{d}w
\int_{\frac{2\alpha+w-1}{w+1}}^{1} \mathrm{d}\beta \right]
\frac{(1-\beta)^{\nu - 1} \Lambda_{2w}^{2\nu} \rho_\mathrm{M}^\nu(w)}{n_\mathrm{M} \sigma^{\nu + 2}} \right\}.
\end{align}
Since the trace acts solely on the Dirac structure $\gamma_5 \gamma \cdot n\, \mathcal{M}_{q,h}$, its evaluation yields
\begin{align}
\mathrm{Tr} \left[ \gamma_5 \gamma\cdot n \mathcal{M}_{q,h}\right]
=&\; 4 N_c\, n\cdot P \left[ M_q + \frac{n \cdot k}{n \cdot P} (M_h - M_q) \right] \,,
\end{align}
and consequently the Mellin moments take the form
\begin{align}\label{ap:eq_MM_pion}
    \langle x^m \rangle_{\psi^q_\mathrm{M}}
    =&\; 4 N_c\, \nu(\nu +1)
    \int_0^1 \mathrm{d}\alpha
    \left[ \int_{-1}^{1-2\alpha} \mathrm{d}w
    \int_{\frac{2 \alpha}{w-1}+1}^1 \mathrm{d}\beta
    + \int_{1-2\alpha}^1 \mathrm{d}w
    \int_{\frac{2\alpha+w-1}{w+1}}^{1} \mathrm{d}\beta \right] \nonumber \\
    &\times \int \frac{\mathrm{d}^2 k_\parallel}{\pi}
    \left( \frac{n \cdot k}{n \cdot P}\right)^m
    \left[ M_q + \frac{n \cdot k}{n \cdot P} (M_h - M_q) \right] 
    \frac{(1-\beta)^{\nu - 1} \Lambda_{2w}^{2\nu} \rho_\mathrm{M}^\nu(w)}
    {n_\mathrm{M} \sigma^{\nu + 2}},
\end{align}
where $\sigma = (k - \alpha P)^2 + \Lambda_{1-2\alpha}^2$. In order to evaluate the $k_\parallel$-integration, we define the auxiliar integral 
\begin{equation}
I = \int \frac{\mathrm{d}^2 k_\parallel}{\pi}
\left[ M_q + \frac{n \cdot k}{n \cdot P} (M_h - M_q) \right]
\frac{(n \cdot k)^m}
{\left[(k - \alpha P )^2 + \Lambda_{1-2\alpha}^2\right]^{\nu + 2}}.
\end{equation}
Introducing the decomposition $k = k_\parallel + k_\perp$, with
$n \cdot k = n \cdot k_\parallel$, $n \cdot k_\perp = 0$,
and shifting $k \rightarrow k + \alpha P$, we obtain the expression
\begin{equation}\label{ap:eq_I}
I = \int \frac{\mathrm{d}^2 k_\parallel}{\pi}
\left[ M_q + \frac{n \cdot k_\parallel + \alpha n\cdot P}{n \cdot P} (M_h - M_q) \right]
\frac{(n \cdot k_\parallel + \alpha n\cdot P)^m}
{(k_\parallel^2 + k_\perp^2 + \Lambda_{1-2\alpha}^2)^{\nu + 2}} \,,
\end{equation}
which can be expanded using the binomial theorem,
\begin{equation}
(a + b)^{m+l} = \sum_{j=0}^{m+l} \binom{m+l}{j} a^j b^{m+l-j},
\end{equation}
which in turn allows us to define
\begin{align}\label{Integraljl}
I_j^l
=& \int \frac{\mathrm{d}^2 k_\parallel}{\pi}
\frac{(n \cdot k_\parallel + \alpha n\cdot P)^{m+l}}
{(k_\parallel^2 + k_\perp^2 + \Lambda_{1-2\alpha}^2)^{\nu + 2}} \nonumber \\
=& \sum_{j=0}^{m+l} \binom{m+l}{j}
(\alpha n\cdot P)^{m+l-j}
\int \frac{\mathrm{d}^2 k_\parallel}{\pi}
\frac{(n \cdot k_\parallel)^{j}}
{(k_\parallel^2 + k_\perp^2 + \Lambda_{1-2\alpha}^2)^{\nu + 2}} \,,
\end{align}
where only the $j=0$ term contributes, since odd-$j$ contributions vanish by translational symmetry of the integration measure, whereas even-$j$ terms, with $j \ge 2$, are proportional to $(n^2)^{j/2}$ and thus vanish because $n^\mu$ is a light-like vector, $n^2=0$. Consequently,
\begin{align}\label{ap:eq_Ij0}
I_{j=0}^l
=& (\alpha n\cdot P)^{m+l}
\int \frac{\mathrm{d}^2 k_\parallel}{\pi}
\frac{1}{(k_\parallel^2 + k_\perp^2 + \Lambda_{1-2\alpha}^2)^{\nu + 2}} \nonumber \\
=& \frac{(\alpha n\cdot P)^{m+l}}
{(\nu + 1)}
\frac{1}{(k_\perp^2 + \Lambda_{1-2\alpha}^2)^{\nu+1}}.
\end{align}
Inserting Eq.~\eqref{ap:eq_Ij0} into Eq.~\eqref{ap:eq_I}, we obtain
\begin{equation}
I = \frac{(\alpha n\cdot P)^{m}}
{(\nu + 1)}
\frac{M_q^2 + \alpha (M_h - M_q)}
{(k_\perp^2 + \Lambda_{1-2\alpha}^2)^{\nu+1}} \,,
\end{equation}
and substituting this result into Eq.~\eqref{ap:eq_MM_pion}, the Mellin moments thus read as
\begin{align}
\langle x^m \rangle_{\psi^q_\mathrm{M}}
=&\; \frac{4 N_c \nu}{n_\mathrm{M}}
\int_0^1 \mathrm{d}\alpha \; \alpha^m
\left[ \int_{-1}^{1-2\alpha} \mathrm{d}w
\int_{\frac{2 \alpha}{w-1}+1}^1 \mathrm{d}\beta
+ \int_{1-2\alpha}^1 \mathrm{d}w
\int_{\frac{2\alpha+w-1}{w+1}}^{1} \mathrm{d}\beta \right] \nonumber \\
&\times \left[\alpha M_h + (1-\alpha)M_q\right]
\frac{(1-\beta)^{\nu - 1} \Lambda_{2w}^{2\nu} \rho_\mathrm{M}^\nu(w)}
{(k_\perp^2 + \Lambda_{1-2\alpha}^2)^{\nu+1}}.
\end{align}
Finally, by recognizing the definition of the Mellin moments, Eq.~\eqref{Mellin moments}, the explicit form of the LFWF is immediately obtained as
\begin{align}\label{ap:eq_LFWF_final}
    \psi_\mathrm{M} (x,k_\perp^2)
=&\; \frac{4 N_c \nu}{n_\mathrm{M}}
\left[ \int_{-1}^{1-2x} \mathrm{d}w
\int_{\frac{2 x}{w-1}+1}^1 \mathrm{d}\beta
+ \int_{1-2x}^1 \mathrm{d}w
\int_{\frac{2x+w-1}{w+1}}^{1} \mathrm{d}\beta \right] \nonumber \\
&\times \left[x M_h + (1-x)M_q\right]
\frac{(1-\beta)^{\nu - 1} \Lambda_{2w}^{2\nu} \rho_\mathrm{M}^\nu(w)}
{(k_\perp^2 + \Lambda_{1-2x}^2)^{\nu+1}}.
\end{align}
This compact expression provides the algebraic model's prediction for the meson's LFWF. Notably, integrating over the transverse momentum dependence of the above LFWF, $\psi_\mathrm{M}(x,k_\perp^2)$, one obtains the corresponding meson's PDA, which encodes the longitudinal structure of the bound state in terms of the light-front momentum fraction $x$. Namely.
\begin{equation}
    f_\mathrm{M}\, \phi_\mathrm{M}^q(x) = \frac{1}{16\pi^3} \int \mathrm{d}^2 k_\perp\, \psi_\mathrm{M}(x, k_\perp^2),
\end{equation}
where $f_\mathrm{M}$ denotes the meson's leptonic decay constant. It is important to emphasize that the LFWF encodes the full three-dimensional momentum-space structure of the hadron, encompassing both the longitudinal momentum fraction $x$ and the transverse momentum $\boldsymbol{k}_\perp$. In contrast, PDA provides a reduced, one-dimensional description in terms of the longitudinal momentum fraction alone. Specifically, $\phi(x)$ represents the probability amplitude for finding a quark and an antiquark within the hadron carrying definite fractions of the longitudinal momentum,  
\begin{equation}
    x_q = x, \qquad x_{\bar{q}} = 1-x.
\end{equation}
According to Eq.~\eqref{ap:eq_LFWF_final}, the only term exhibiting dependence on the transverse momentum $k_\perp$ is the denominator $(k_\perp^2 + \Lambda_{1-2x}^2)^{\nu+1}$. The integration over the transverse degrees of freedom therefore reduces to an analytic expression
\begin{align}
\frac{1}{16\pi^3} \int \mathrm{d}^2k_\perp \frac{1}{(k_\perp^2 + \Lambda_{1-2x}^2)^{\nu+1}} 
&= \frac{1}{16\pi^2}\, \frac{1}{\nu\, (\Lambda_{1-2x}^2)^{\nu}}.
\end{align}
Substituting the explicit form of the LFWF and performing the integration over $k_\perp$, one obtains a closed expression for the PDA:
\begin{align}
\phi_\mathrm{M}^q (x) = &\; \frac{4 N_c \nu}{n_\mathrm{M} f_\mathrm{M}} 
\left[
\int_{-1}^{1-2x} \mathrm{d}w \int_{\frac{2 x}{w-1}+1}^1 \mathrm{d}\beta
+ 
\int_{1-2x}^1 \mathrm{d}w \int_{\frac{2x+w-1}{w+1}}^{1} \mathrm{d}\beta
\right]
\nonumber\\
&\times \left[x M_h + (1-x) M_q \right]
(1-\beta)^{\nu - 1} 
\Lambda_{2w}^{2\nu} 
\rho_\mathrm{M}^\nu(w)
\left[\frac{1}{16\pi^2} \frac{1}{\nu (\Lambda_{1-2x}^2)^{\nu}}\right].
\end{align}
The comparison of this expression with Eq.~\eqref{ap:eq_LFWF_final} reveals an elegant algebraic connection between the LFWF and the corresponding PDA, leading to the following compact relation
\begin{align}\label{LFWF}
\psi_\mathrm{M} (x,k_\perp^2) = 16\pi^2 f_\mathrm{M} \frac{\nu\, \Lambda_{1-2x}^{2\nu}}{(k_\perp^2 + \Lambda_{1-2x}^2)^{\nu+1}} \, \phi_\mathrm{M}^q(x).
\end{align}
This identity establishes that, within the algebraic model, the complete light-front structure of the meson can be reconstructed directly from its longitudinal amplitude, ensuring a unified connection between the LFWF and the PDA.

\subsection{Generalized Parton Distributions}\label{ap:AM_GPD}
A fundamental aspect in the investigation of hadron structure is the connection between GPDs and LFWFs. While LFWFs encode the full three-dimensional momentum-space information, GPDs provide a complementary perspective by correlating the longitudinal momentum fractions of partons with their transverse spatial distributions \cite{Polyakov:2002yz}.
The GPD provides a unified description of the hadron’s internal structure, simultaneously encoding information on both the longitudinal momentum and the transverse spatial distributions of partons. Within the light-front framework, GPDs can be expressed in terms of overlap integrals of LFWFs corresponding to different light-front momentum fractions and transverse momenta. Following Refs.~\cite{Diehl:2003ny, Brodsky:2000xy, Albino:2022gzs},In the DGLAP region $\xi < x < 1$, the valence-quark GPD can be written a
\begin{equation}\label{ap.GPDdef}
\mathcal{H}_\mathrm{M}^q(x,\xi,t)
= \int \frac{\mathrm{d}^2 k_\perp}{(2\pi)^4} \;
\psi_\mathrm{M}^{q*}\!\left[x^-, (k_\perp^-)^2\right]
\psi_\mathrm{M}^{q}\!\left[x^+, (k_\perp^+)^2\right],
\end{equation}
where the variables
\begin{equation}\label{ap.def_xkpm}
x^{\pm} = \frac{x \pm \xi}{1 \pm \xi}, \qquad
k_\perp^{\pm} = k_\perp \mp \frac{\Delta_\perp}{2} \frac{1-x}{1 \pm \xi},
\end{equation}
encode the longitudinal and transverse momentum shifts between the incoming and outgoing quark states, with $\xi$ representing the longitudinal skewness and $t = -\Delta^2$ the invariant momentum transfer. 

Substituting the LFWF given by Eq.~\eqref{LFWF} into Eq.~\eqref{ap.GPDdef}, one obtains
\begin{align}\label{ap:eq_gpd1}
\mathcal{H}_\mathrm{M}^q(x,\xi,t)
=&\; \frac{(16\pi^2 f_\mathrm{M} \nu)^2}{16\pi^3}
\Lambda_{1-2x^-}^{2\nu}
\Lambda_{1-2x^+}^{2\nu}
\phi_\mathrm{M}^q(x^-)\phi_\mathrm{M}^q(x^+)
\nonumber\\
&\times
\int
\frac{\mathrm{d}^2 k_\perp}{
\left[(k_\perp^-)^2+\Lambda_{1-2x^-}^2\right]^{\nu+1}
\left[(k_\perp^+)^2+\Lambda_{1-2x^+}^2\right]^{\nu+1}}.
\end{align}

To evaluate the integral, we rewrite the denominators using Eq.~\eqref{ap.def_xkpm}:
\begin{align}
D =&\;\left[(k_\perp^-)^2+ \Lambda_{1-2x^-}^2\right]^{\nu+1}\left[(k_\perp^+)^2+ \Lambda_{1-2x^+}^2\right]^{\nu+1} \nonumber \\
=&\;\left[ k_\perp^2 + \frac{1-x}{1-\xi} k_\perp \cdot \Delta_\perp+ \frac{\Delta_\perp^2}{4} \left(\frac{1-x}{1-\xi}\right)^2 + \Lambda_{1-2x^-}^2 \right]^{\nu+1} \times \nonumber \\ 
& \times\left[k_\perp^2 - \frac{1-x}{1+\xi} k_\perp \cdot \Delta_\perp+ \frac{\Delta_\perp^2}{4} \left(\frac{1-x}{1+\xi}\right)^2 + \Lambda_{1-2x^+}^2 \right]^{\nu+1}.
\end{align}
We then employ Feynman's parametrization,
\begin{equation}
\frac{1}{A^{\nu+1} B^{\nu+1}}
= \frac{\Gamma(2\nu+2)}{\Gamma(\nu+1)^2}
\int_0^1 \mathrm{d}u \;
\frac{u^\nu (1-u)^\nu}{
\left[uA+(1-u)B\right]^{2\nu+2}},
\end{equation}
where
\begin{subequations}
\begin{align}
A &= k_\perp^2 + \frac{1-x}{1-\xi}\, k_\perp\!\cdot\!\Delta_\perp
+ \frac{\Delta_\perp^2}{4} \!\left(\frac{1-x}{1-\xi}\right)^{\!2}
+ \Lambda_{1-2x^+}^2, \\
B &= k_\perp^2 - \frac{1-x}{1+\xi}\, k_\perp\!\cdot\!\Delta_\perp
+ \frac{\Delta_\perp^2}{4} \!\left(\frac{1-x}{1+\xi}\right)^{\!2}
+ \Lambda_{1-2x^-}^2.
\end{align}
\end{subequations}
After combining and rearranging terms, the denominator becomes
\begin{align}
uA + (1-u)B
&= \left[k_\perp + c\,\Delta_\perp \right]^2
+ \frac{(1-x)^2}{(1-\xi^2)^2}\Delta_\perp^2\, u(1-u)
\nonumber\\
&\quad + u(\Lambda_{1-2x^+}^2 - \Lambda_{1-2x^-}^2)
+ \Lambda_{1-2x^-}^2,
\end{align}
with the definitions
\begin{equation}
a = \frac{1-x}{1-\xi}, \qquad
b = \frac{1-x}{1+\xi}, \qquad
c = \frac{1}{2}\left[a u + (1-u)b\right].
\end{equation}
Substituting back into Eq.~\eqref{ap:eq_gpd1} and performing the shift $k_\perp \rightarrow k_\perp - c\Delta_\perp$, one finds
\begin{align}
\mathcal{H}_\mathrm{M}^q(x,\xi,t)
=&\; \frac{(16\pi^2 f_\mathrm{M} \nu)^2}{16\pi^3}
\frac{\Gamma(2\nu+2)}{\Gamma(\nu+1)^2}
\Lambda_{1-2x^-}^{2\nu}
\Lambda_{1-2x^+}^{2\nu}
\phi_\mathrm{M}^q(x^-)\phi_\mathrm{M}^q(x^+)\int_0^1 \mathrm{d}u
\nonumber\\
&\times
\int \mathrm{d}^2 k_\perp
\frac{u^\nu (1-u)^\nu}{
\left[k_\perp^2 +
\frac{(1-x)^2}{(1-\xi^2)^2}\Delta_\perp^2 u(1-u)
+u(\Lambda_{1-2x^+}^2 - \Lambda_{1-2x^-}^2)
+ \Lambda_{1-2x^-}^2\right]^{2\nu+2}}.
\end{align}
Carrying out the integration over $k_\perp$ and absorbing numerical constants into a normalization factor $\mathcal{N}$, the final expression for the GPD reads
\begin{align}\label{GPD_pseudo}
    \mathcal{H}_\mathrm{M}^q(x,\xi,t)
= \mathcal{N}\,
\Lambda_{1-2x^-}^{2\nu}
\Lambda_{1-2x^+}^{2\nu}
\phi_\mathrm{M}^q(x^-)\phi_\mathrm{M}^q(x^+)
\frac{\Gamma(2\nu+2)}{\Gamma(\nu+1)^2}
\int_0^1 \frac{\mathrm{d}u\, u^\nu (1-u)^\nu}{
\left[\mathbb{M}^2(u,x,\xi,t)\right]^{2\nu+1}},
\end{align}
where
\begin{equation}
\mathbb{M}^2(u,x,\xi,t)
= -\frac{(1-x)^2}{(1-\xi^2)^2} t\,u^2
+ \left[\frac{(1-x)^2}{(1-\xi^2)^2} t
+ \Lambda_{1-2x^+}^2 - \Lambda_{1-2x^-}^2 \right] u
+ \Lambda_{1-2x^-}^2,
\end{equation}
and $\Delta_\perp^2 = -t$. This compact expression enables the extraction of various physical observables from the same analytic framework.

\subsection{Kinematic Limits and Spatial Structure in GPDs}
 GPDs encapsulate a three-dimensional structure of hadrons, providing a unified description that connects EFFs, PDFs, and spatial parton imaging. These complementary observables can be extracted from GPDs through specific kinematic limits or suitable integral projections \cite{Albino:2022gzs,Raya:2024ejx,Diehl:2003ny}:
 
\begin{enumerate}
    \item \textbf{Forward Limit: Parton Distribution Functions}\\
    In the limit $\xi \to 0$ and $t \to 0$, the GPD reduces to the conventional parton distribution function:
    \begin{equation}
         q_\mathrm{M}(x) = \mathcal{H}_\mathrm{M}^q(x,0,0).
    \end{equation}
It illustrates that GPDs not only extend the concept of PDFs but also consistently reduce to them when no momentum is transferred to the hadron. 

It is worth noting that, within the parton model, the structure functions are expected to depend solely on the Bjorken variable $x$ and not on the squared momentum transfer $Q^2$, a property referred to as \textit{Bjorken scaling}. Nonetheless, high-precision experiments revealed logarithmic deviations from this behavior. Such scaling violations are explained through the property of asymptotic freedom, which states that the strong coupling constant decreases at high energies, thereby enabling perturbative corrections arising from gluon emission and quark–antiquark pair production. These quantum corrections are systematically described by the DGLAP evolution equations \cite{Dokshitzer:1977sg,Gribov:1972ri,Lipatov:1974qm,Altarelli:1977zs}, which govern the $Q^2$-dependence of PDFs.
\item \textbf{Zeroth Mellin Moment: Elastic Form Factor (EFF)}\\
    The contribution of the quark $q$ to the meson’s elastic electromagnetic form factor is given by the zeroth Mellin moment:
    \begin{equation}
         F_\mathrm{M}(t) = \int_{-1}^{1} \mathrm{d}x \; \mathcal{H}_\mathrm{M}^q(x,\xi,t),
    \end{equation}
 For a spinless meson, EFF can be expressed as the sum of independent quark and antiquark contributions. This decomposition arises naturally from the definition of GPD $\mathcal{H}(x, \xi, t)$, which is defined in the domain $x \in [-1,1]$. In this framework, the region $x > 0$ corresponds to quark degrees of freedom, while $x < 0$ represents antiquarks. The total meson form factor can therefore be written as
\begin{equation}
    F_\mathrm{M}(t) = e_q F_\mathrm{M}^q(t) + e_{\bar{h}} F_\mathrm{M}^{\bar{h}}(t),
\end{equation}
where $e_{q,\bar{h}}$ denote the electric charges of the valence quark and antiquark, respectively, in units of the positron charge. Consequently, the meson form factor can be expressed in terms of its quark GPDs as
\begin{equation}
    F_\mathrm{M}(t) = e_q \int_{0}^{1} dx\, \mathcal{H}_\mathrm{M}^q(x, \xi, t)
    + e_{\bar{h}} \int_{0}^{1} dx\, \mathcal{H}_\mathrm{M}^{\bar{h}}(x, \xi, t),
\end{equation}
where the antiquark contribution is related to the negative-$x$ region of the quark GPD via
\begin{equation}
    \mathcal{H}_\mathrm{M}^{\bar{h}}(x, \xi, t) = \mathcal{H}_\mathrm{M}^q(1-x, \xi, t).
\end{equation}

Although GPDs depend explicitly on the skewness parameter $\xi$, their integrals over $x$ are independent of it. This property follows from Lorentz invariance and the so-called \textit{polynomiality condition}, which requires that the Mellin moments of GPDs be polynomials in $\xi$. Specifically,
\begin{equation}
     \int_{-1}^{1} dx\, x^n \,\mathcal{H}_\mathrm{M}^q(x, \xi, t)
     = \sum_{k=0}^{n+1} A_{nk}(t)\, \xi^k,
\end{equation}
implying that $\mathcal{H}_\mathrm{M}^q(x, \xi, t)$ is at most of degree $n+1$ in $\xi$. In particular, the zeroth moment, corresponding to the elastic form factor, is constant and therefore independent of $\xi$, as expected from a physical observable that depends only on the invariant momentum transfer $t=-Q^2$. Consequently, one obtains
\begin{equation}
    F_\mathrm{M}^q(t) = \int_{0}^{1} dx\, \mathcal{H}_\mathrm{M}^q(x, 0, t).
\end{equation}

The charge radius of a hadron provides a measure of the spatial distribution of its electric charge and is directly related to the electric form factor. In terms of the charge density, the meson electric form factor (EFF) is defined as
\begin{equation}\label{eq:EFF_integral}
  F_\mathrm{M}(Q^2)  = \int d^3r \, \rho(\mathbf{r}) \, e^{i \mathbf{q} \cdot \mathbf{r}},
\end{equation}
where $\rho(\mathbf{r})$ is the charge density and $Q^2 = \abs{\mathbf{q}}^2$. For sufficiently small $Q^2$, a Taylor expansion of the exponential yields
\begin{equation}
    e^{i \mathbf{q} \cdot \mathbf{r}} \approx 1 + i (\mathbf{q} \cdot \mathbf{r}) - \frac{1}{2} (\mathbf{q} \cdot \mathbf{r})^2 + \mathcal{O}(Q^4),
\end{equation}
and, upon substitution into Eq.~\eqref{eq:EFF_integral}, the linear term vanishes by symmetry, giving
\begin{equation}
   F_\mathrm{M}(Q^2) \approx 1 - \frac{1}{6} Q^2 \langle r^2_\mathrm{M} \rangle + \mathcal{O}(Q^4),
\end{equation}
where the mean-square charge radius is defined as
\begin{equation}
    \langle r^2_\mathrm{M} \rangle = \int d^3r \, \rho(\mathbf{r}) \, r^2.
\end{equation}
Consequently, at low momentum transfer, the EFF encodes the charge radius through its slope at $Q^2 = 0$:
\begin{equation}\label{eq:chargeradii}
 \langle r_\mathrm{M}^2 \rangle = -6 \left. \frac{d F_\mathrm{M}(Q^2)}{d Q^2} \right|_{Q^2=0}.
\end{equation}

Summing the contributions from the quark and antiquark, the squared meson charge radius can be written as
\begin{equation}
\label{eq:crgeneral}
    r_{\mathrm{M}}^2 = e_q \, (r_{\mathrm{M}}^q)^2 + e_{\bar{h}} \, (r_{\mathrm{M}}^{\bar{h}})^2,
\end{equation}
where $e_q$ and $e_{\bar{h}}$ denote the electric charges of the valence quark and antiquark. In the isospin-symmetric limit, $e_q + e_{\bar{h}} = 1$, such that $F_{\mathrm{M}}(t) = F_{\mathrm{M}}^q(t)$ and, consequently, $r_{\mathrm{M}} = r_{\mathrm{\mathrm{M}}}^q$. This demonstrates that, for symmetric systems, the meson charge radius coincides with the charge radius of the constituent quark.

\item \textbf{Impact-Parameter Dependent GPDs (IPS-GPDs)}\\
    For zero skewness ($\xi = 0$), the Fourier transform of the GPD with respect to the transverse momentum transfer $\Delta_\perp$ yields the distribution of partons in transverse position space:
    \begin{equation}
        u_\mathrm{M}(x,b_\perp)
    = \int_0^\infty \frac{\mathrm{d}\Delta_\perp}{2\pi}\,
    \Delta_\perp J_0(b_\perp \Delta_\perp)\,
    \mathcal{H}_\mathrm{M}^q(x,0,t),
    \end{equation}
    where $J_0(z)$ is the zeroth-order Bessel function and $b_\perp$ denotes the transverse impact parameter. This compact expression shows that the spatial distribution of partons in the transverse plane is dictated by the radial dependence of the GPD on the squared transverse momentum transfer, with the Bessel function acting as the kernel that modulates this behavior.

\end{enumerate}
This construction provides a symmetry-preserving and analytically tractable framework that connects PDFs, form factors, and spatial distributions through a single underlying amplitude, highlighting the versatility of the algebraic model in describing hadron structure across complementary regimes.


\section{Results}

\subsection{Pseudoscalar mesons: light sector}

In this section, we present the results obtained for the light pseudoscalar mesons, namely the pion and kaon, within the framework of the Algebraic Model (AM). These systems serve as an essential testing ground for any approach to hadron structure, as they embody the interplay between dynamical chiral symmetry breaking and confinement—two fundamental features of QCD. The analysis includes the computation of light-front wave functions (LFWFs), parton distribution amplitudes (PDAs), parton distribution functions (PDFs), electromagnetic form factors (EFFs), and generalized parton distributions (GPDs), all derived consistently from the same underlying Bethe–Salpeter amplitude. This unified treatment provides a coherent and comprehensive picture of the internal structure of light pseudoscalar mesons across different representations of hadronic dynamics.

\begin{figure}[t]
    \centering
    \includegraphics[width=0.8\linewidth]{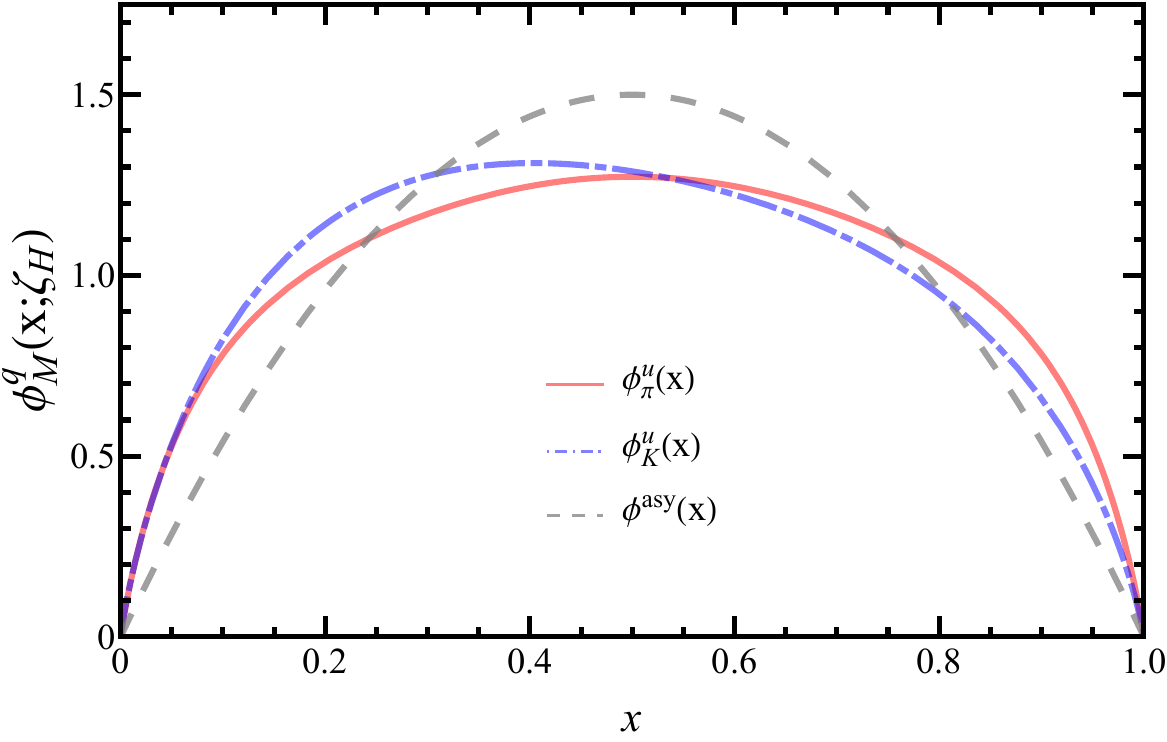}
    \caption{Pion and kaon parton distribution amplitudes evaluated at the hadronic scale $\zeta_H$. All distributions were obtained within the Schwinger–Dyson equations framework~\cite{Cui:2020tdf,Ding:2015rkn} and parametrized following Eqs.~\eqref{eq:PDAsSDE}. For reference, the asymptotic form $\phi_{\text{asy}}(x) = 6x(1-x)$ is also displayed.}
    \label{fig:PDA1}
\end{figure}
After deriving a consistent set of algebraic expressions for several parton distributions and related observables, we turn our attention to specifying the input elements of the AM. Our starting point is Eq.~\eqref{LFWF}, which establishes a direct link between the leading-twist LFWF and the PDA. This relation allows one to reconstruct the LFWF straightforwardly once $\phi_{\mathrm{M}}^q(x)$ is known, providing a representation of the meson’s internal structure at the characteristic hadronic scale $\zeta_H$. Considering the demonstrated predictive power of the SDE approach in calculating PDAs, we employ as model inputs the results obtained within that framework~\cite{Cui:2020tdf,Ding:2015rkn}. The particular forms of the PDAs used in this work are listed below, with $\bar{x}=1-x$
\begin{eqnarray}
\nonumber
\phi_\pi^u(x)&=&20.226\, x\bar{x}\,[1-2.509 \sqrt{x\bar{x}}+2.025x\bar{x}]\;,\\
\phi_K^u(x)&=&18.04\,x\bar{x}\,[1+5x^{0.032}\bar{x}^{0.024}-5.97x^{0.064}\bar{x}^{0.048}]\;.\label{eq:PDAsSDE}
\end{eqnarray}

As illustrated in Fig.~\ref{fig:PDA1}, the PDAs of light mesons such as the pion and kaon appear broader than the asymptotic distribution, reflecting the nonperturbative dynamics that dominate at low energy scales. In contrast, PDAs associated with mesons containing heavier quarks exhibit a more compressed profile. Additionally, the mass asymmetry between the $s$ and $u$ quarks in the kaon introduces a visible skewness in the distribution, whereas the PDAs of flavor-symmetric systems remain symmetric about $x=0.5$.

\begin{figure}[t]
\begin{adjustwidth}{-\extralength}{-2.5cm}
\centering
\subfloat[\centering]{\includegraphics[width=10.0cm]{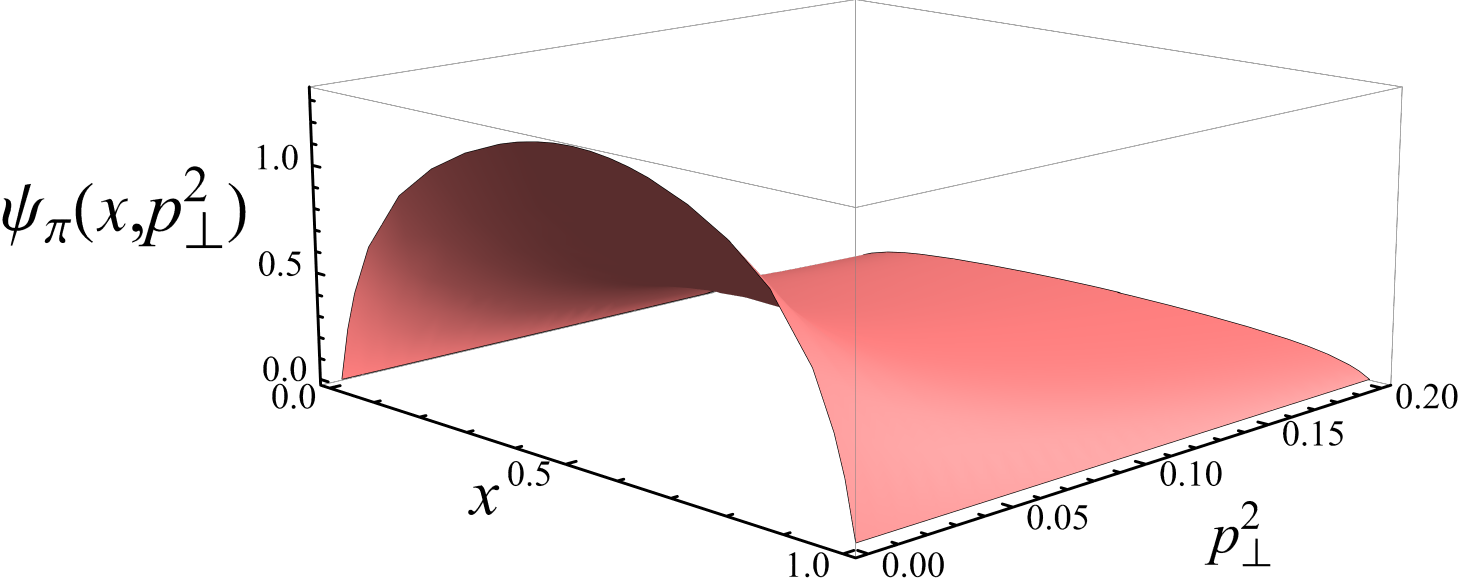}}\\
\subfloat[\centering]{\includegraphics[width=10.0cm]{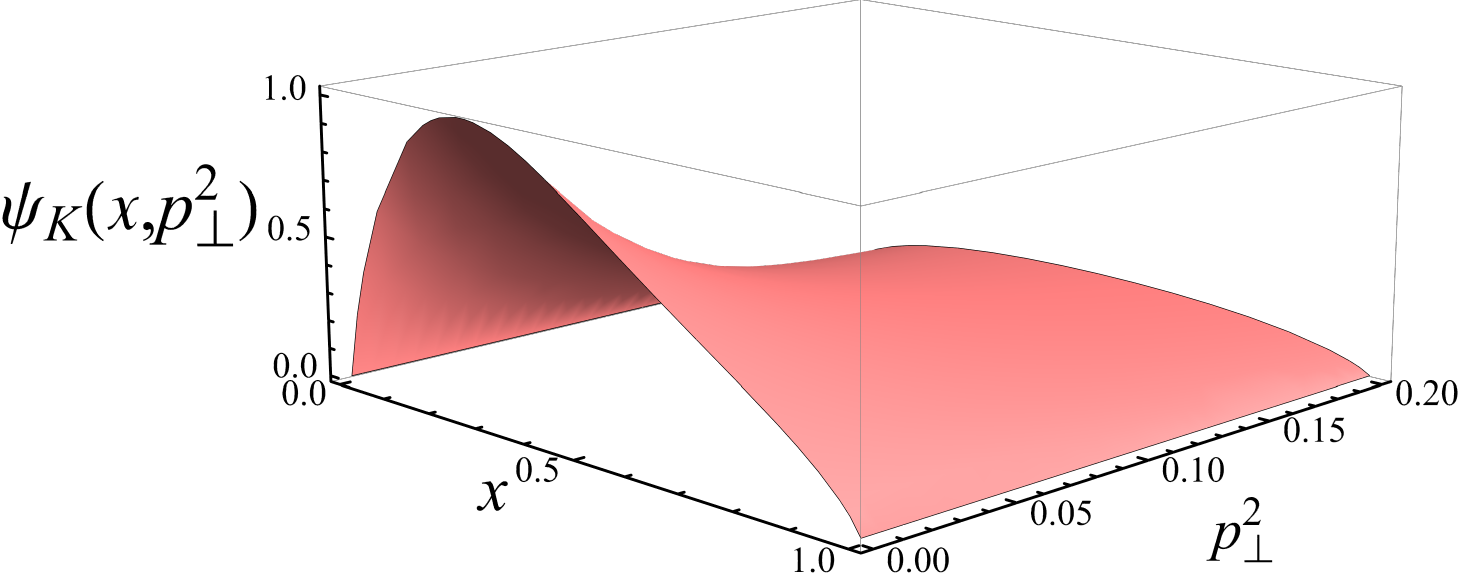}}
\end{adjustwidth}
\caption{Light-front wave functions of the pion and kaon calculated from Eq.~\eqref{LFWF}. The wave functions are shown as $\psi_{\mathrm{M}}(x,k_\perp^2) \to \psi_{\mathrm{M}}(x,k_\perp^2)/(16\pi^2 f_{\mathrm{M}})$. All masses are expressed in GeV.
}
\label{PionLFWF}
\end{figure} 

To compute the LFWFs we must specify the parameter $\nu$ and the constituent quark masses $M_q$. We adopt $\nu = 1$, which guarantees the correct asymptotic behaviour of the Bethe--Salpeter wave function and is consistent with prior DSE--BSE analyses \cite{Roberts:1994dr}. The constituent masses are fixed by benchmarking against experimental determinations and theory constraints: empirical charge radii \cite{ParticleDataGroup:2020ssz}, Dyson--Schwinger equation predictions \cite{Chang:2013nia,Eichmann:2019bqf,Bhagwat:2006xi,Miramontes:2021exi,Raya:2022ued}, and lattice-QCD results \cite{Dudek:2007zz,Dudek:2006ej}. In practice we determine each $M_q$ via the relation in Eq.~\eqref{eq:crgeneral}, using the aforementioned results as inputs. The set of constituent-quark masses that define the algebraic model, together with the corresponding charge radii, is collected in Table~\ref{tab:params}.

\begin{table}[h]
\centering
\caption{Model parameters: meson and constituent-quark masses (in GeV). The values of $M_q$ are determined using Eq.~\eqref{eq:crgeneral} together with the specified charge radii. The corresponding distribution amplitudes employed in the calculations are listed in Eq.~\eqref{eq:PDAsSDE}.}
\label{tab:params}
\begin{tabular}[t]{c|c|l||c|c}
\hline
Meson\;\; & $m_{\mathrm{M}}$ \;\;& \;$r_{\mathrm{M}}$ (in fm) \;& Quark\;\; & $M_q$ \;\;\\
\hline
$\pi^+$ & 0.14 &\; 0.659~\cite{ParticleDataGroup:2020ssz,Chang:2013nia}\; & $u$ & 0.317\\
$K^+$ & 0.49 &\; 0.600~\cite{Eichmann:2019bqf,Miramontes:2021exi,Raya:2022ued} \;& $s$ & 0.574\\
\hline
\end{tabular}
\end{table}%

The three-dimensional surfaces displayed in Fig.~\ref{PionLFWF} illustrate the LFWFs $\psi_{\pi}(x,p_\perp^2)$ and $\psi_{K}(x,p_\perp^2)$ for the pion and kaon, respectively. Each LFWF encodes the probability amplitude to find a quark and an antiquark inside the meson carrying longitudinal momentum fractions $x$ and $1-x$, and relative transverse momentum $\vec{p}_\perp$. The LFWF thus provides a direct representation of the internal structure of mesons in mixed longitudinal-transverse momentum space.  

In both cases, the LFWFs exhibit a maximum at intermediate values of $x$, reflecting the most probable momentum sharing between the valence quark and antiquark. The distributions vanish at the endpoints, $x=0$ and $x=1$, in accordance with QCD expectations and the behavior of leading-twist distribution amplitudes. As $p_\perp^2$ increases, the wave functions decrease smoothly, illustrating how configurations with large transverse momentum are suppressed. This decay encodes confinement physics—large relative momenta between constituents are less probable due to the strong interaction binding them.  

Comparing the two panels, one observes a noticeable asymmetry in the kaon’s LFWF with respect to $x=0.5$. This behavior is a direct manifestation of flavor-symmetry breaking: the kaon contains a heavier strange quark ($s$) and a lighter up quark ($u$), so the longitudinal momentum is preferentially carried by the heavier constituent. In contrast, the pion—composed of quarks with nearly equal masses—displays a symmetric profile centered around $x=0.5$.

\begin{figure}[t]
\begin{adjustwidth}{-\extralength}{-2.5cm}
\centering
\subfloat[\centering]{\includegraphics[width=10.0cm]{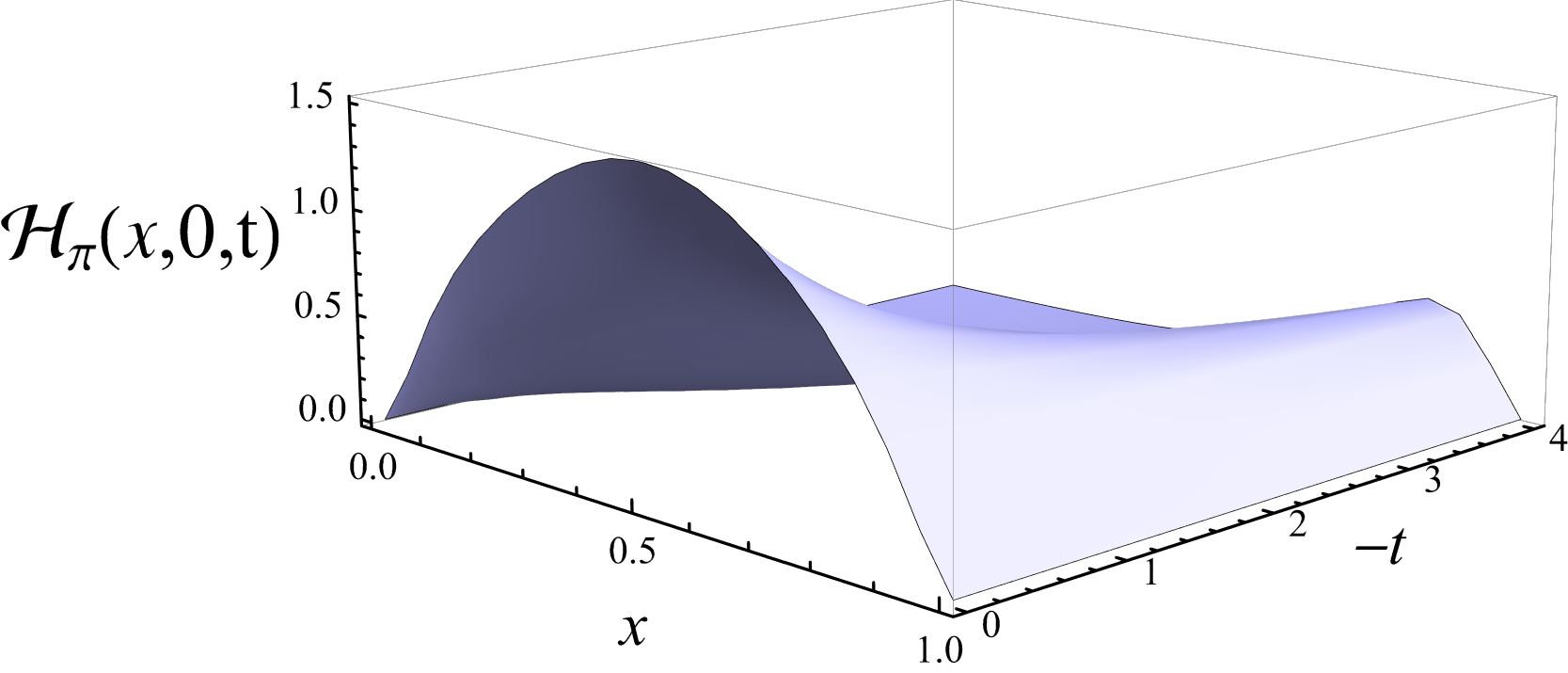}}\\
\subfloat[\centering]{\includegraphics[width=10.0cm]{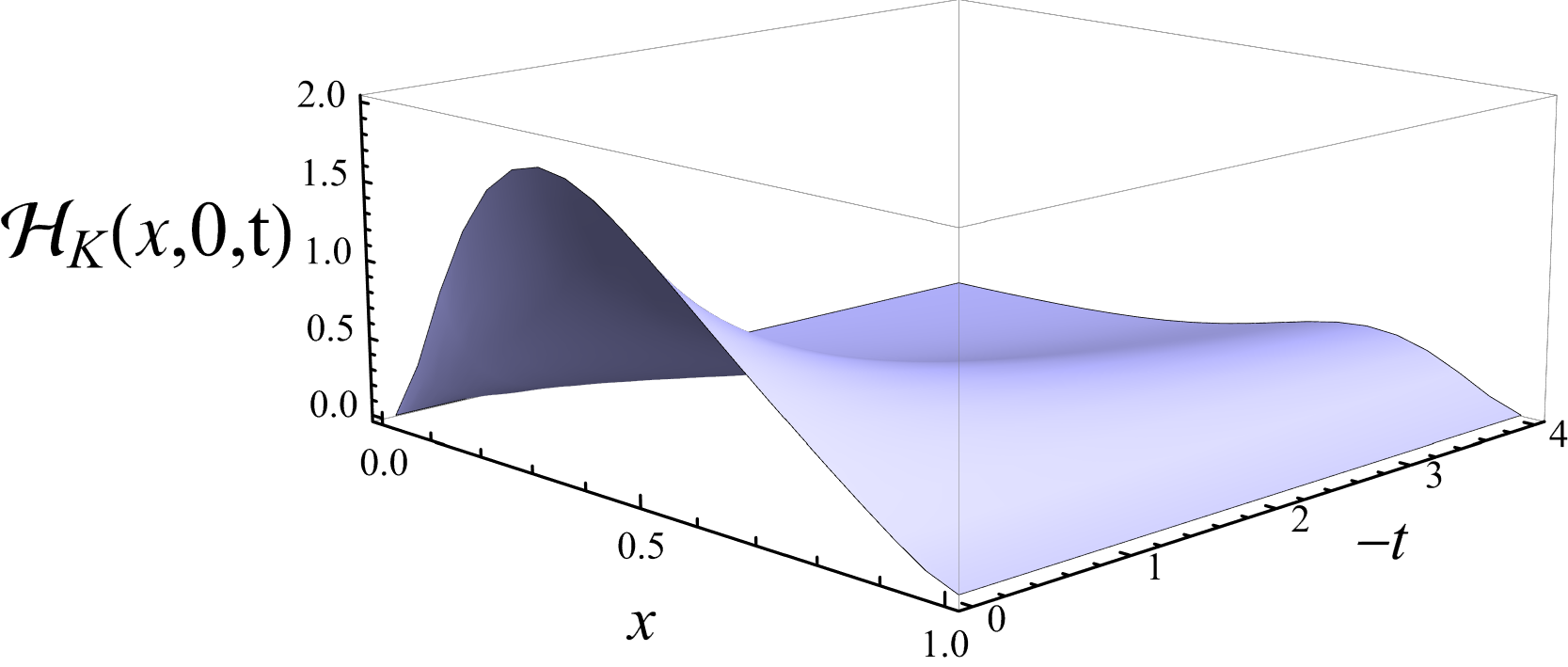}}
\end{adjustwidth}
\caption{Valence quark GPDs for the $D$, $D_s$, $B$, $B_s$, and $B_c$ mesons obtained from Eq.~\eqref{GPD_pseudo} for $\xi=0$. Mass units are given in GeV.}
\label{fig:Pion-Kaon-GPDs}
\end{figure}

Figure~\ref{fig:Pion-Kaon-GPDs} displays the valence-quark GPDs $\mathcal{H}_\pi(x,0,t)$ and $\mathcal{H}_K(x,0,t)$ for the pion and kaon, respectively. These quantities encode correlated information about the longitudinal momentum fraction $x$ and the transverse momentum transfer squared $t=-\Delta_\perp^2$, thus providing a three-dimensional view of hadron structure in mixed momentum and position space.

In both mesons, the GPDs exhibit the characteristic bell-shaped profile along $x$, peaking around the region where the quark and antiquark share the longitudinal momentum most equally. For the pion, which is an isospin-symmetric system, the distribution is symmetric around $x = 0.5$. In contrast, the kaon GPD displays a noticeable asymmetry toward larger $x$ values. This skewness arises from the explicit breaking of flavor symmetry due to the mass difference between the $u$ and $s$ quarks: the heavier strange quark tends to carry a larger fraction of the total longitudinal momentum.

These surfaces illustrate the dual role of GPDs as a bridge between momentum and coordinate space descriptions of hadron structure. They unify form-factor and parton-density information, allowing one to visualize how partons are distributed both in longitudinal momentum and in the transverse plane. The observed asymmetry in the kaon, compared with the pion, highlights the sensitivity of GPDs to explicit flavor-symmetry breaking and to the mass hierarchy among the constituent quarks.

At $t=0$, the GPDs reduce to the corresponding PDFs, recovering the longitudinal momentum structure of the valence quarks. As the momentum transfer $|t|$ increases, the magnitude of $\mathcal{H}_\mathrm{M}(x,0,t)$ decreases, reflecting the decreasing probability of finding partons with a large transverse separation, an effect directly tied to the spatial localization of the quarks inside the meson. This falloff behavior encapsulates the underlying confinement mechanism: hadrons appear as compact bound states whose internal structure becomes less correlated at higher resolution.

\begin{figure}[t]
    \centering
    \includegraphics[width=0.8\linewidth]{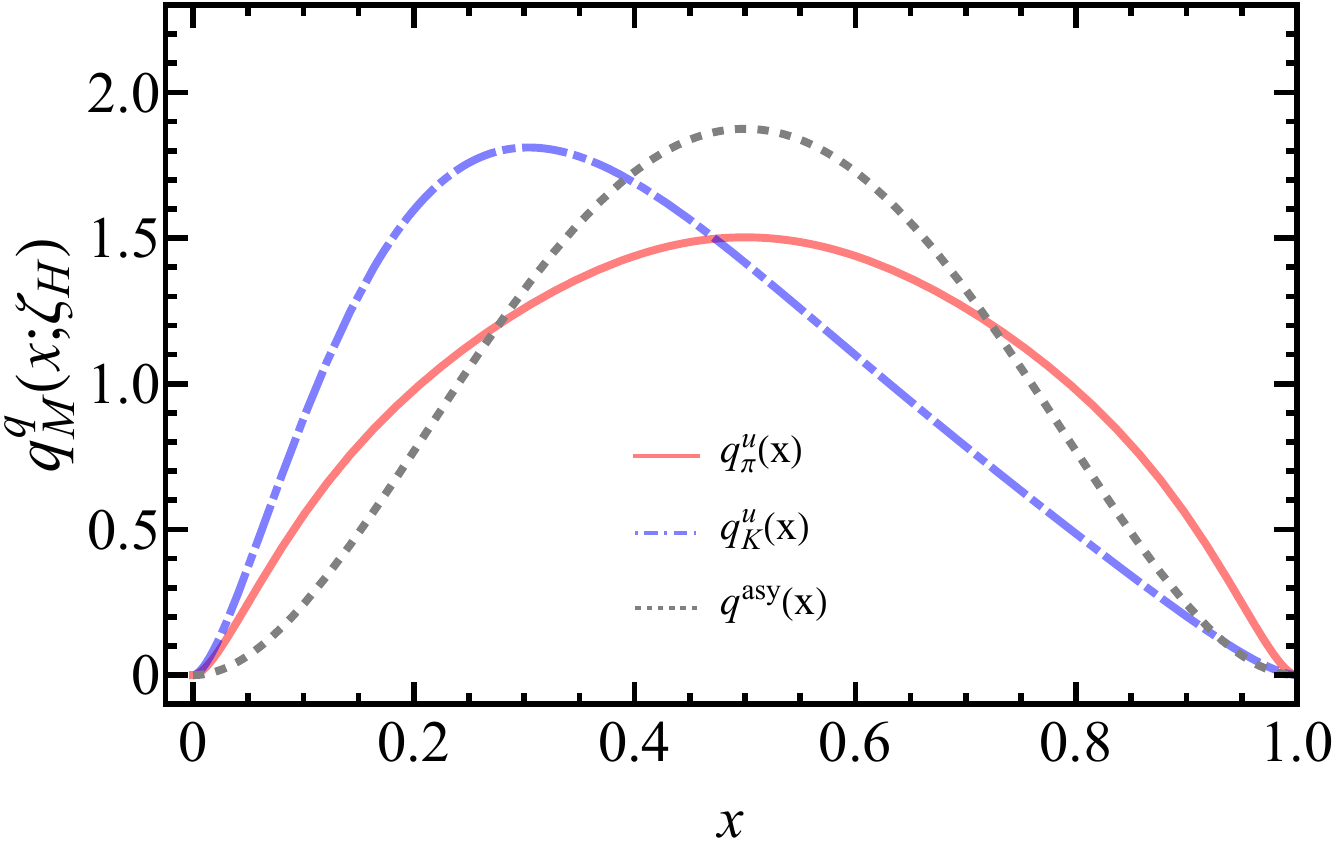}
    \caption{Valence-quark PDFs at the hadronic scale $\zeta_H$. The solid (red) line represents the pion, while the dashed-dotted (blue) line corresponds to the light-quark distribution in the kaon. The dashed (gray) line shows the scale-free, parton-like distribution $q_{sf}(x) = 30 x^2 (1-x)^2$ for comparison.}
    \label{PDFs}
\end{figure}

Figure~\ref{PDFs} compares the valence $u$--quark PDFs of the pion and kaon at the hadronic scale $\zeta_H$. The red solid curve, $q_\pi^{u}(x)$, is symmetric under $x\!\leftrightarrow\! 1-x$ as expected in the isospin limit, with a broad maximum around $x\simeq 0.5$ and vanishing behavior at the endpoints, $x\to 0,1$. This bell-shaped profile reflects the nearly equal sharing of longitudinal momentum between the pion’s valence constituents and embodies the generic endpoint suppression predicted for leading-twist, valence-like distributions.

The blue dot–dashed curve, $q_K^{u}(x)$, is visibly shifted toward smaller $x$ and is narrower than $q_\pi^{u}(x)$. This distortion is a direct manifestation of explicit flavor-symmetry breaking in the kaon: because the strange quark is heavier than the up quark, the $\bar s$ typically carries a larger fraction of the meson’s longitudinal momentum, forcing the $u$ distribution to be softer (peak at $x<0.5$) and more strongly suppressed as $x\to 1$. Complementarily, one expects $q_K^{\bar s}(x)$ to be harder, with its peak displaced toward larger $x$ and a slower falloff near the $x\to 1$ endpoint. These features mirror the corresponding PDAs and LFWFs discussed earlier: the heavier constituent biases the longitudinal momentum partition and compresses the $x$-profile.

The gray dashed reference curve, $q^{\mathrm{asy}}(x)$, illustrates the canonical valence-like shape; the pion distribution is comparatively more dilated, while the kaon $u$-PDF is comparatively more compressed.
The relative hardness/softness at large $x$ connects with perturbative counting rules: heavier constituents generally correlate with a harder partner distribution and a softer light-quark distribution as $x\to 1$.
Upon DGLAP evolution to higher $\zeta$, both curves broaden toward small $x$ and soften at large $x$, yet the qualitative ordering (kaon $u$ softer than pion $u$) persists.

\begin{figure}[t]
\centering
\subfloat[\centering]{\includegraphics[width=6.0cm]{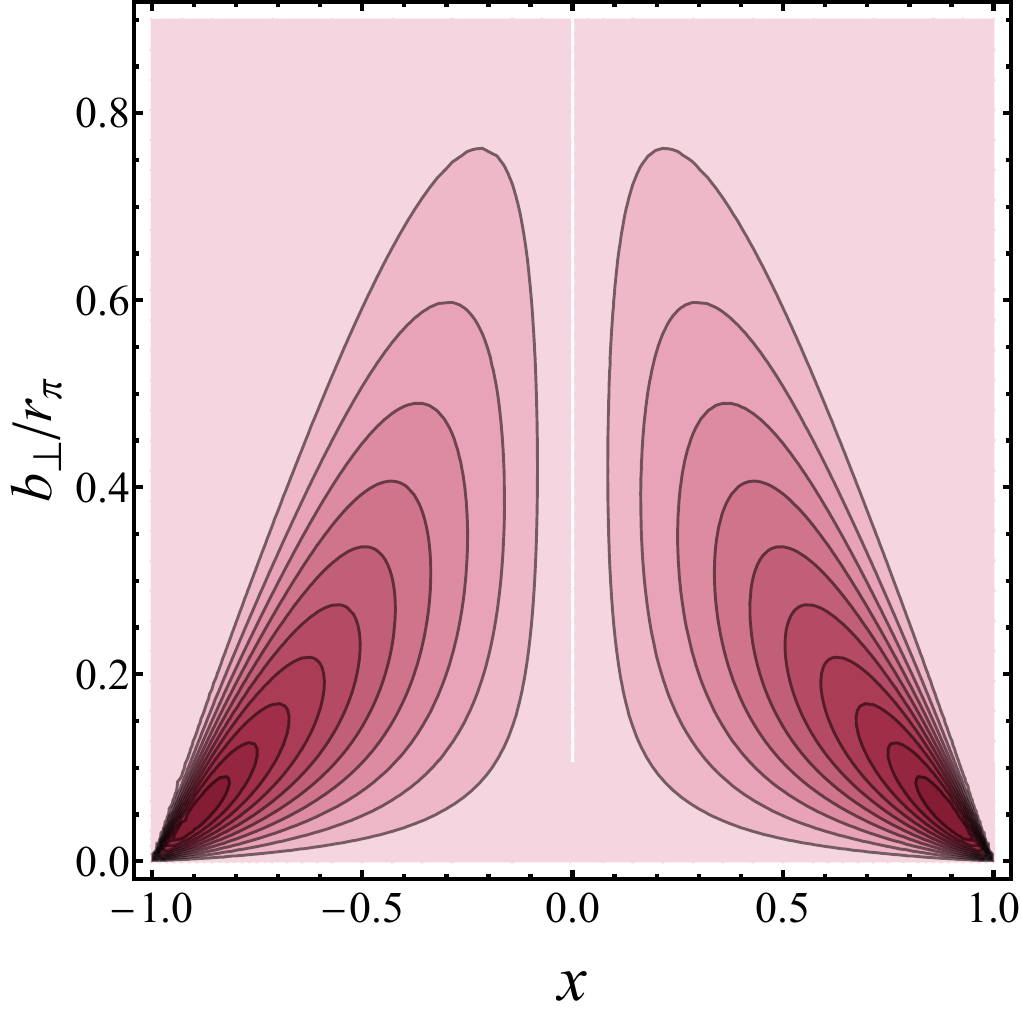}}
\hfill
\subfloat[\centering]{\includegraphics[width=6.0cm]{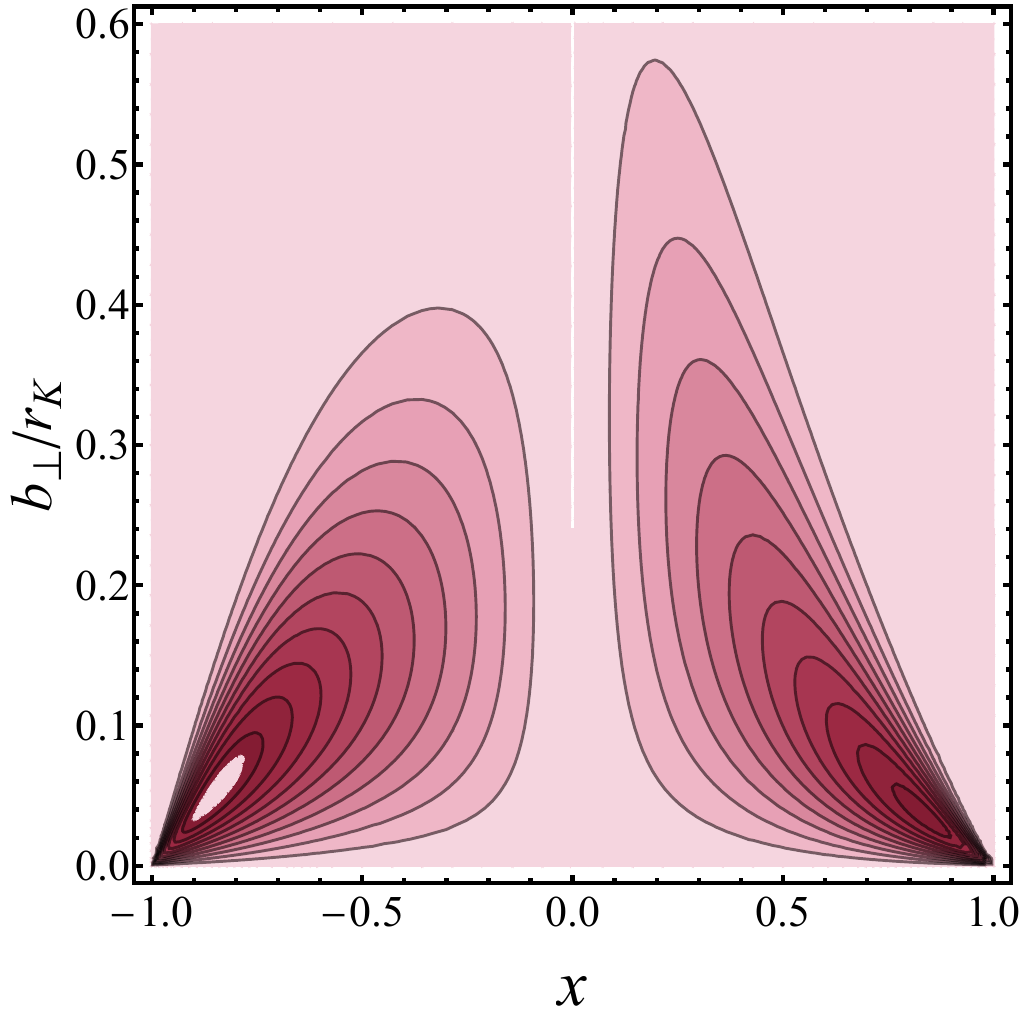}}
\caption{\label{fig:IPS-GPD_mesonspionkaon}
Impact-parameter dependent GPDs for pion and kaon, where the quark and antiquark contributions are assigned to the regions $x>0$ and $x<0$, respectively. The resulting left--right asymmetry reflects the mass imbalance between the dressed valence constituents, with the heavier quark exerting a dominant influence on the transverse center of momentum.}
\end{figure}

The IPS-GPDs for the pion and kaon provide direct insight into the transverse spatial distribution of partons correlated with their longitudinal momentum fraction. These distributions are obtained through the Fourier transform of the zero-skewness GPD are shown in Fig.~\ref{fig:IPS-GPD_mesonspionkaon}. In this representation, the quark (antiquark) contributions populate the regions $x>0$ ($x<0$), enabling a transparent visualization of flavor-dependent spatial asymmetries.

For the pion, isospin symmetry ensures a left--right symmetric IPS-GPD, reflecting the equal contribution of the dressed $u$ and $d$ valence quarks to the transverse center of momentum. In contrast, the kaon exhibits a pronounced asymmetry in impact-parameter space: the heavier $s$ quark is more tightly localized and dominates the transverse center of momentum, while the lighter $u$ quark displays a broader spatial distribution with its maximum shifted toward larger transverse distances. This behavior arises from the explicit breaking of flavor symmetry and is consistent with the skewed structure observed in the kaon PDA and PDF.

Overall, these results demonstrate that increasing the constituent-quark mass leads to a contraction of the transverse spatial distribution and a reduction of its spatial extent, while simultaneously modifying the longitudinal momentum profile. The IPS-GPDs thus provide a natural and intuitive link between the momentum-space information encoded in GPDs and the spatial imaging of partons inside light pseudoscalar mesons.

\subsection{Pseudoscalar Mesons: Heavy-light sector}
In this section, we turn our attention to the heavy–light meson sector, which encompasses bound states formed by one heavy quark ($c$ or $b$) and one light partner ($u$, $d$, or $s$). These systems, such as the $D$, $D_s$, $B$, $B_s$, and $B_c$ mesons, serve as an essential bridge between the dynamics of light and heavy quark physics. They offer a unique window into the interplay between nonperturbative confinement effects and heavy-quark symmetry, allowing for a unified description within the same algebraic framework introduced earlier.

At this point, all necessary components are in place to compute the LFWFs from the PDAs within the framework of the algebraic model. Equipped with these tools, we proceed to determine the LFWFs for the lowest-lying heavy–light pseudoscalar mesons, namely the $D$, $D_s$, $B$, $B_s$, and $B_c$ states. The calculations rely on input PDAs previously determined in the literature.

\begin{figure}[t]
\begin{adjustwidth}{-\extralength}{-2.50cm}
\centering
\subfloat[\centering]{\includegraphics[width=8.5cm]{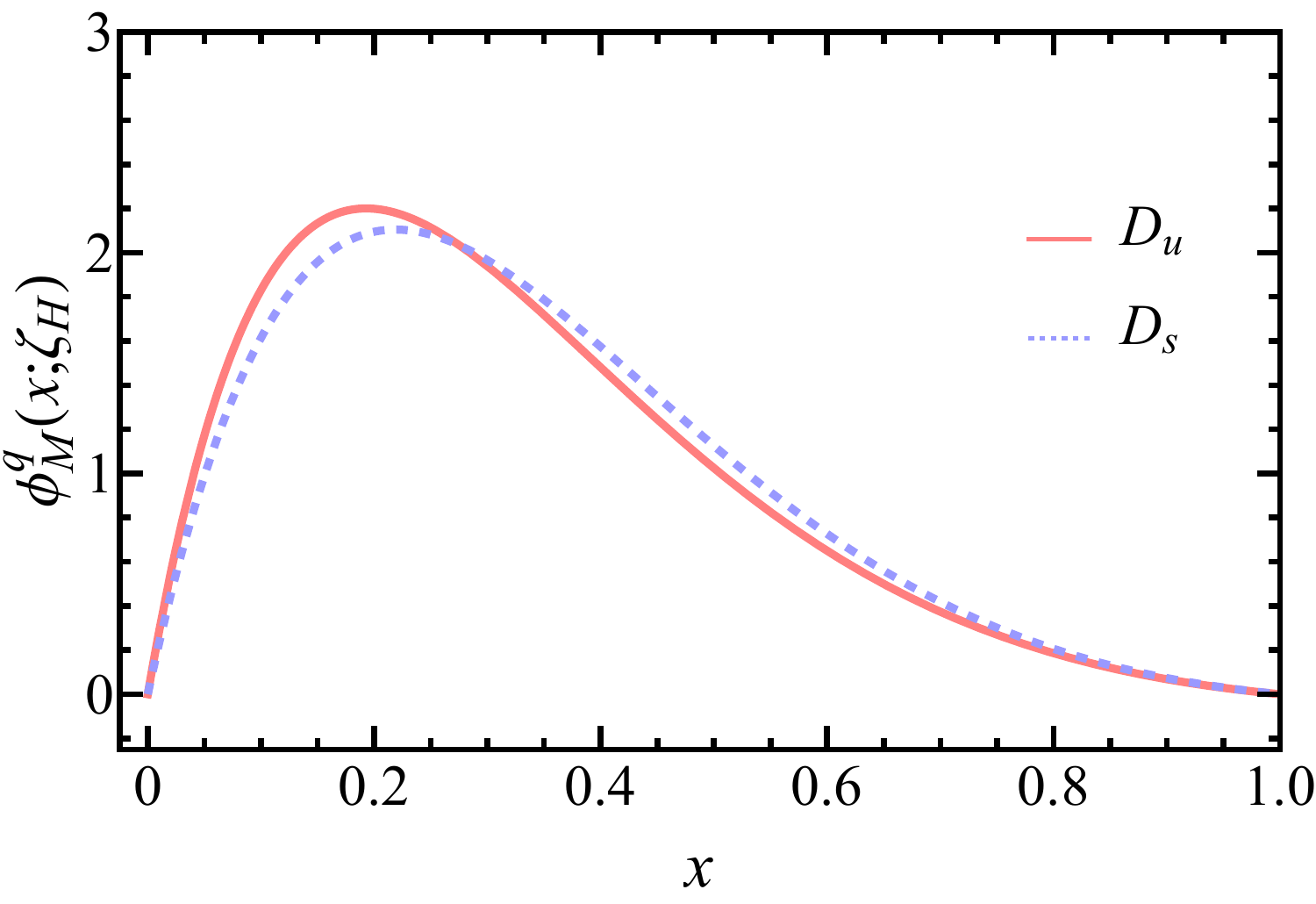}}\\
\subfloat[\centering]{\includegraphics[width=8.5cm]{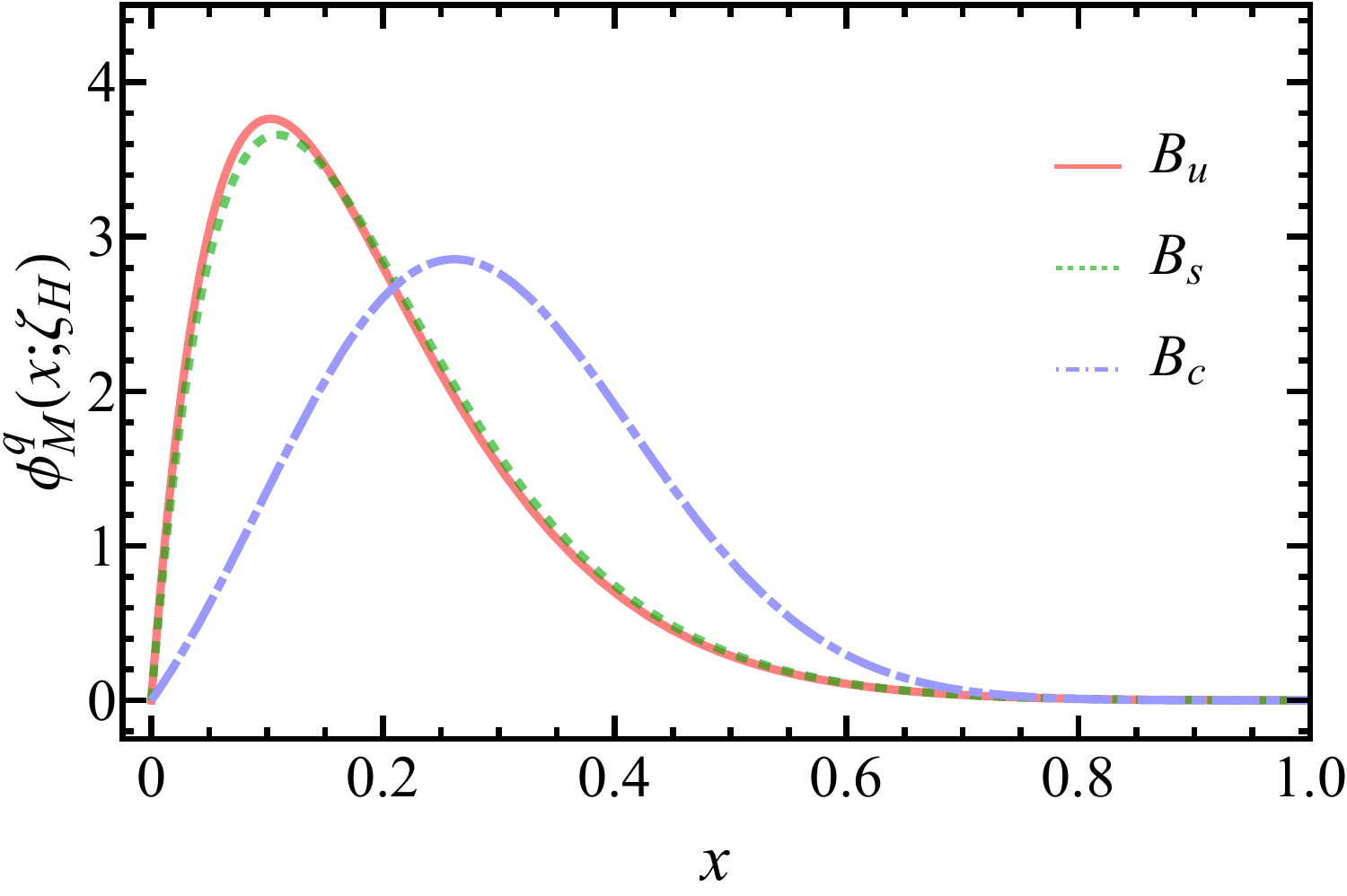}}
\end{adjustwidth}
\caption{Parton distribution amplitudes for heavy-light mesons evaluated at the hadronic scale $\zeta_H$.}
\label{fig:PDAs}
\end{figure}

Over the past decade, the detailed pointwise behavior of light-meson PDAs has been determined with remarkable precision \cite{Chang:2013pq,Chang:2013nia,Segovia:2013eca,Gao:2014bca,Braun:2015axa,Shi:2015esa,Raya:2015gva,Li:2016dzv,Li:2016mah,Gao:2016jka,Chang:2016ouf,Zhang:2017bzy,Gao:2017mmp,Zhang:2017zfe,Chen:2018rwz,Ding:2018xwy}. In contrast, heavy–light systems remain comparatively less explored, as empirical and lattice constraints are still limited \cite{Braun:2003wx,Lee:2005gza,Grozin:2005iz,Beneke:2011nf,Braun:2012kp,Bell:2013tfa,Beneke:2018wjp}. To date, only a few comprehensive studies have provided unified analyses of all heavy–light pseudoscalar PDAs \cite{Binosi:2018rht,Serna:2020txe}. In this work, we employ the rainbow--ladder parametrizations from Ref.~\cite{Binosi:2018rht}, which are well-defined within their applicable domain for the $D$, $D_s$, $B$, $B_s$, and $B_c$ mesons:
\begin{equation}
\phi_{0^-}(x) = 4 N_{\alpha\beta} x \bar{x} \, e^{4\alpha^2x\bar{x} - \beta^2 (x-\bar{x})} \,,
\label{eq:PDApara}
\end{equation}
where $N_{\alpha\beta}$ ensures proper normalization, and $\bar{x}=1-x$. The fitted $(\alpha,\beta)$ parameters used to characterize each PDA are listed in Table~\ref{tab:AlphaBeta}, and the resulting distributions are shown in Fig.~\ref{fig:PDAs}. As the mass asymmetry between the valence quarks increases, the PDAs become more asymmetric and sharply peaked toward the heavier constituent. This trend reflects the pronounced momentum imbalance between the heavy and light quarks, which plays a decisive role in shaping the internal structure of heavy–light mesons.

\begin{table}[h]
\caption{\label{tab:AlphaBeta} The $(\alpha,\beta)$-pairs that specify the PDAs of heavy-light mesons via Eq. \eqref{eq:PDApara}.}
  \centering
  \begin{tabular}{cccccc}
    \toprule
    & $D$ & $D_s$ & $B$ & $B_s$ & $B_c$ \\
    \midrule
    $\alpha$ & $0.265(30)$ & $0.508(30)$ & $0.497(70)$ & $0.669(60)$ & $1.901(70)$ \\
    $\beta$  & $1.435(30)$ & $1.391(30)$ & $2.166(60)$ & $2.177(60)$ & $2.163(60)$ \\
    \bottomrule
  \end{tabular}
\end{table}

\begin{table}[h]
\centering
\caption{\label{tab:Parameters} Meson masses ($m_\mathrm{M}$) and leptonic decay constants ($f_\mathrm{M}$) for the lowest-lying heavy–light pseudoscalar states, expressed in GeV. The dressed quark masses ($M_q$) are also given in GeV, while the electric charges ($e_q$) are quoted in units of the positron charge. The corresponding antiquarks possess equal masses but opposite electric charges.
}
\begin{tabular}{lccccc}
\toprule
& \multicolumn{2}{c}{Mesons} & & \multicolumn{2}{c}{Quarks} \\
& $m_\mathrm{M}$ & $f_\mathrm{M}$ & & $M_q$ & $e_q$ \\
\midrule
$D$   & $1.88(5)$  & $0.158(8)$  & $u$ & $0.533$  & $2/3$     \\
$D_s$ & $1.94(4)$  & $0.171(6)$  & $d$ & $0.533$  & $-1/3$     \\
$B$   & $5.30(15)$ & $0.142(13)$ & $s$ & $0.717$ & $-1/3$     \\
$B_s$ & $5.38(13)$ & $0.179(12)$ & $c$ & $1.65$    & $2/3$     \\
$B_c$ & $6.31(1)$  & $0.367(1)$  & $b$ & $5.09$    & $-1/3$  \\
\bottomrule
\end{tabular}
\end{table}

With all inputs established, we proceed to compute the LFWFs for the $D$, $D_s$, $B$, $B_s$, and $B_c$ mesons, employing their connection to the corresponding PDAs as defined in Eq.~\eqref{LFWF}. The parameters required for these calculations are listed in Table~\ref{tab:Parameters}. As shown therein, the static properties of the heavy–light pseudoscalar mesons obtained within this framework exhibit excellent agreement with available experimental data reported by the Particle Data Group~\cite{ParticleDataGroup:2022pth}, as well as with recent Lattice QCD determinations~\cite{Chiu:2007bc}. This consistency underscores the reliability of the algebraic model in describing systems that interpolate between the light and heavy quark regimes, effectively capturing the essential nonperturbative dynamics governing their internal structure.

\begin{figure}[H]
\begin{adjustwidth}{-\extralength}{-2.50cm}
\centering
\subfloat[\centering]{\includegraphics[width=9.0cm]{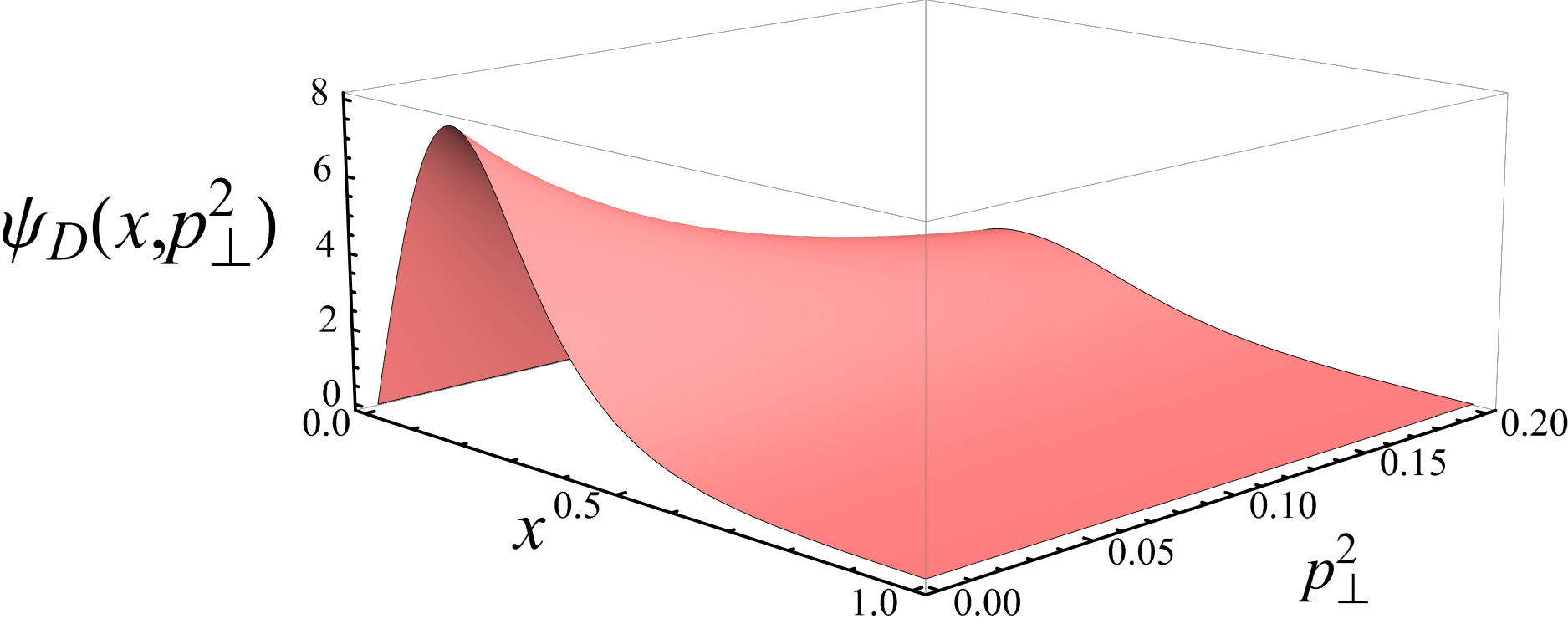}}\\
\subfloat[\centering]{\includegraphics[width=9.0cm]{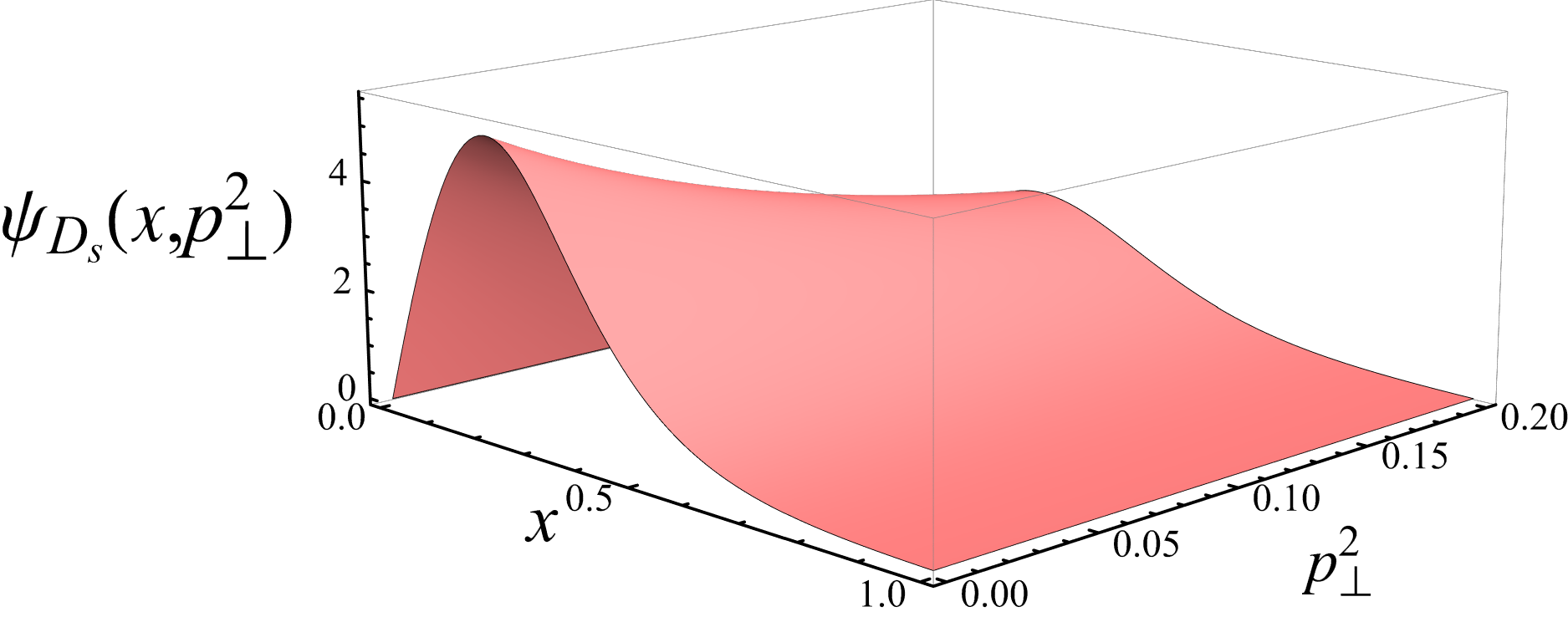}}\\
\subfloat[\centering]{\includegraphics[width=9.0cm]{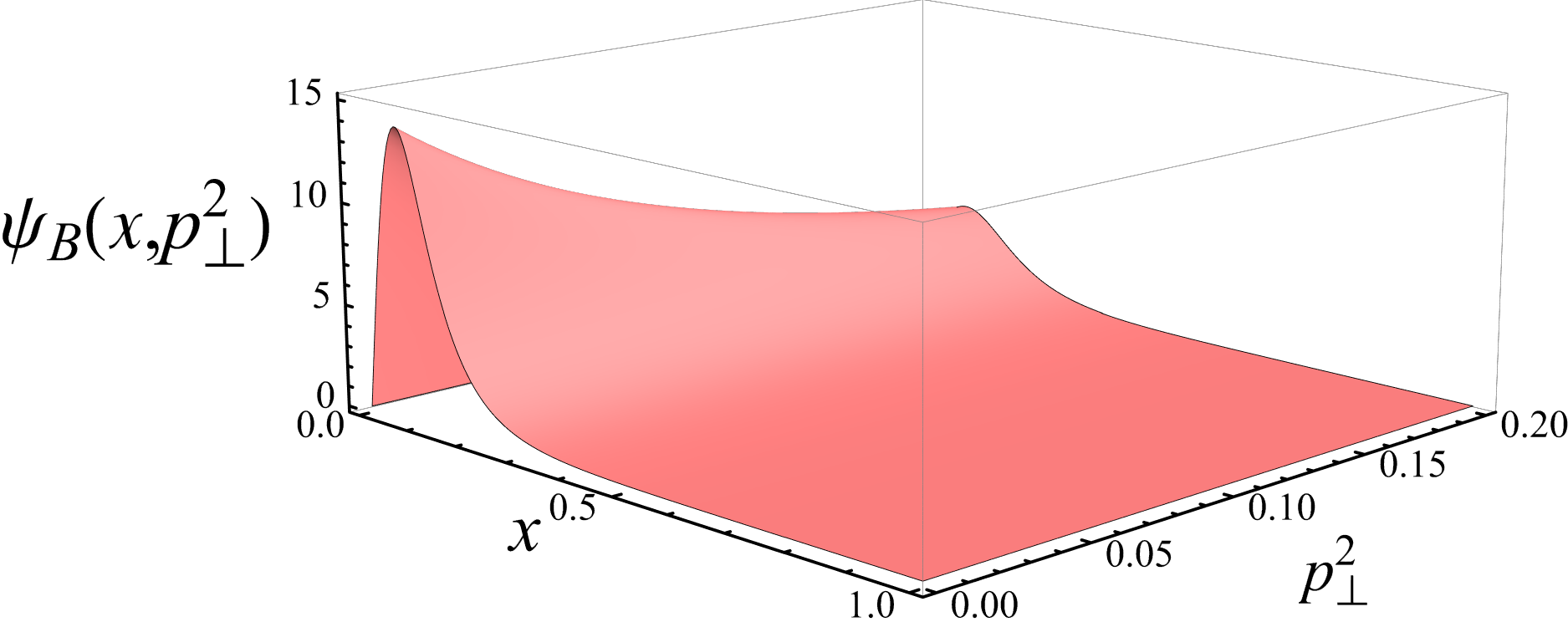}}\\
\subfloat[\centering]{\includegraphics[width=9.0cm]{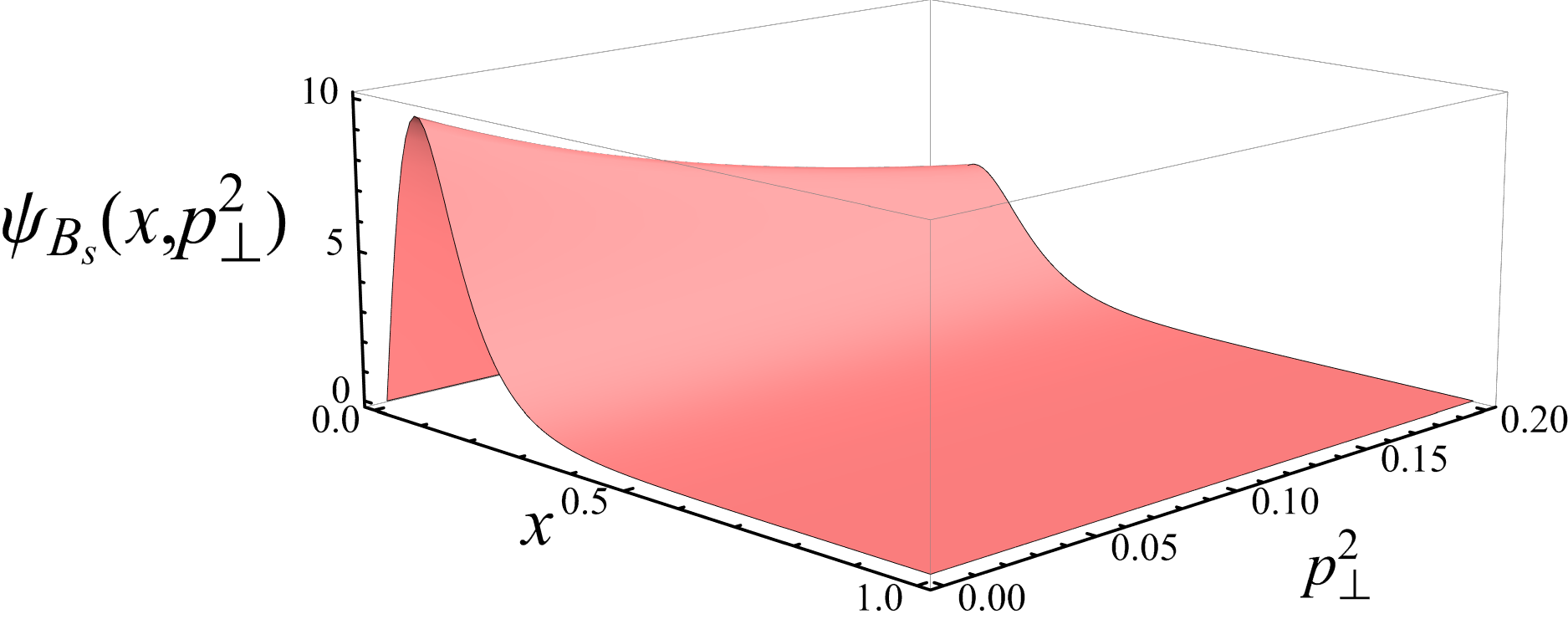}}\\
\subfloat[\centering]{\includegraphics[width=9.0cm]{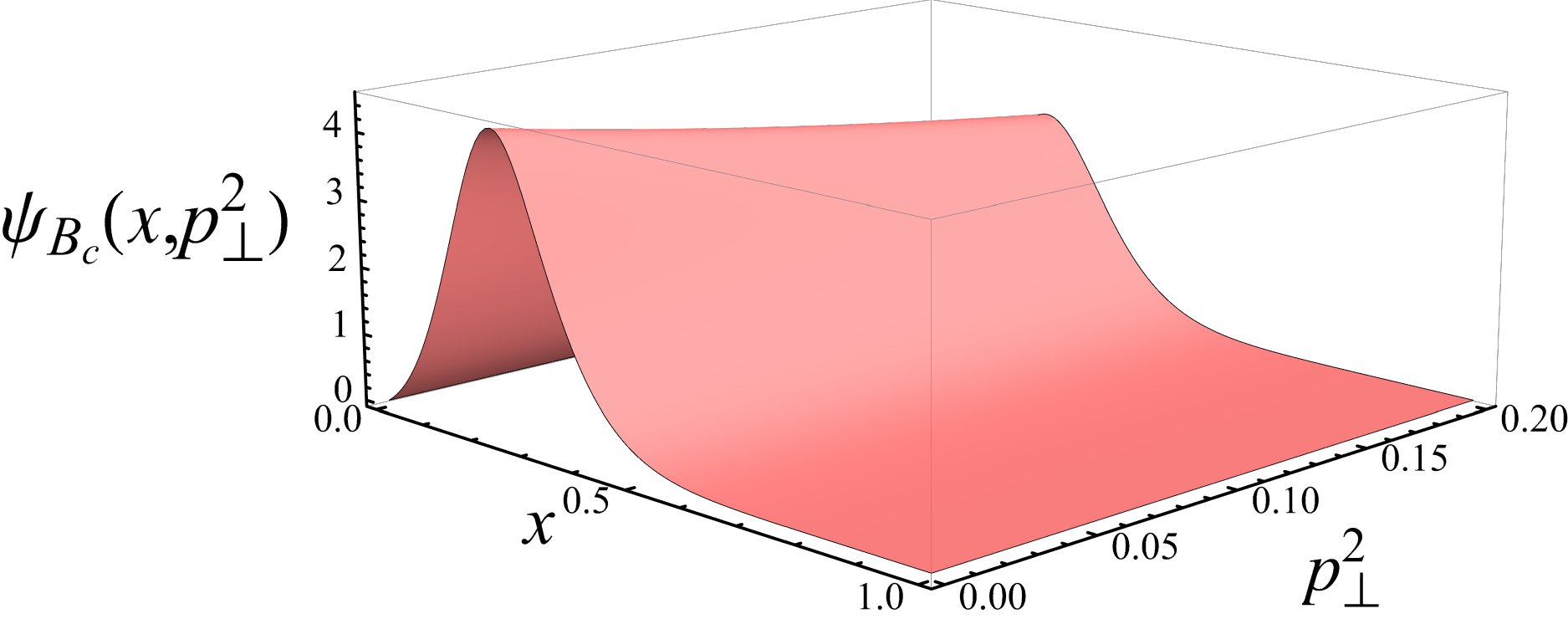}}
\end{adjustwidth}
\caption{Light-front wave functions of heavy--light pseudoscalar mesons as functions of the longitudinal momentum fraction \( x \) and the transverse momentum squared \( p_\perp^2 \). Panel (a) shows the LFWF of the \( D \) meson, while panel (b) corresponds to the \( D_s \) meson; (c) corresponds to the $B$, (d) $B_s$, and (e) $B_c$.}
\label{fig:LFWFsDmesons}
\end{figure}

Fig. ~\ref{fig:LFWFsDmesons} present the leading-twist LFWFs for the lowest-lying heavy–light pseudoscalar mesons, namely $D$, $D_s$, $B$, $B_s$, and $B_c$. A prominent feature of all these distributions is their pronounced asymmetry, which originates from the mass difference between the valence quark and antiquark constituents. Examining their dependence on the longitudinal momentum fraction $x$ and the transverse momentum squared $p_\perp^2$, distinct patterns emerge. 

In the charm sector, the $D$-meson LFWF exhibits a sharper dependence on $x$, yielding a narrower distribution compared to the broader and more symmetric profile of the $D_s$ meson. Conversely, in the bottom sector, the LFWFs display a systematic evolution: the $B$ meson shows the narrowest distribution in $x$, while the $B_s$ and $B_c$ wave functions become progressively wider. This indicates that as the light valence quark becomes heavier, the longitudinal momentum is distributed more evenly between the two constituents. In contrast, for lighter valence quarks, the probability distribution becomes increasingly localized around smaller $x$, reflecting a stronger momentum imbalance within the bound state.

Regarding the transverse momentum dependence, all wave functions display the expected damping with increasing $p_\perp^2$. For the $D$ meson, the LFWF amplitude decreases sharply as $p_\perp^2$ increases from $0$ to $0.2\,\text{GeV}^2$, while in the $D_s$ meson this decline is more gradual. A similar trend is observed for the bottom mesons, although the suppression with $p_\perp^2$ becomes even milder as the heavy-quark mass increases. Hence, as the mass of the valence light quark rises, the LFWF becomes less sensitive to transverse momentum, revealing that systems with heavier constituents are more compact in configuration space but extend further in momentum space.

Having established the LFWFs for the heavy–light pseudoscalar mesons, we now proceed to compute their corresponding GPDs. Following the formalism introduced in Eq.~\eqref{GPD_pseudo}, we evaluate the valence-quark GPDs for the $D$- and $B$-mesons. This analysis allows us to explore how the internal partonic correlations evolve with the heavy-quark mass and how the three-dimensional structure of these mesons manifests in both momentum and impact-parameter space.

\begin{figure}[H]
\begin{adjustwidth}{-\extralength}{-2.50cm}
\centering
\subfloat[\centering]{\includegraphics[width=9.0cm]{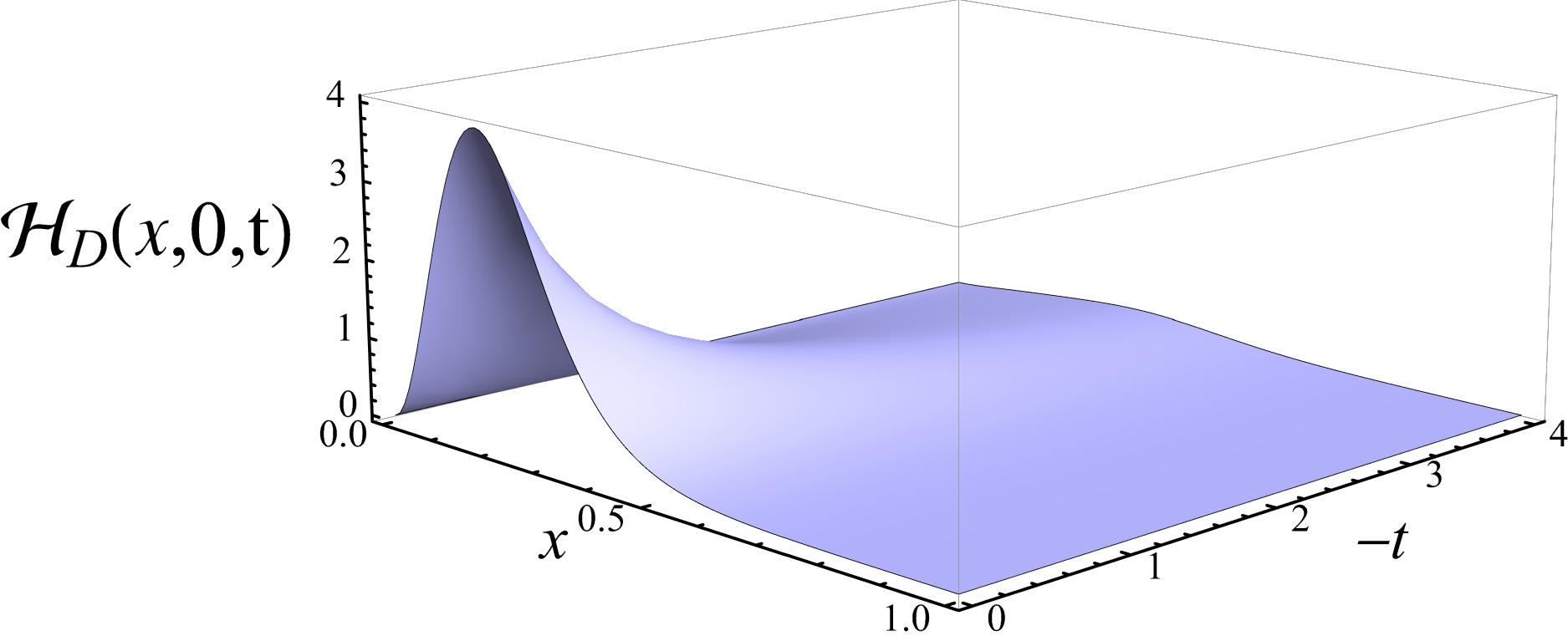}}\\
\subfloat[\centering]{\includegraphics[width=9.0cm]{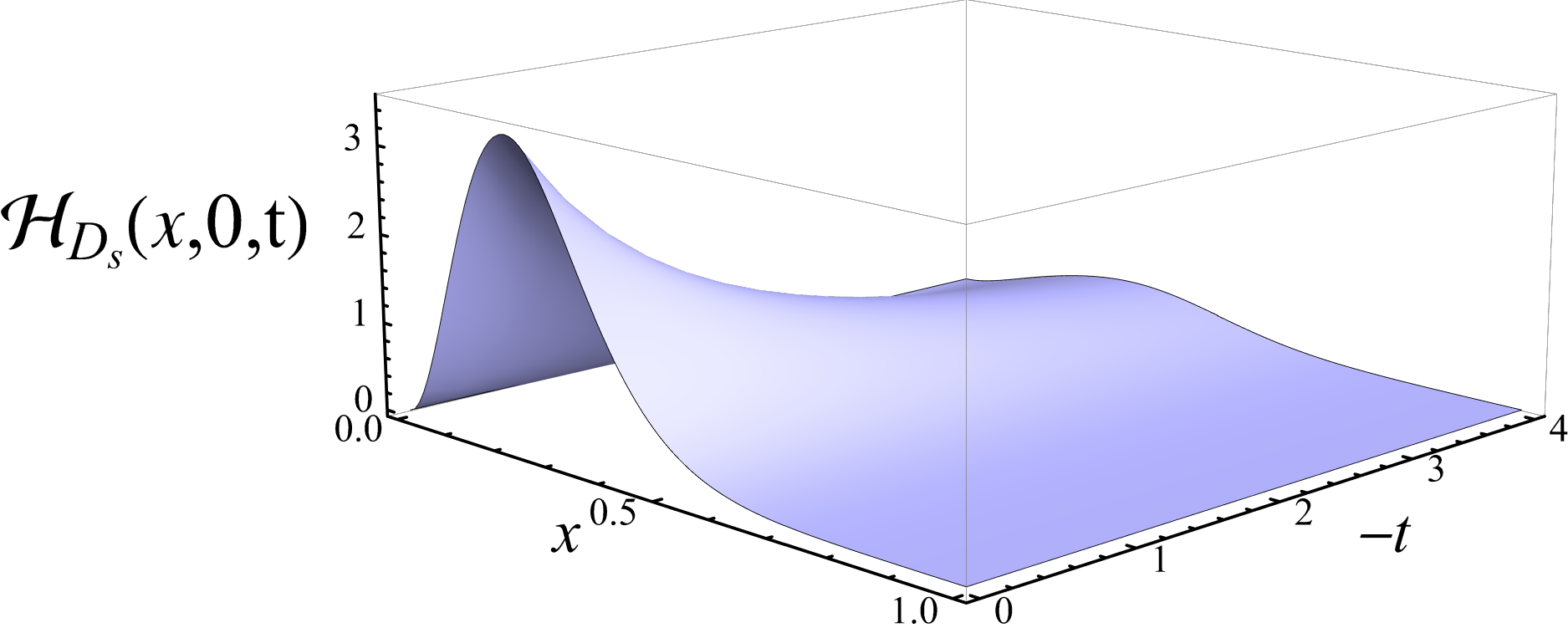}}\\
\subfloat[\centering]{\includegraphics[width=10.0cm]{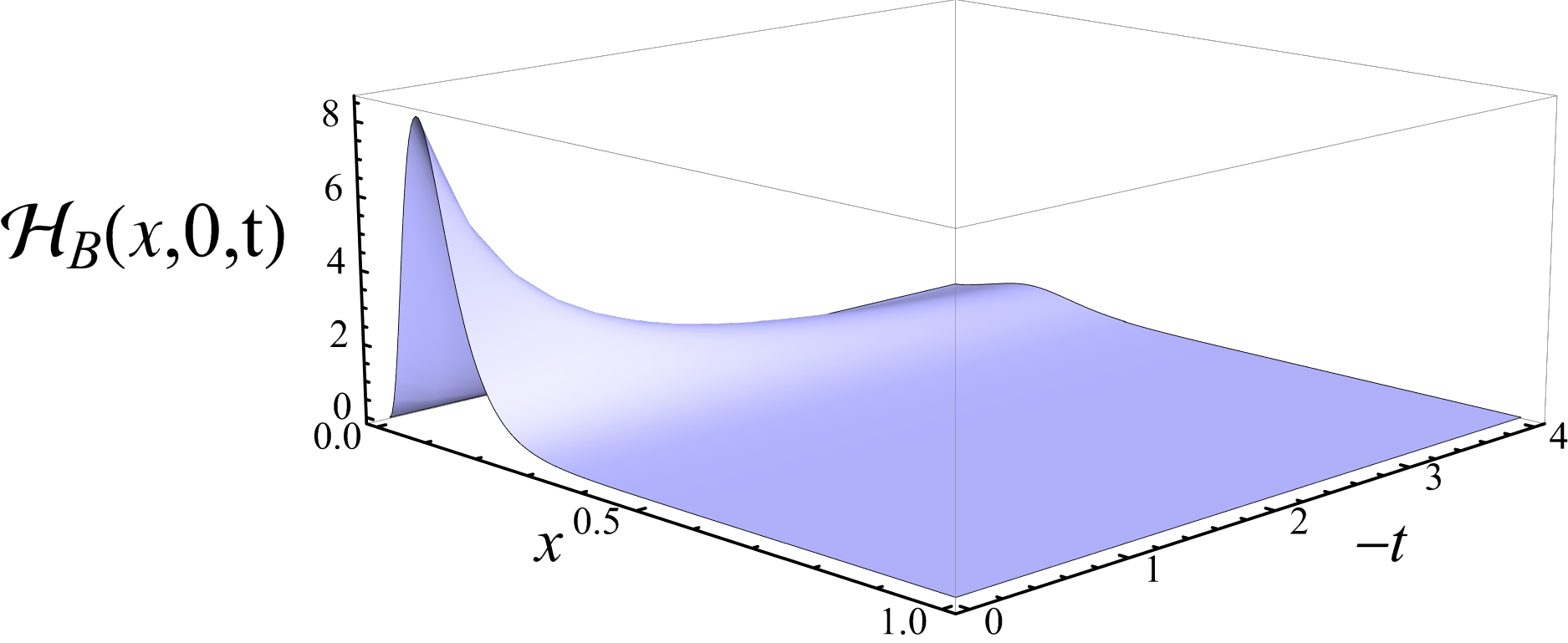}}\\
\subfloat[\centering]{\includegraphics[width=9.0cm]{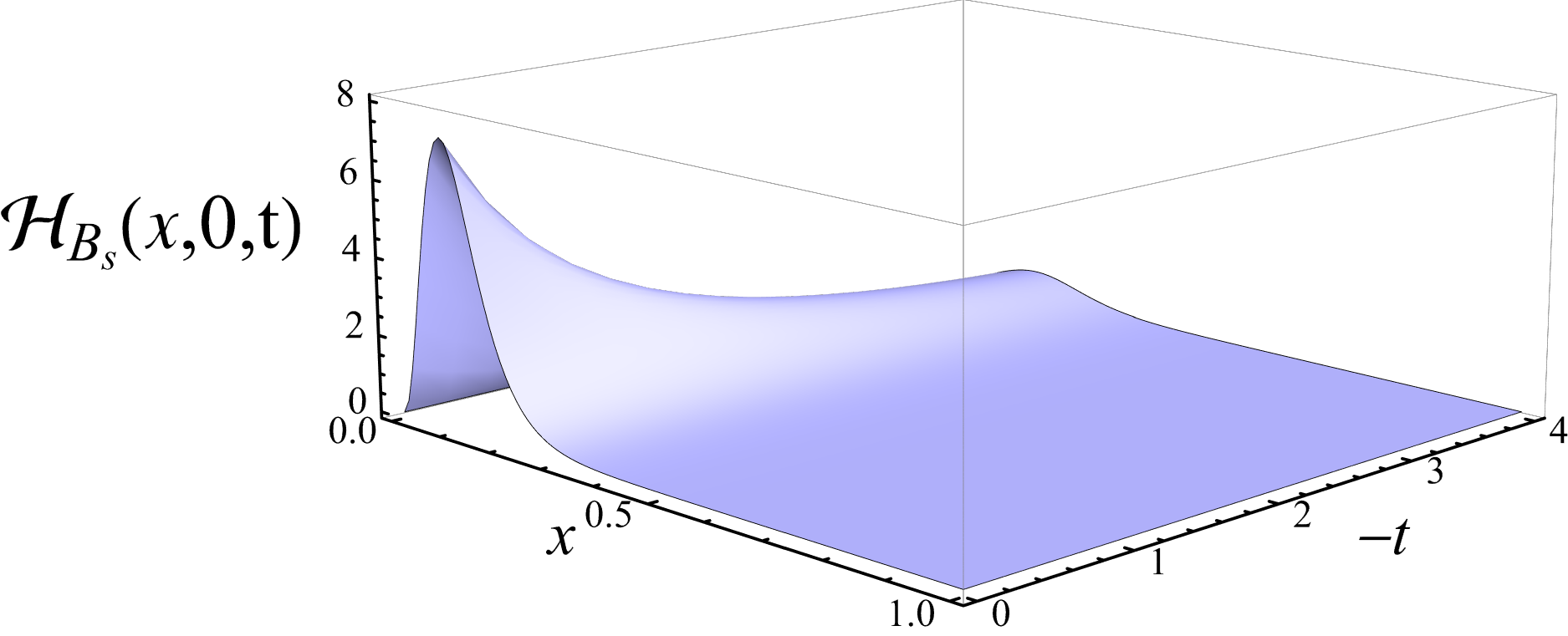}}\\
\subfloat[\centering]{\includegraphics[width=9.0cm]{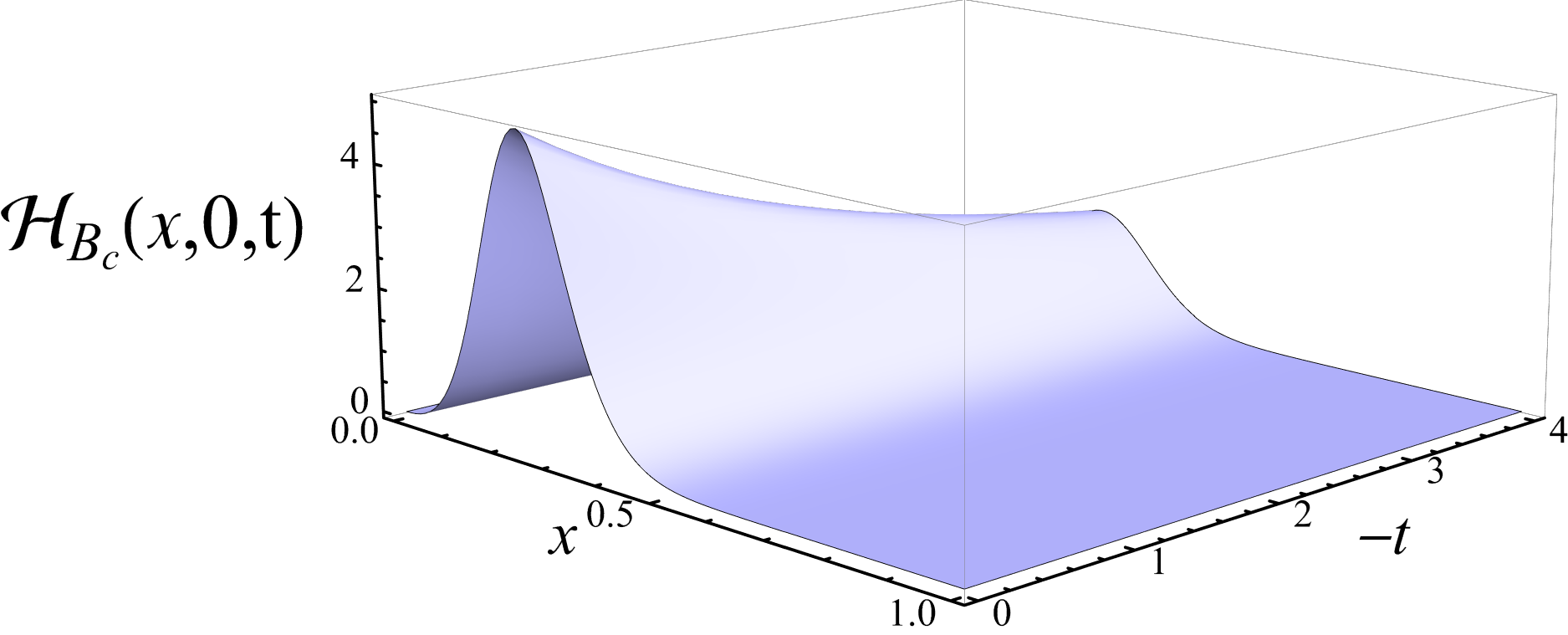}}
\end{adjustwidth}
\caption{Generalized parton distributions of heavy--light pseudoscalar mesons as functions of the longitudinal momentum fraction \( x \) and the momentum transfer squared \( -t \). Panel (a) shows the GPD of the \( D \) meson, while panel (b) corresponds to the \( D_s \) meson; (c) corresponds to $B$, (d) $B_s$ and (e) $B_c$.}
\label{fig:DGPDs}
\end{figure}

Figure~\ref{fig:DGPDs} shows the valence-quark GPDs. For the case of $\mathcal{H}_{D}(x,0,t)$ and $\mathcal{H}_{D_s}(x,0,t)$, both distributions are strongly asymmetric in $x$, reflecting the characteristic momentum imbalance of heavy–light systems in which the charm quark carries most of the longitudinal momentum. The $D_s$ GPD peaks at slightly larger values of $x$ than the $D$ meson, consistent with the heavier strange quark sharing a larger fraction of the total momentum.

In the $t$-dependence, both GPDs decrease monotonically as $-t$ increases, but the $D_s$ meson exhibits a noticeably slower falloff. This behavior indicates a more compact transverse structure for the $D_s$, in contrast with the more diffuse spatial profile of the $D$ meson driven by its lighter $u/d$ valence quark. Overall, the comparison highlights how increasing the mass of the light quark leads to a GPD that is less asymmetric in $x$ and harder in $t$, signaling enhanced binding and reduced transverse extent.

The GPDs of the bottom mesons displayed in Fig.~\ref{fig:DGPDs} exhibit the characteristic signatures of heavy--light systems dominated by the large bottom-quark mass. In all three cases, the GPD is strongly peaked at small values of the light-quark momentum fraction $x$, reflecting the pronounced momentum imbalance between the heavy $b$ quark and its lighter partner. This peak is sharpest for the $B$ meson, where the valence light quark is the lightest, indicating that the bulk of the longitudinal momentum is carried by the bottom quark. As the mass of the light quark increases, moving from $B$ to $B_s$ and finally to $B_c$, the peak shifts slightly toward larger $x$ and becomes less abrupt, signaling a more balanced longitudinal momentum sharing within the meson.

The dependence on the momentum transfer $t$ follows a similar pattern across the three systems. The $B$-meson GPD decreases rapidly with increasing $-t$, consistent with a compact transverse spatial distribution. In contrast, the falloff becomes progressively slower for the $B_s$ and especially the $B_c$ meson, indicating increasingly localized partonic distributions in impact-parameter space as the combined quark mass grows. Altogether, these results reveal a coherent and systematic influence of both the heavy- and light-quark masses on the longitudinal and transverse structure encoded in the GPDs, providing a unified physical picture across the spectrum of heavy--light pseudoscalar mesons.

In the forward limit, corresponding to vanishing momentum transfer and skewness ($t = 0$, $\xi = 0$), the GPD reduces to the valence-quark PDF:
\begin{equation}
q_{0^-}(x) = H_{0^-}(x,0,0).
\end{equation}

At the hadronic scale $\zeta_H$, where the meson is described solely in terms of its dressed valence degrees of freedom, the associated antiquark distribution is simply $\bar{Q}_{0^-}(\zeta_H;x) = q_{0^-}(\zeta_H;1-x)$. This identity highlights that, at this scale, the valence constituents exhaust the longitudinal momentum of the bound state.

\begin{figure}[t]
\begin{adjustwidth}{-\extralength}{-2.50cm}
\centering
\subfloat[\centering]{\includegraphics[width=10.0cm]{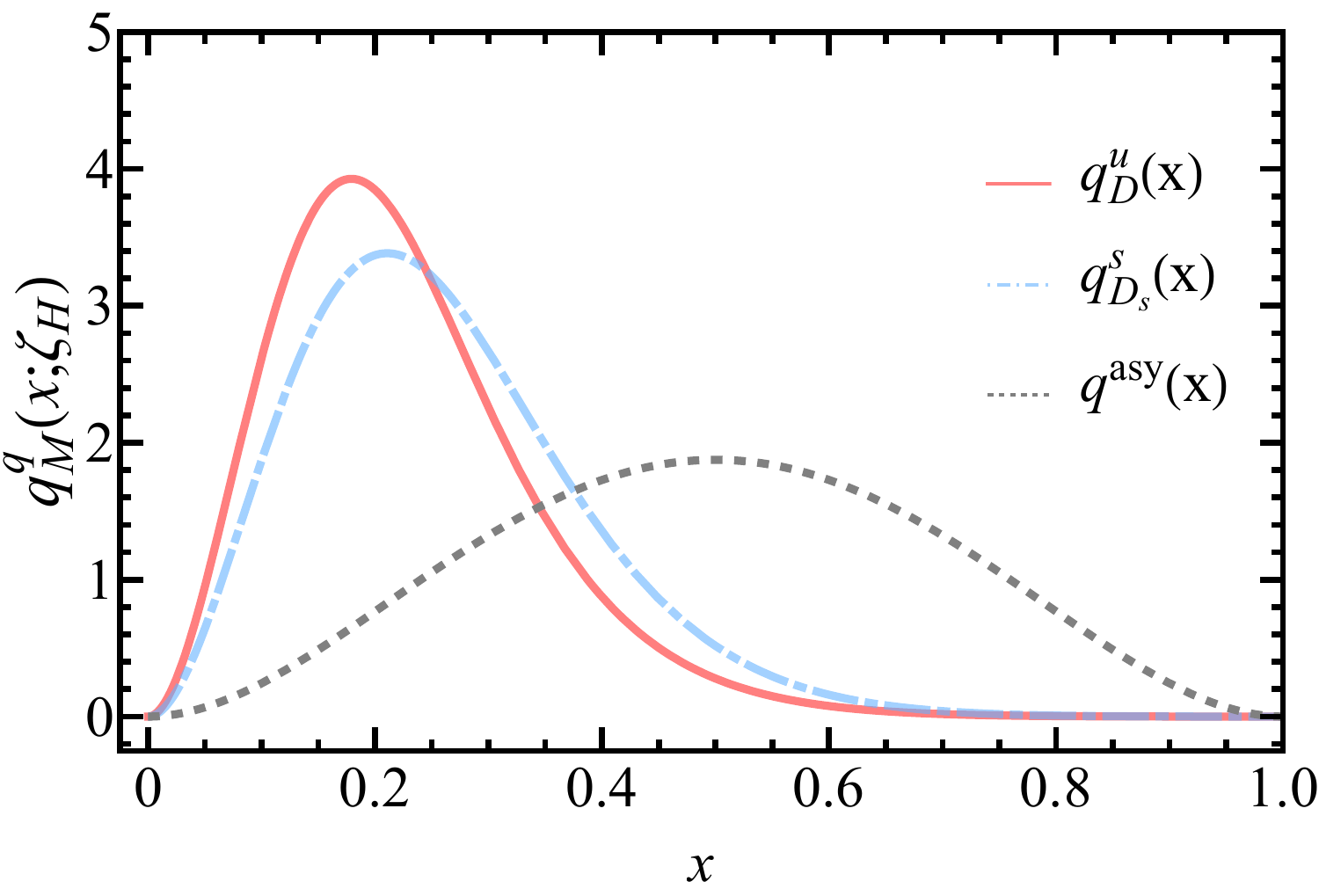}}\\
\subfloat[\centering]{\includegraphics[width=10.0cm]{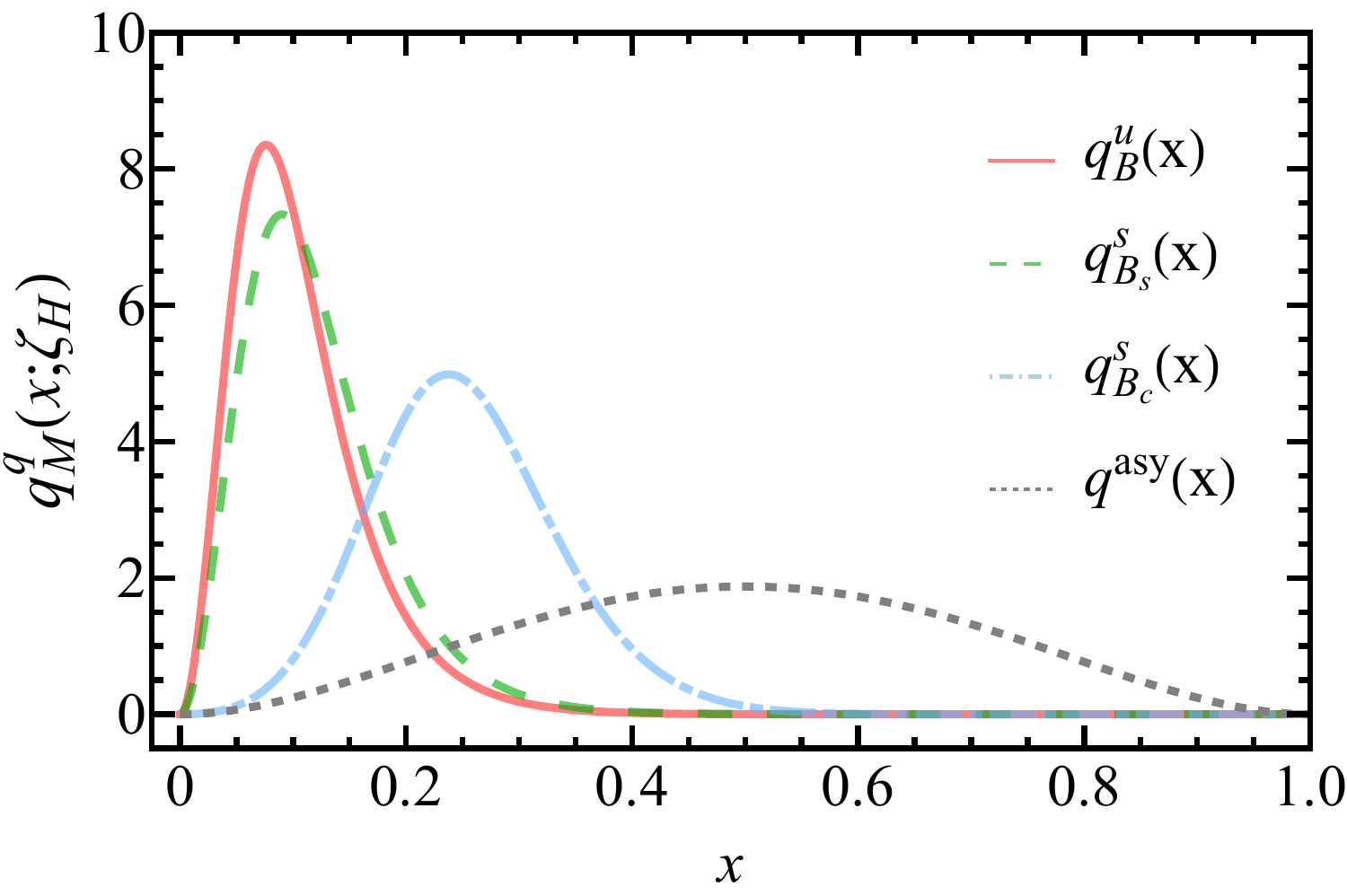}}
\end{adjustwidth}
\caption{Parton distribution functions of pseudoscalar mesons. Panel (a) shows the PDFs of the \( D \) and \( D_s \) mesons, while panel (b) displays those of the \( B \), \( B_s \), and \( B_c \) mesons. For completeness, the results are compared with the asymptotic distribution \( q(x) = 30\,x^2(1-x)^2 \). The comparison illustrates the impact of quark-mass effects on the longitudinal momentum distributions within the adopted light-front framework.}
\label{fig:PDFs}
\end{figure}

Figure~\ref{fig:PDFs} displays the resulting valence light-quark PDFs for the lowest-lying pseudoscalar charmed mesons and their bottom analogues. For reference, the dashed black curve shows the conformal parton-like profile $q(x)=30x^2(1-x)^2$. All model predictions are noticeably more sharply peaked than this benchmark and exhibit the characteristic $x$-asymmetry expected in heavy–light systems. A systematic flavor dependence is also evident: as the mass of the light valence quark increases, the PDFs become broader and their maxima shift toward larger momentum fractions. This effect is most pronounced in the $B_c$ meson, whose valence light-quark distribution is significantly displaced relative to those of the $B$ and $B_s$ mesons.

\begin{figure}[t]
\begin{adjustwidth}{-\extralength}{-2.50cm}
\centering
\subfloat[\centering]{\includegraphics[width=10.0cm]{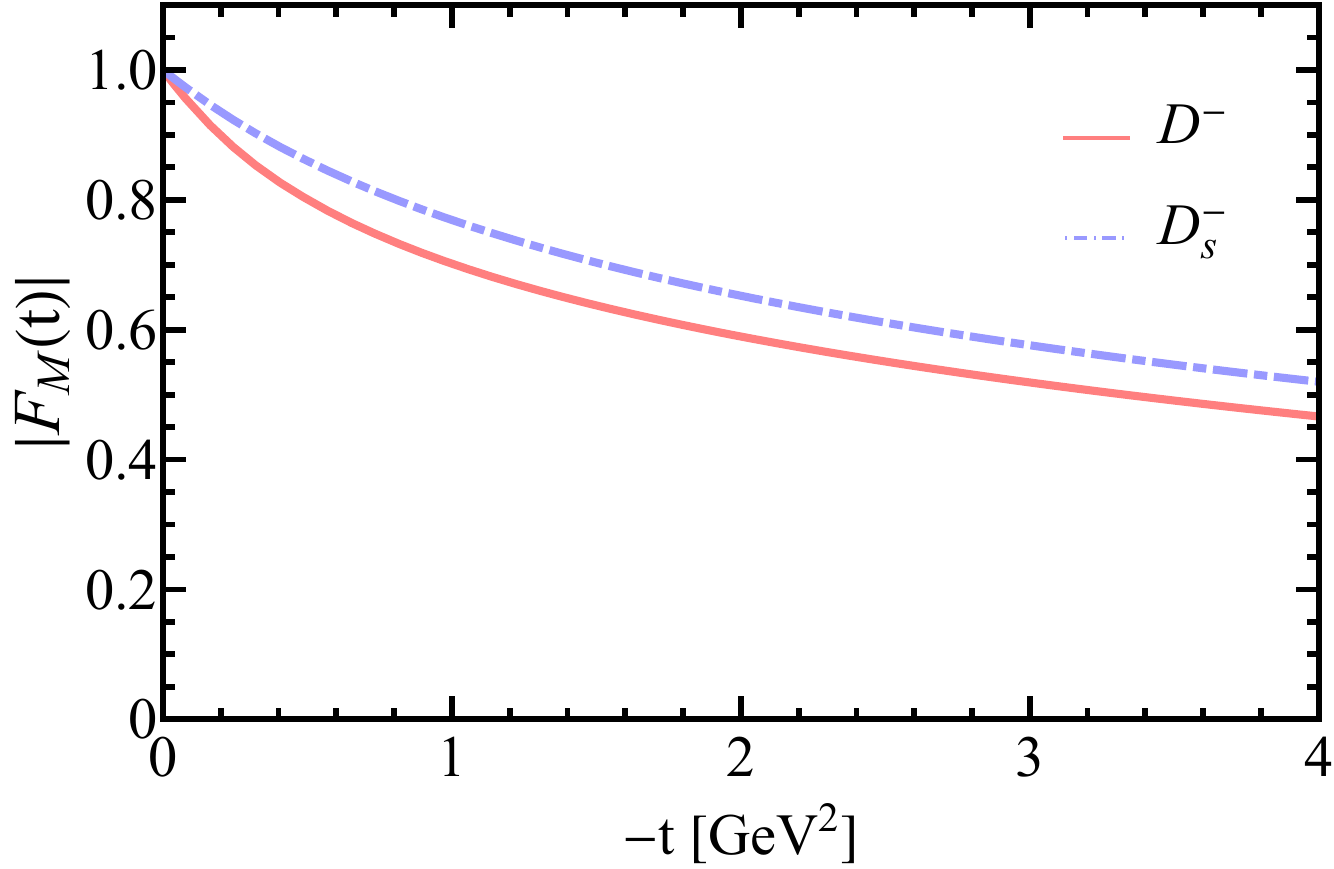}}\\
\subfloat[\centering]{\includegraphics[width=10.0cm]{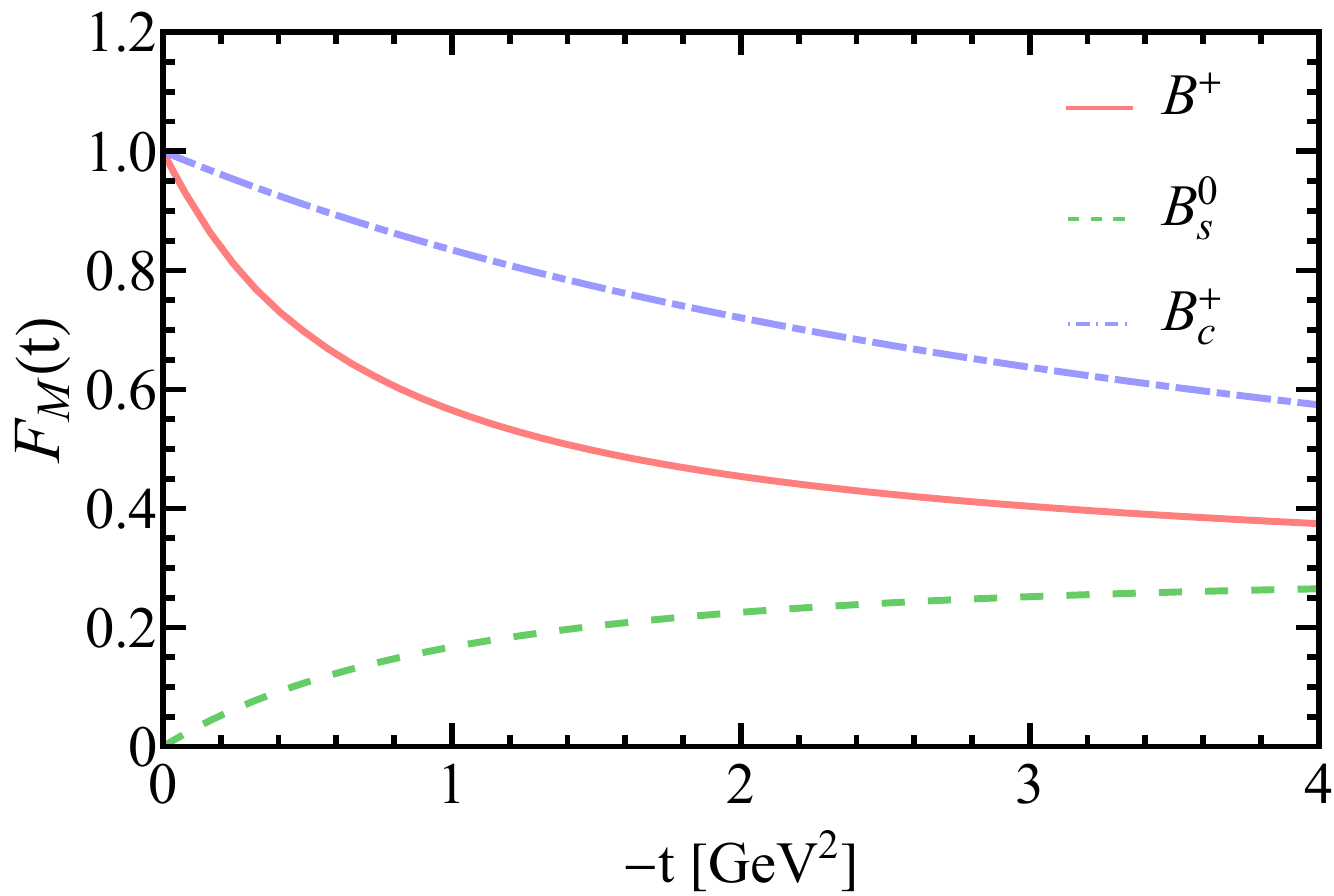}}
\end{adjustwidth}
\caption{Elastic electromagnetic form factors of pseudoscalar mesons. Panel (a) shows the EFFs of the \( D \) and \( D_s \) mesons, while panel (b) displays those of the \( B \), \( B_s \), and \( B_c \) mesons.}
\label{fig:EFFs}
\end{figure} 

Figure~\ref{fig:EFFs} presents the elastic EFFs obtained for the lowest-lying pseudoscalar charmed mesons and their bottom counterparts. In all cases, the curves correspond to the charged mesons composed of a light valence quark $q$ and a heavy antiquark $\bar{Q}$, namely
$D^- = d\bar{c}$, $D_s^- = s\bar{c}$, $B^+ = u\bar{b}$, $B_s^+ = s\bar{b}$, and $B_c^+ = c\bar{b}$.

A clear systematic behavior is observed: the EFF exhibits a slower decrease with increasing momentum transfer when the mass difference between the valence constituents is reduced. Moreover, within each heavy-quark sector, the various mesons display a similar asymptotic falloff of their form factors. This qualitative behavior is consistent with earlier investigations \cite{Moita:2021xcd,Das:2016rio,Hwang:2001th,Hernandez-Pinto:2023yin}. Nevertheless, a direct comparison with those results is not pursued here, since the corresponding hadronic scales are not specified, preventing a meaningful quantitative assessment.

\begin{table}[h]
\centering
\caption{\label{tab:EMradii} Charge radii (in fm) of heavy--light pseudoscalar mesons. Owing to the absence of a universally defined hadronic scale in many alternative approaches, a direct one-to-one comparison is not always possible. Nevertheless, results from Refs.~\cite{Moita:2021xcd,Das:2016rio,Hwang:2001th,Hernandez-Pinto:2023yin,Li:2017eic,Can:2012tx} are quoted for qualitative comparison and context.}
\begin{tabular}{lrrrrrr}
  \toprule
   & & $D^{-}$ & $D_{s}^{-}$ & $B^{+}$ & $B_s^{0}$ & $B_c^{+}$ \\
  \midrule
  $|r_{0^{-}}|$ & & 0.374 & 0.289 & 0.478 & 0.264 & 0.217 \\
  Covariant CQM & \cite{Moita:2021xcd} & 0.505 & 0.377 & – & – & – \\
  PM             & \cite{Das:2016rio}    & –     & 0.460 & 0.730 & 0.460 & – \\
  LFQM           & \cite{Hwang:2001th}   & 0.429 & 0.352 & 0.615 & 0.345 & 0.208 \\
  CI             & \cite{Hernandez-Pinto:2023yin} & – & 0.260 & 0.340 & 0.240 & 0.170 \\
  Lattice        & \cite{Li:2017eic}     & 0.450(24) & 0.465(57) & – & – & – \\
  Lattice        & \cite{Can:2012tx}      & 0.390(33) & – & – & – & – \\
  \bottomrule
\end{tabular}
\end{table}

The charge radii extracted from the slopes of the EFFs at $Q^2=0$ are collected in Table~\ref{tab:EMradii}, together with available lattice QCD results and predictions from other theoretical approaches. For both the $D$- and $B$-meson sectors, our results show good overall agreement with existing studies. It is also instructive to contrast these predictions with those obtained using the contact interaction (CI) framework. Within the algebraic model, the charge radii (in fm) are found to be
$0.374$, $0.289$, $0.478$, $0.264$, and $0.217$, respectively, whereas the CI yields systematically smaller values: no result for $D^-$, and $0.260$, $0.340$, $0.240$, and $0.170$ for the remaining mesons.

This systematic enhancement of the charge radii in the algebraic model can be traced back to its momentum-dependent interaction, which produces softer Bethe--Salpeter amplitudes and consequently broader spatial distributions. In contrast, the momentum-independent kernel of the CI leads to harder amplitudes and more localized charge distributions.

To further quantify these differences, we evaluate the relative deviations between the charge radii predicted by the algebraic model and those obtained with the CI. For the $D_s^-$, $B^+$, $B_s^+$, and $B_c^+$ mesons, the algebraic model yields radii that are approximately $11.2\%$, $40.6\%$, $10.0\%$, and $27.6\%$ larger, respectively. These sizeable discrepancies highlight the crucial role played by momentum dependence in determining the spatial structure of heavy--light mesons. In particular, the larger deviations observed for the $B^+$ and $B_c^+$ systems suggest that the algebraic model provides a more realistic description of internal dynamics when heavy bottom quarks are involved.

\begin{figure}[H]
\centering
\subfloat[\centering]{\includegraphics[width=7.0cm]{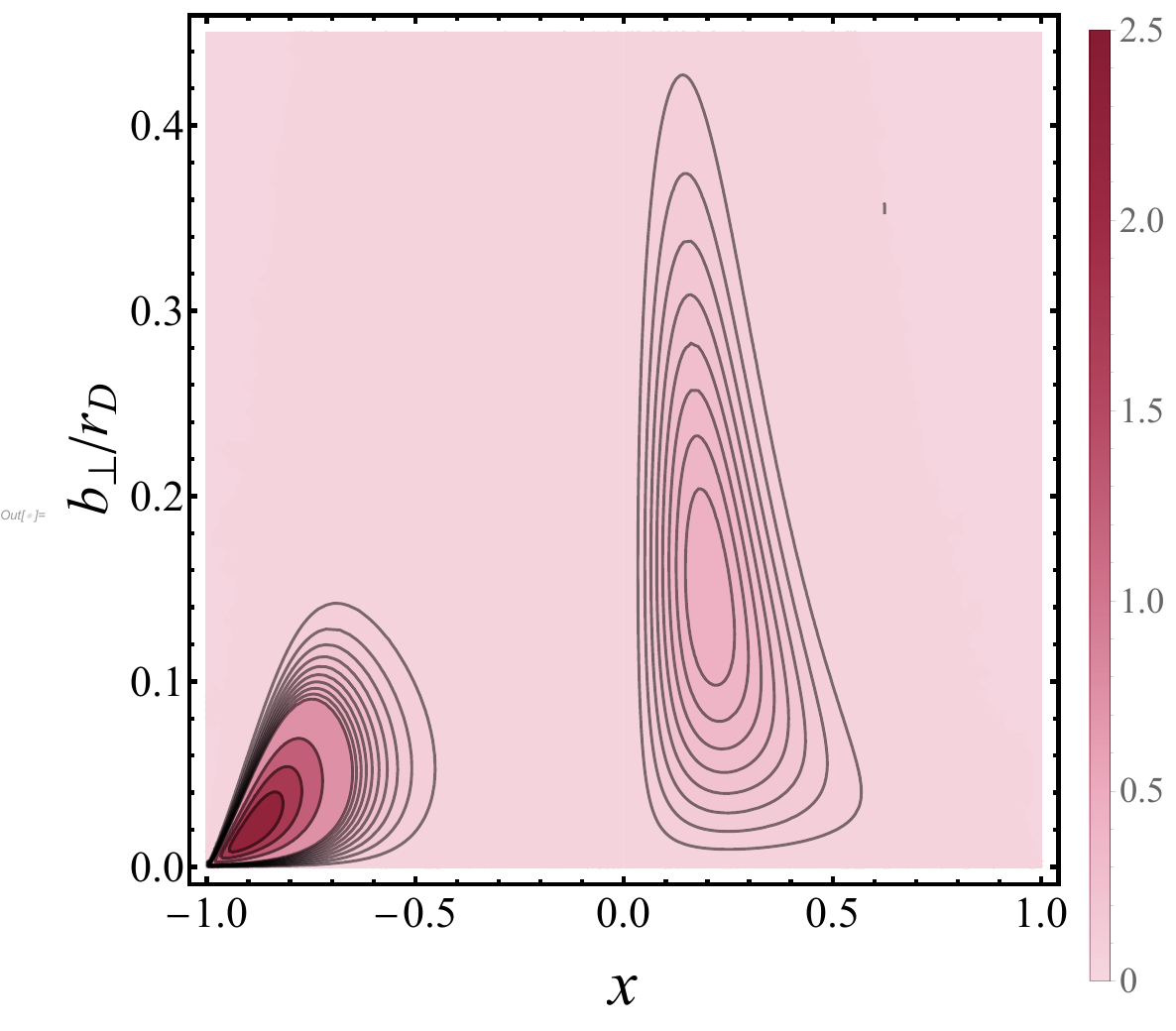}}
\subfloat[\centering]{\includegraphics[width=7.0cm]{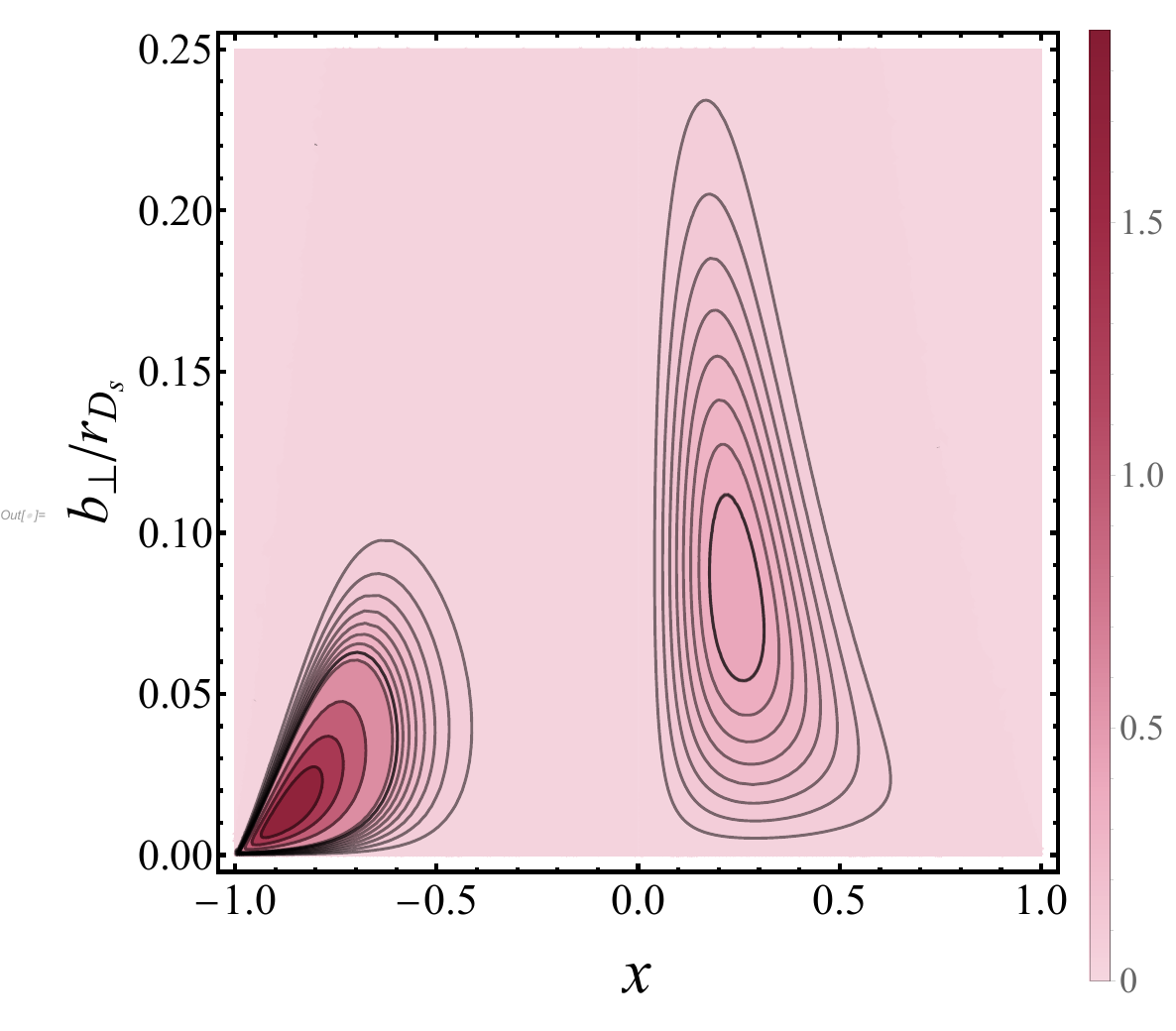}}\\
\subfloat[\centering]{\includegraphics[width=7.0cm]{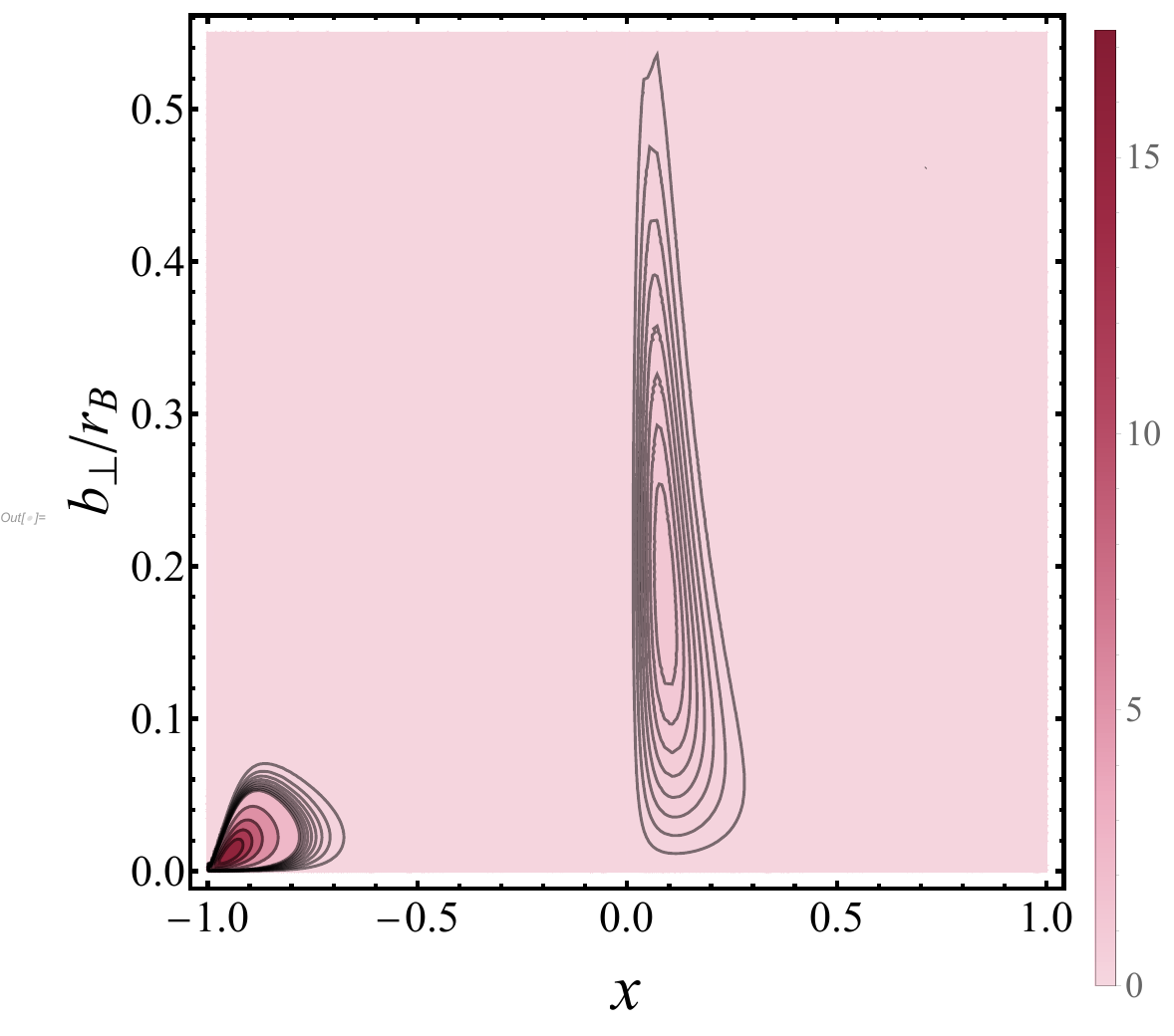}}
\subfloat[\centering]{\includegraphics[width=7.0cm]{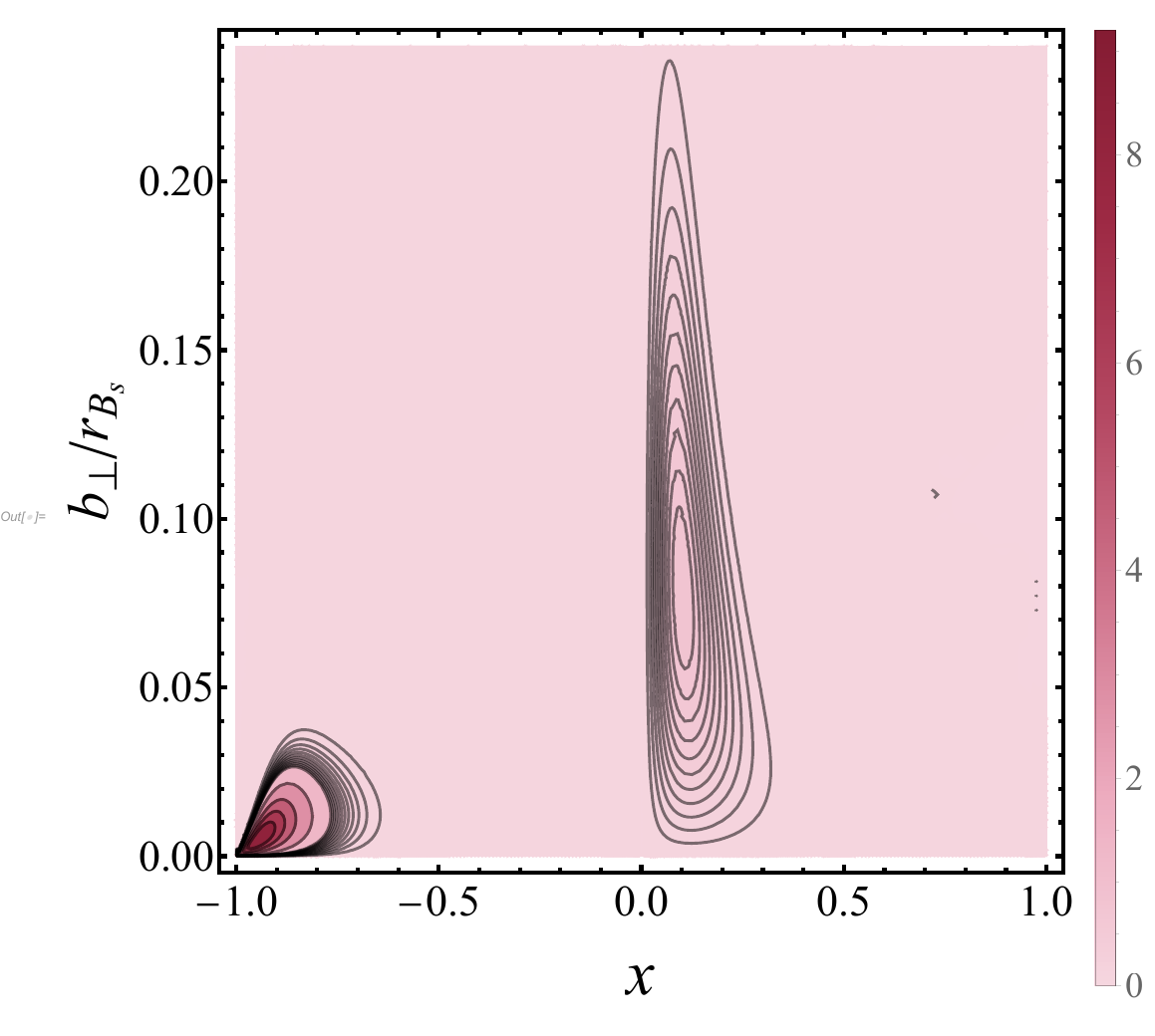}}\\
\subfloat[\centering]{\includegraphics[width=7.0cm]{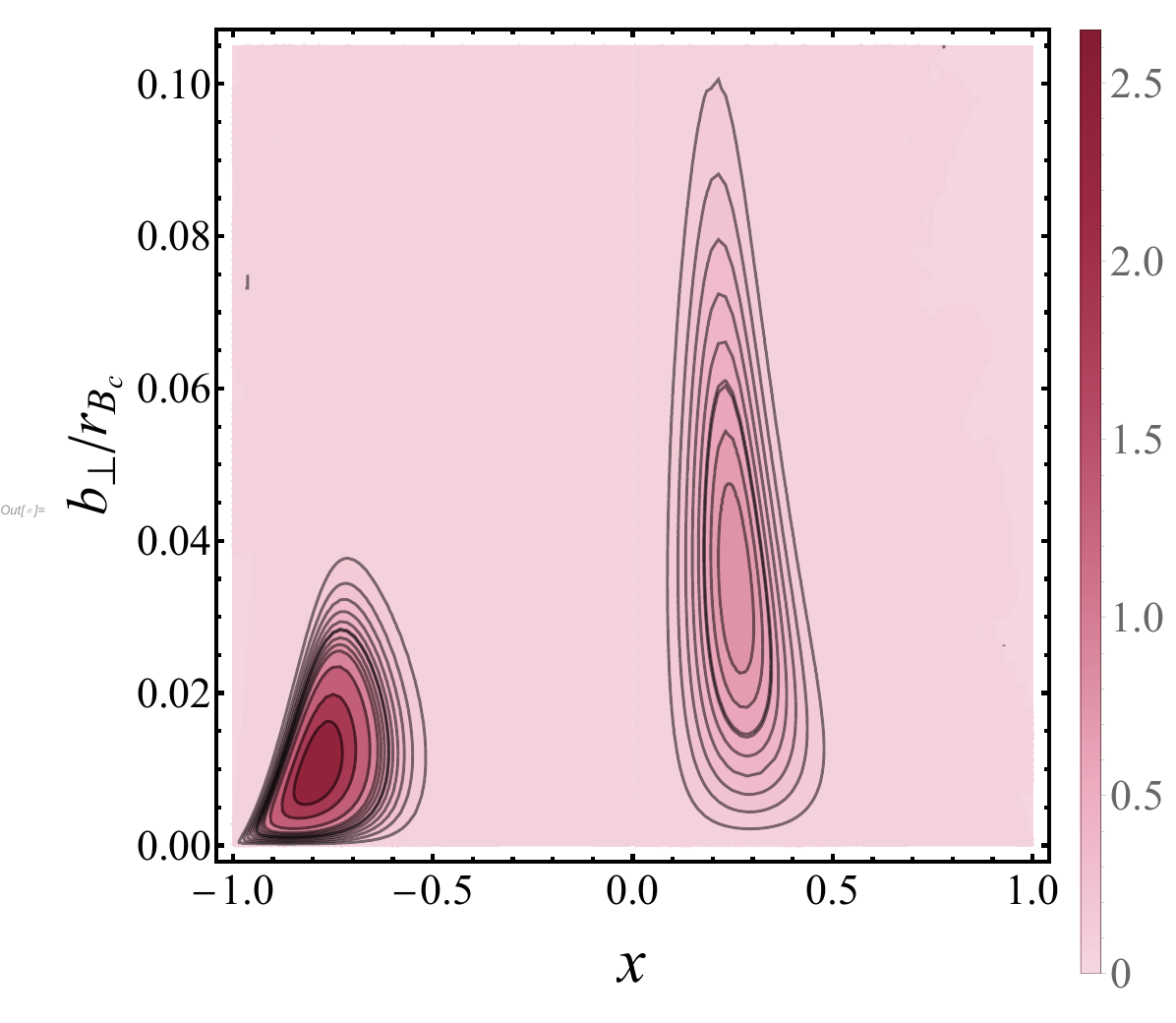}}
\caption{\label{fig:IPS-GPD_mesons}
Impact-parameter dependent GPDs, where the quark and antiquark contributions are assigned to the regions $x>0$ and $x<0$, respectively. The resulting left--right asymmetry reflects the mass imbalance between the dressed valence constituents, with the heavier quark exerting a dominant influence on the transverse center of momentum.}
\end{figure}

Figure~\ref{fig:IPS-GPD_mesons} presents the IPS-GPDs for the $D$, $D_s$, $B$, $B_s$, and $B_c$ mesons, where the quark and antiquark contributions populate the regions $x>0$ and $x<0$, respectively. For the charmed systems, the heavy antiquark is predominantly localized near the transverse center of momentum, whereas the light quark exhibits its highest probability at transverse distances of approximately $0.14\,r_D$ and $0.08\,r_{D_s}$. As the mass of the light valence quark increases from $d$ to $s$, the corresponding IPS-GPD becomes broader in the longitudinal momentum fraction $x$ while simultaneously narrowing in impact-parameter space, accompanied by a reduction in the peak magnitude. This behavior signals a stronger transverse localization associated with heavier constituents.

An analogous but more pronounced trend is observed in the bottom sector, reflecting the larger mass hierarchy between the valence quarks. In these systems, the heavy antiquark dominates the transverse center of momentum to an even greater extent. The most probable transverse locations of the light quark occur at approximately $0.18\,r_B$, $0.07\,r_{B_s}$, and $0.035\,r_{B_c}$, respectively. In particular, the $B_c$ meson displays a highly compact transverse structure, with the charm quark and bottom antiquark located in close proximity within the transverse plane. Overall, increasing the constituent quark masses leads to IPS-GPDs that are increasingly concentrated toward the center of transverse momentum, broader in $x$, more localized in $b_\perp$, and characterized by a reduced peak height.

\subsection{Pseudoscalar Mesons: Heavy-heavy sector}

We now turn to the heavy--heavy pseudoscalar sector, focusing on the charmonium and bottomonium ground states, $\eta_c$ and $\eta_b$. Owing to their equal-mass valence constituents, these systems provide a particularly clean environment to investigate heavy-quark dynamics, free from flavor asymmetry effects that characterize heavy--light mesons. In this case, all distributions are symmetric under $x \leftrightarrow 1-x$, reflecting the identical momentum sharing between the quark and antiquark.

After establishing the algebraic framework for light pseudoscalar mesons, we now specialize the inputs of the algebraic model to the heavy--heavy sector, focusing on the charmonium and bottomonium ground states, $\eta_c$ and $\eta_b$. As in the light-meson case, our starting point is Eq.~\eqref{LFWF}, which provides a direct algebraic connection between the leading-twist LFWF and the corresponding PDA. Once the PDA is specified, the LFWF follows straightforwardly, furnishing a consistent representation of the internal structure of heavy quarkonia at the hadronic scale $\zeta_H$.

Given the demonstrated reliability of the Schwinger--Dyson equation (SDE) approach in determining heavy-quark PDAs, we adopt as inputs the results obtained within that framework~\cite{Cui:2020tdf,Ding:2015rkn,Albino:2022gzs}. In contrast to light and heavy--light mesons, the equal-mass nature of the valence constituents in $\eta_c$ and $\eta_b$ ensures that their PDAs are symmetric under $x\leftrightarrow 1-x$ and strongly concentrated around $x=1/2$, reflecting the suppression of endpoint configurations characteristic of heavy quarkonia. The explicit parametrizations of the PDAs employed in this work are summarized below, with $\bar{x}=1-x$,
\begin{eqnarray}
\nonumber
\phi_{\eta_c}^c(x)&=&9.222x\bar{x}\exp\left[-2.89(1-4x\bar{x}) \right],\\
\phi_K^u(x)&=&12.264x\bar{x}\exp\left[-6.25(1-4x\bar{x}) \right].
\end{eqnarray}

\begin{figure}[t]
\begin{adjustwidth}{-\extralength}{-2.50cm}
\centering
\subfloat[\centering]{\includegraphics[width=10.0cm]{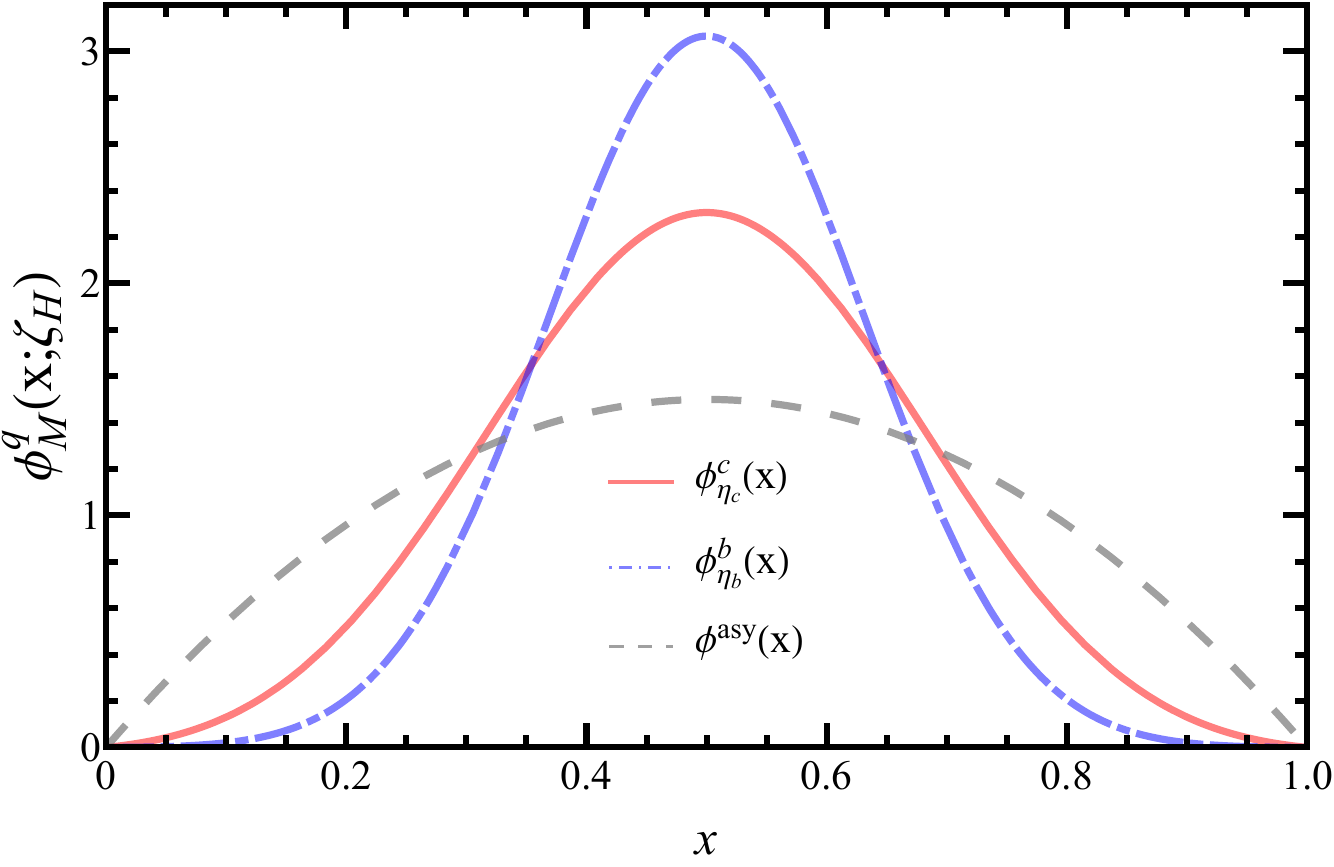}}
\end{adjustwidth}
\caption{Leading-twist parton distribution amplitudes of the heavy--heavy pseudoscalar mesons $\eta_c$ and $\eta_b$ at the hadronic scale $\zeta_H$. The distributions are symmetric under $x\leftrightarrow 1-x$ due to the equal-mass valence constituents and exhibit a pronounced localization around $x=1/2$, reflecting the increasingly nonrelativistic nature of heavy quarkonia as the quark mass increases.}
\label{fig:PDAs_etas}
\end{figure}

As illustrated in Fig.~\ref{fig:PDAs_etas}, the PDAs of $\eta_c$ and $\eta_b$ are significantly narrower than those of light pseudoscalar mesons, exhibiting a progressive compression as the quark mass increases. This behavior signals the transition toward the nonrelativistic regime, in which the heavy quark and antiquark share the longitudinal momentum almost equally. This behavior is well documented in modern continuum and lattice analyses and is faithfully captured within the algebraic model framework employed here \cite{Cui:2020tdf,Ding:2015rkn,Albino:2022gzs}.

To compute the heavy-quark LFWFs, we again fix the parameter $\nu=1$, ensuring the correct ultraviolet behavior of the Bethe-Salpeter wave function and maintaining consistency with previous DSE--BSE studies~\cite{Roberts:1994dr}. The dressed heavy-quark masses are determined by benchmarking against experimental and theoretical constraints, including lattice-QCD determinations of quarkonium charge radii~\cite{Dudek:2007zz,Dudek:2006ej} and continuum analyses within the SDE framework. The resulting set of constituent masses and the corresponding static observables that define the algebraic model in the heavy--heavy sector are reported in Table~\ref{tab:params_etas}.

\begin{table}[h]
\centering
\caption{Model parameters: meson and constituent-quark masses (in GeV). The values of $M_q$ are determined using Eq.~\eqref{eq:crgeneral} together with the specified charge radii. The corresponding distribution amplitudes employed in the calculations are listed in Eq.~\eqref{eq:PDAsSDE}.}
\label{tab:params_etas}
\begin{tabular}[t]{c|c|l||c|c}
\hline
Meson\;\; & $m_{\mathrm{M}}$ \;\;& \;$r_{\mathrm{M}}$ (in fm) \;& Quark\;\; & $M_q$ \;\;\\
\hline
$\eta_c$ & 0.14 &\; 0.255~\cite{ParticleDataGroup:2020ssz,Chang:2013nia}\; & $c$ & 1.65\\
$\eta_b$ & 0.49 &\; 0.088~\cite{Eichmann:2019bqf,Miramontes:2021exi,Raya:2022ued} \;& $b$ & 5.09\\
\hline
\end{tabular}
\end{table}

\begin{figure}[H]
\begin{adjustwidth}{-\extralength}{-2.50cm}
\centering
\subfloat[\centering]{\includegraphics[width=9.0cm]{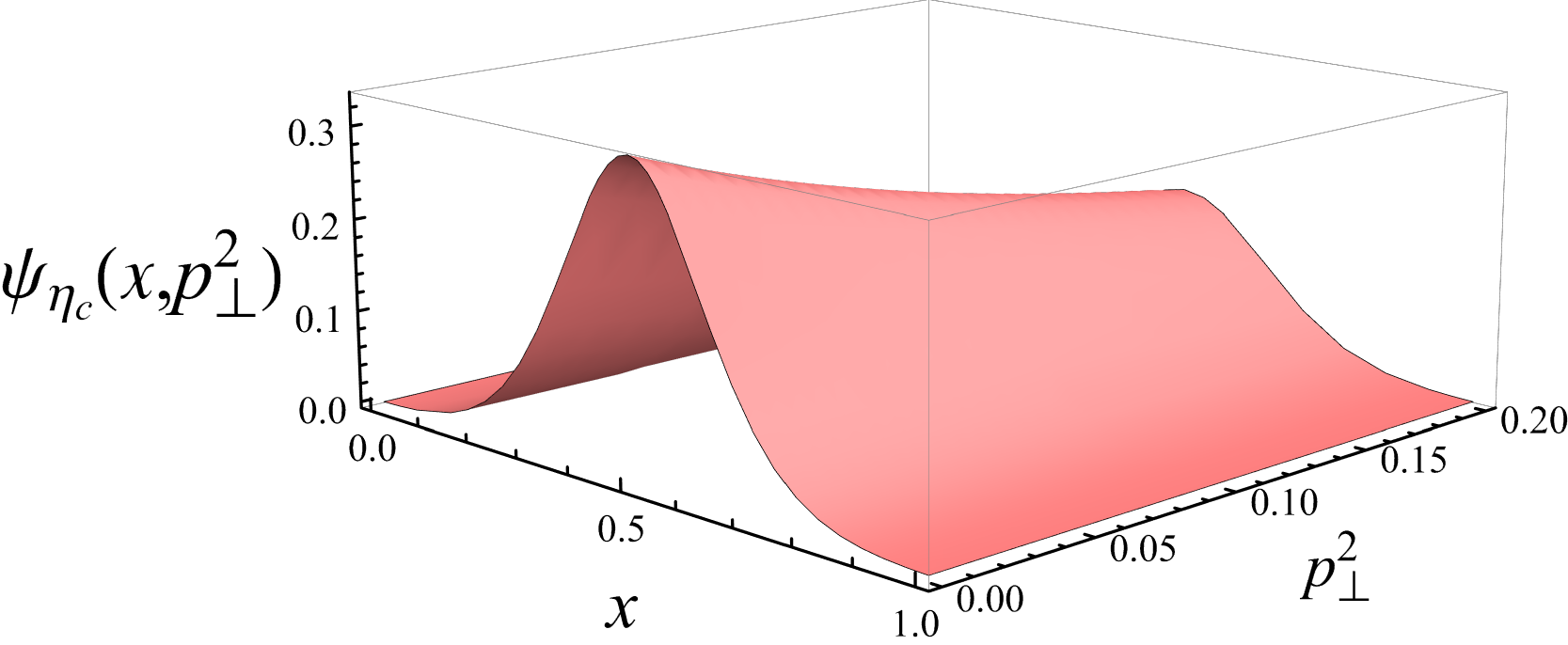}}\\
\subfloat[\centering]{\includegraphics[width=9.0cm]{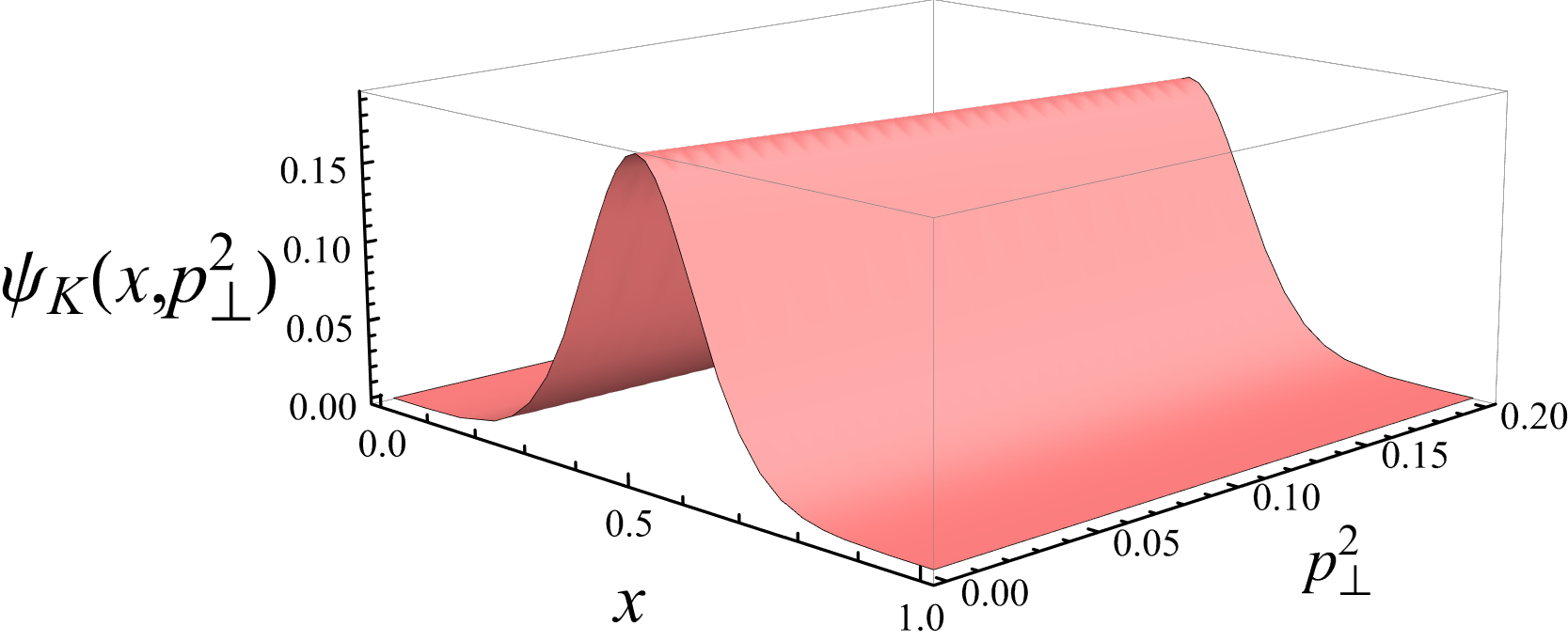}}
\end{adjustwidth}
\caption{ Light-front wave functions of the heavy--heavy pseudoscalar mesons $\eta_c$ (upper panel) and $\eta_b$ (lower panel), computed from Eq.~\eqref{LFWF} using the corresponding PDAs as inputs. Owing to the equal-mass valence constituents, the distributions are symmetric under $x\leftrightarrow 1-x$ and display strong localization around $x=1/2$. The increasing quark mass from charm to bottom leads to a narrower longitudinal profile and a slower falloff with increasing transverse momentum $k_\perp^2$, signaling the transition toward a more compact and less relativistic bound state.}
\label{fig:LFWF_etas}
\end{figure}

The resulting LFWFs for $\eta_c$ and $\eta_b$, shown in Fig. \ref{fig:LFWF_etas}, display a strong localization in both longitudinal momentum fraction $x$ and transverse momentum $k_\perp^2$. As the quark mass increases from charm to bottom, the LFWFs become increasingly compressed in $x$ and exhibit a markedly slower falloff in $k_\perp^2$, indicating a more compact transverse structure. This behavior reflects the reduced relativistic motion of heavy quarks and aligns with expectations from nonrelativistic QCD-inspired descriptions \cite{Albino:2022gzs}.

\begin{figure}[H]
\begin{adjustwidth}{-\extralength}{-2.50cm}
\centering
\subfloat[\centering]{\includegraphics[width=9.0cm]{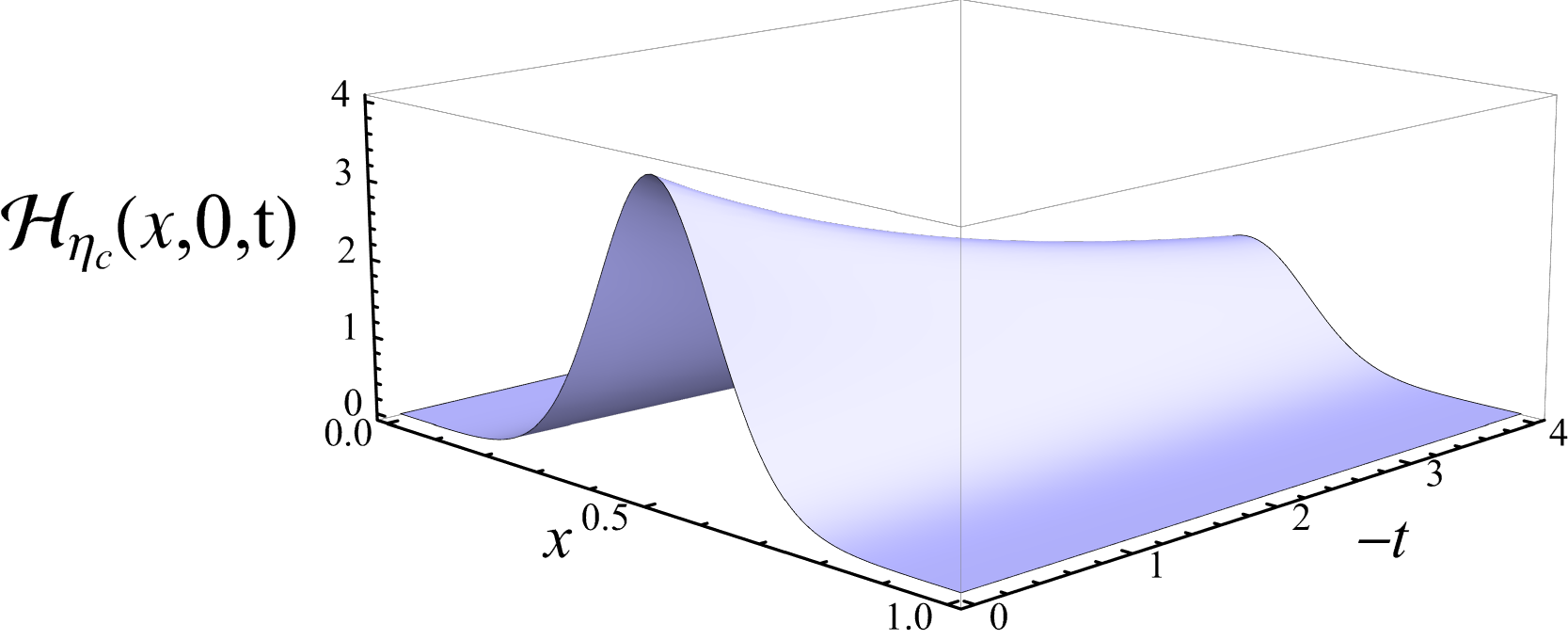}}\\
\subfloat[\centering]{\includegraphics[width=9.0cm]{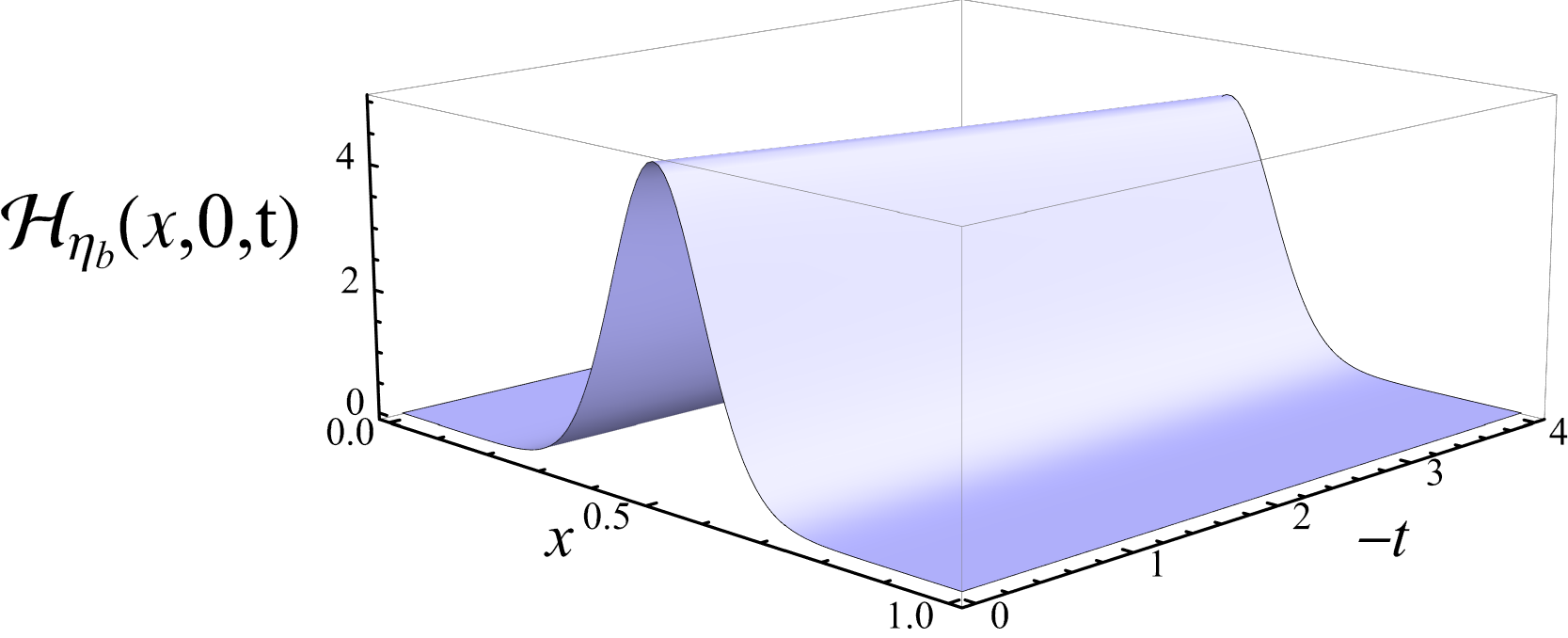}}
\end{adjustwidth}
\caption{ Valence-quark generalized parton distributions of the heavy--heavy pseudoscalar mesons $\eta_c$ and $\eta_b$ at zero skewness, $\xi=0$, obtained from the overlap representation of the corresponding LFWFs. Owing to the equal-mass valence constituents, the GPDs are symmetric under $x\leftrightarrow 1-x$ and sharply peaked around $x=1/2$. As the quark mass increases from charm to bottom, the distributions become narrower in $x$ and exhibit a harder dependence on the momentum transfer $t$, reflecting the increasingly compact spatial structure of heavy quarkonia.}
\label{fig:GPD_etas}
\end{figure}

The valence-quark GPDs for $\eta_c$ and $\eta_b$, Fig.\ref{fig:GPD_etas} are obtained through the overlap representation of the corresponding LFWFs. At zero skewness, the resulting GPDs are symmetric in $x$ and exhibit a narrow longitudinal profile centered at $x=1/2$. Furthermore, the $t$-dependence becomes increasingly hard as the quark mass grows, with $\eta_b$ showing a significantly slower decrease with $-t$ than $\eta_c$. This reflects the smaller spatial extent of bottomonium compared to charmonium and is consistent with previous continuum and lattice-based studies \cite{Albino:2022gzs}.

\begin{figure}[H]
\begin{adjustwidth}{-\extralength}{-2.50cm}
\centering
\subfloat[\centering]{\includegraphics[width=10.0cm]{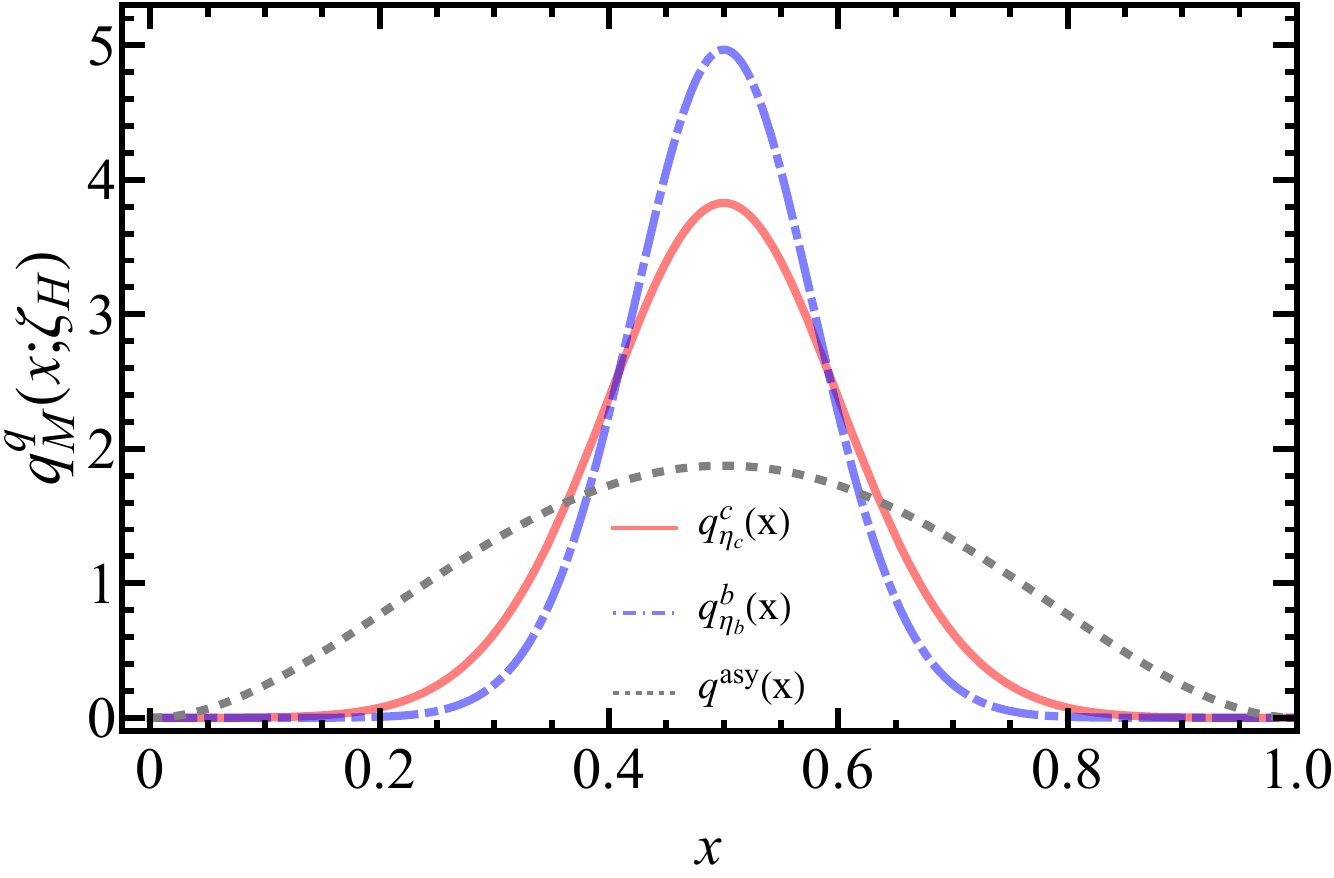}}
\end{adjustwidth}
\caption{ Valence-quark parton distribution functions of the heavy--heavy pseudoscalar mesons $\eta_c$ and $\eta_b$, obtained as the forward limit of the corresponding generalized parton distributions, $q(x)=H(x,0,0)$. Due to the equal-mass valence constituents, the PDFs are symmetric under $x\leftrightarrow 1-x$ and sharply peaked around $x=1/2$. The distribution becomes progressively narrower as the quark mass increases from charm to bottom, reflecting the suppression of endpoint configurations and the transition toward the nonrelativistic regime in heavy quarkonia.}
\label{fig:PDFs_etas}
\end{figure}

In the forward limit, the GPDs reduce to the valence-quark PDFs for $\eta_c$ and $\eta_b$. These PDFs, Fig. \ref{fig:PDFs_etas} are sharply peaked around $x=1/2$ and are substantially narrower than those of light and heavy--light mesons. The narrowing increases with the heavy-quark mass, signaling the dominance of configurations in which the quark and antiquark equally share the longitudinal momentum. This behavior highlights the transition toward the nonrelativistic regime in heavy quarkonia \cite{Albino:2022gzs}.

\begin{figure}[H]
\begin{adjustwidth}{-\extralength}{-2.50cm}
\centering
\subfloat[\centering]{\includegraphics[width=10.0cm]{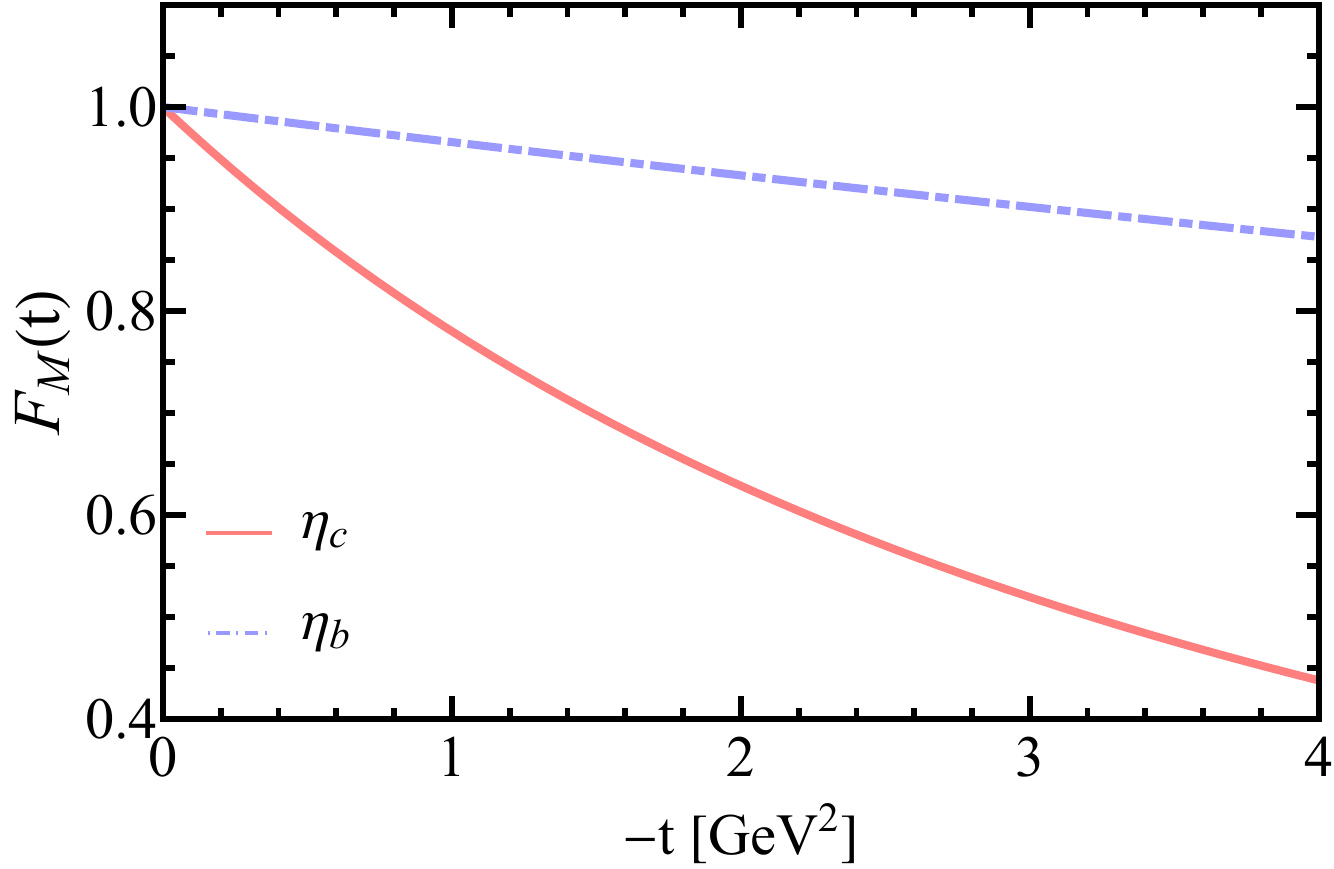}}
\end{adjustwidth}
\caption{Elastic electromagnetic form factors of the heavy--heavy pseudoscalar mesons $\eta_c$ and $\eta_b$, obtained from the zeroth Mellin moment of the corresponding GPD. The form factors exhibit a slow decrease with increasing momentum transfer $Q^2$, reflecting the compact spatial structure of heavy quarkonia. As the quark mass increases from charm to bottom, the EFF becomes harder, indicating a further reduction of the transverse spatial extent and a smaller charge radius for the $\eta_b$ compared to the $\eta_c$.}
\label{fig:EFFs_etas}
\end{figure}

The EFFs are obtained from the zeroth Mellin moment of the GPDs and are shown in Fig. \ref{fig:EFFs_etas}. For $\eta_c$ and $\eta_b$, the form factors decrease slowly with increasing momentum transfer, reflecting their compact spatial structure. The corresponding charge radii are found to be small and decrease significantly from $\eta_c$ to $\eta_b$, in agreement with lattice QCD determinations and continuum model predictions. The algebraic model results reproduce these trends quantitatively, reinforcing the role of quark mass in shaping the spatial size of heavy quarkonia \cite{Dudek:2006ej,Dudek:2007zz,Albino:2022gzs}.

Finally, the IPS-GPDs provide a spatial imaging of the heavy quarkonia in the transverse plane. Due to flavor symmetry, the quark and antiquark distributions are identical and centered at the transverse center of momentum. As the quark mass increases, the IPS-GPDs become increasingly localized in $b_\perp$, with $\eta_b$ exhibiting a notably narrower transverse profile than $\eta_c$. This confirms that heavier quarkonia are more compact objects, both longitudinally and transversely, and further illustrates the strong correlation between momentum and spatial distributions encoded in GPDs \cite{Albino:2022gzs}.

\begin{figure}[H]
\centering
\subfloat[\centering]{\includegraphics[width=5.8cm]{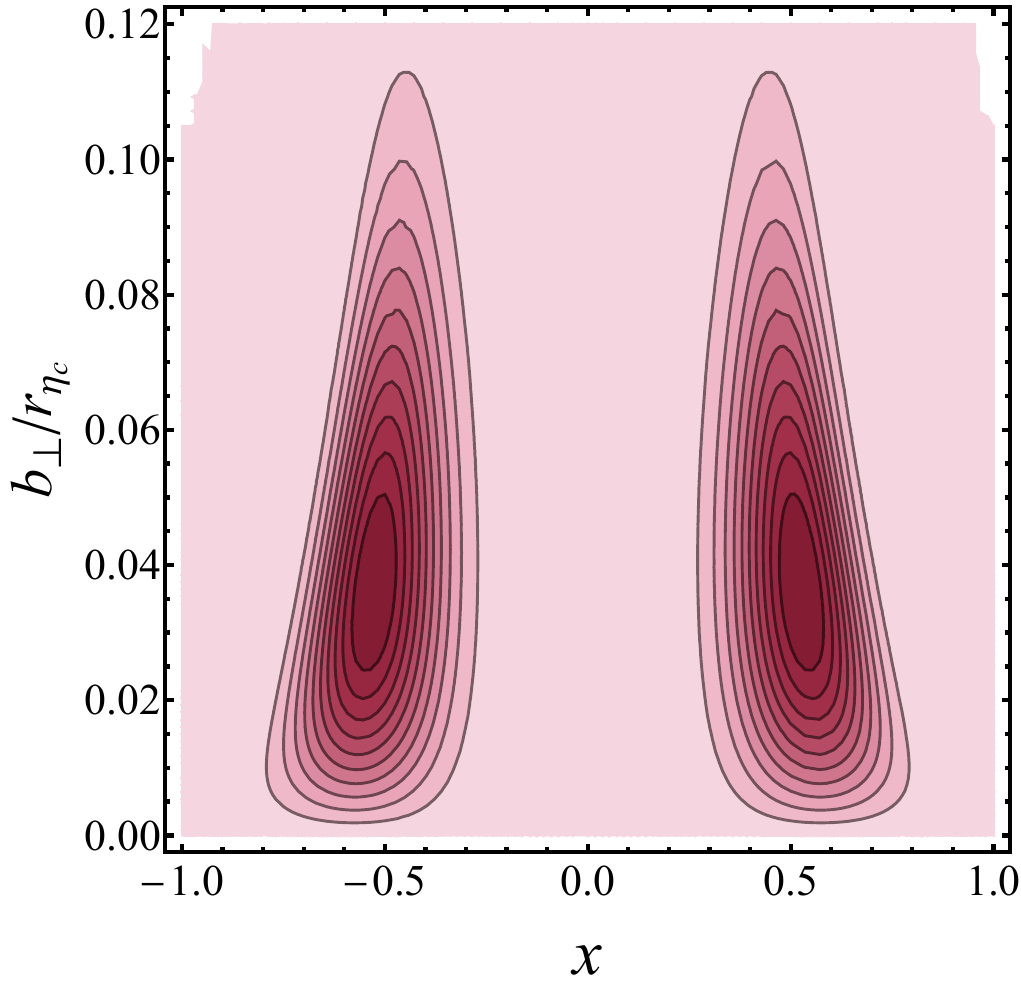}}
\hfill
\subfloat[\centering]{\includegraphics[width=6.0cm]{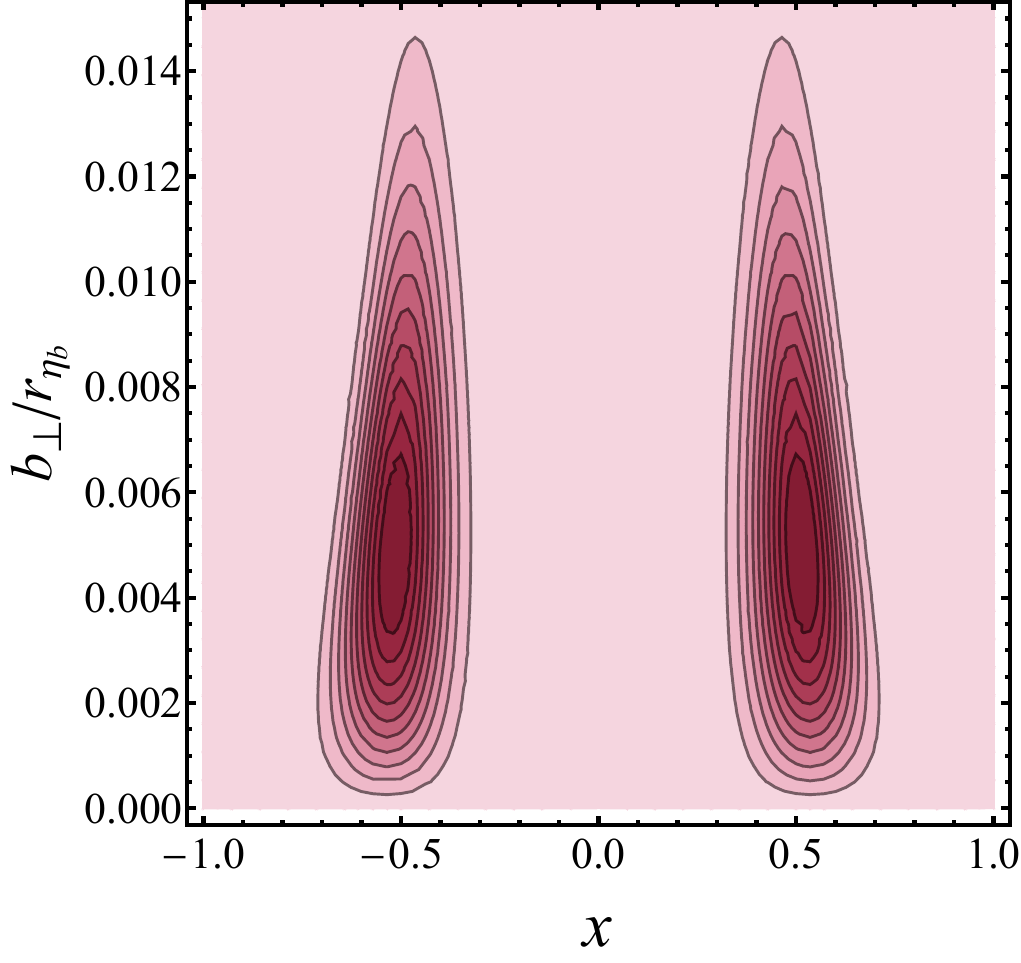}}
\caption{Impact-parameter space generalized parton distributions of the heavy--heavy pseudoscalar mesons $\eta_c$ and $\eta_b$, obtained from the Fourier transform of the zero-skewness GPDs. Owing to the equal-mass valence constituents, the distributions are symmetric in $x$ and centered at the transverse center of momentum. As the quark mass increases from charm to bottom, the IPS-GPD becomes more localized in impact-parameter space, signaling a progressively more compact transverse spatial structure in heavy quarkonia.}
\end{figure}

\section{Conclusions}
In this review we have presented a unified analysis of the internal structure of pseudoscalar mesons across all quark-mass regimes—light, heavy--light, and heavy--heavy—within an algebraic model formulated in the light-front framework. The approach provides a transparent and consistent connection between leading-twist parton distribution amplitudes, light-front wave functions, generalized parton distributions, parton distribution functions, elastic electromagnetic form factors, charge radii, and impact-parameter dependent GPDs.

For light pseudoscalar mesons, such as the pion and kaon, the results highlight the central role of nonperturbative QCD dynamics and flavor-symmetry breaking. The broad PDAs and LFWFs, together with the pronounced asymmetries observed in the kaon sector, reflect the influence of explicit quark-mass differences. These features propagate consistently through the corresponding PDFs, GPDs, and IPS-GPDs, providing a coherent picture that links longitudinal momentum distributions with transverse spatial structure \cite{Cui:2020tdf,Ding:2015rkn,Diehl:2003ny}.

In the heavy--light sector, encompassing the $D$, $D_s$, $B$, $B_s$, and $B_c$ mesons, the framework captures the systematic transition induced by increasing quark-mass asymmetry. As the heavy-quark mass grows, the distributions become increasingly asymmetric in the longitudinal momentum fraction, while simultaneously exhibiting stronger localization in impact-parameter space. The corresponding electromagnetic form factors become harder and the extracted charge radii decrease, in agreement with lattice-QCD calculations and continuum studies \cite{Dudek:2007zz,Dudek:2006ej,Chang:2013nia}. These results emphasize the importance of momentum-dependent interactions in reproducing realistic spatial distributions, in contrast with momentum-independent approaches such as the contact interaction.

For heavy--heavy pseudoscalar mesons, $\eta_c$ and $\eta_b$, the equal-mass nature of the valence constituents restores symmetry under $x\leftrightarrow 1-x$ in all partonic distributions. The PDAs, LFWFs, and PDFs are sharply peaked around $x=1/2$, signaling the suppression of endpoint configurations and the onset of nonrelativistic dynamics. Correspondingly, the GPDs exhibit a hard dependence on the momentum transfer and the IPS-GPDs reveal highly compact transverse spatial profiles, with $\eta_b$ being significantly more localized than $\eta_c$. These findings are consistent with expectations from lattice QCD and Dyson--Schwinger equation studies \cite{Dudek:2007zz,Dudek:2006ej,Albino:2022gzs}.

Overall, the algebraic model reviewed here offers a flexible and symmetry-consistent framework that bridges momentum-space and coordinate-space descriptions of hadron structure. By enabling the simultaneous description of a wide range of observables within a single formalism, it provides valuable insight into how quark masses and symmetries shape the three-dimensional structure of pseudoscalar mesons. The results reviewed in this work underscore the utility of such algebraic approaches as complementary tools to lattice QCD and continuum functional methods, and they offer a solid foundation for future studies of more complex hadronic systems, including vector mesons and baryons.

\vspace{6pt} 

\authorcontributions{Conceptualization, A.B., J.J.C.M. and J.S.; methodology, A.B. and J.S.; software, B.A. and L.A.; validation, B.A., L.A. and J.S.; formal analysis, A.C.-A, T.A. and M.C.-C; investigation, B.A. and L.A.; resources, B.A., L.A., A.B., J.J.C.M. and J.S.; writing---original draft preparation, B.A. and J.S.; writing---review and editing, B.A., L.A., A.B., J.J.C.M. and J.S.; visualization, B.A., L.A., A.B., J.J.C.M. and J.S.; supervision, A.B., J.J.C.M. and J.S.; project administration, A.B., J.J.C.M. and J.S.; funding acquisition, A.B., J.J.C.M. and J.S. All authors have read and agreed to the published version of the manuscript.}

\funding{B. Almeida Zamora acknowledges SECIHTI (CVU No.~935777) for PhD and Postdoctoral fellowships.
L.A. acknowledges financial support from SECIHTI (CVU No.~446053) through the program ``Posdoctorados por México'', and Ayuda B3 ``Ayudas para el desarrollo de líneas de investigación propias" del V Plan Propio de Investigación y Transferencia 2018-2020 de la Universidad Pablo de
Olavide, de Sevilla. 
A.B. wishes to acknowledge the Coordinación de la Investigación Científica of the Universidad Michoacana de San Nicolás de Hidalgo, Morelia, Mexico, grant no. 4.10, the Consejo Nacional de Humanidades, Ciencias y Tecnologías, Mexico, project CBF2023-2024-3544 as well as the Beatriz Galindo support during his scientific stay at the University of Huelva, Huelva, Spain. 
J.J. Cobos-Martínez acknowledges financial support from the University of Sonora under grant USO315007861. 
Otherwise, this work has been partially financed by Ministerio Español de Ciencia e Innovación under grant No. PID2019-107844GBC22; Junta de Andalucía under contract Nos. PAIDI FQM-370 and PCI+D+i under the title: ``Tecnologías avanzadas para la exploración del universo y sus componentes" (Code AST22-0001).}

\institutionalreview{Not applicable.}

\informedconsent{Not applicable.}

\dataavailability{The data are not publicly available. The data are
available from the authors upon reasonable request.} 

\acknowledgments{The authors acknowledges the use of the computer facilities of C3UPO at the Universidad Pablo de Olavide, de Sevilla.}

\begin{adjustwidth}{-\extralength}{0cm}

\reftitle{References}





\bibitem{Schmidt:1994di}
S.~M.~Schmidt, D.~Blaschke and Y.~L.~Kalinovsky,
Phys. Rev. C \textbf{50} (1994), 435-446
doi:10.1103/PhysRevC.50.435

\bibitem{Roberts:2012sv}
C.~D.~Roberts,
IRMA Lect. Math. Theor. Phys. \textbf{21} (2015), 355-458
[arXiv:1203.5341 [nucl-th]].

%


\PublishersNote{}
\end{adjustwidth}
\end{document}